\documentclass[preprint,12pt,authoryear]{elsarticle}

\usepackage{amssymb}
\usepackage{amsmath}
\usepackage{booktabs}
\usepackage{amsthm}
\usepackage{graphicx}
\usepackage{subfig} 
\usepackage{bbm} 
\usepackage{enumerate}
\usepackage{algorithm}
\usepackage{algpseudocode}
\usepackage[table]{xcolor} 

\usepackage{url} 
\usepackage{float}
\usepackage{xcolor}
\usepackage{comment}
\usepackage{multirow}
\usepackage{amsmath,bm}

\DeclareMathOperator*{\argmin}{arg\,min}

\journal{Computational Statistics \& Data Analysis}
\usepackage{xcolor}
\usepackage{amsmath}
\usepackage{amsthm}
\usepackage{verbatim}
\usepackage[font=small,labelfont=bf]{caption}
\usepackage{setspace}
\usepackage{hyperref}
\usepackage{multicol}

\definecolor{olive}{rgb}{0.3, 0.4, .1}
\definecolor{fore}{RGB}{249,242,215}
\definecolor{back}{RGB}{51,51,51}
\definecolor{title}{RGB}{255,0,90}
\definecolor{dgreen}{rgb}{0.,0.6,0.}
\definecolor{gold}{rgb}{1.,0.84,0.}
\definecolor{JungleGreen}{cmyk}{0.99,0,0.52,0}
\definecolor{BlueGreen}{cmyk}{0.85,0,0.33,0}
\definecolor{RawSienna}{cmyk}{0,0.72,1,0.45}
\definecolor{Magenta}{cmyk}{0,1,0,0}

\newcounter{savedenum}

\newcommand\bzero{\mbox{\boldmath${0}$}}
\newcommand\bbe{\mbox{\boldmath${ \beta}$}}

\newcommand\bpsi{\mbox{\boldmath${\psi}$}}
\newcommand\bgPhi{\mbox{\boldmath${\Phi}$}}

\newcommand\btheta{\mbox{\boldmath${\theta}$}}

\newcommand\bdel{\mbox{\boldmath${\delta}$}}
\newcommand\bep{\mbox{\boldmath${\epsilon}$}}
\newcommand\bomega{\mbox{\boldmath${\omega}$}}

\newcommand\bSig{\mbox{\boldmath${\Sigma}$}}

\newcommand\bB{{\bf B}}

\newcommand\bI{{\bf I}}

\newcommand\mcN{{\mathcal N}}

\newcommand\bS{{\bf S}}

\newcommand\bs{{\bf s}}

\newcommand\bW{{\bf W}}

\newcommand\bX{{\bf X}}
\newcommand\bx{{\bf x}}

\newcommand\bZ{{\bf Z}}

\newcommand\mbE{{\mathbb E}}

\newcommand{\tb}[1]{\textbf{#1}}

\newcommand{\tblue}[1]{\textcolor{blue}{#1}}

\newcommand{\ssq}{\sigma^{2}}

\newcounter{saveenumi}

\begin{document}

\begin{frontmatter}


\author[PU]{Jin Hyung Lee} 
\author[GMU]{Ben Seiyon Lee\corref{slee}}

\affiliation[PU]{organization={Purdue University},
            addressline={250 N. University Street}, 
            city={West Lafayette},
            postcode={47907}, 
            state={IN},
            country={United States}}
\affiliation[GMU]{organization={George Mason University},
            addressline={4400 University Drive}, 
            city={Fairfax},
            postcode={22030}, 
            state={VA},
            country={United States}}

\cortext[slee]{Corresponding author. \textit{Postal address:} 4400 University Drive, MS 4A7, Fairfax, VA 22030, USA. 
\textit{Email:} slee287@gmu.edu. \textit{Phone:} 703-993-9113}

\title{A Scalable Variational Bayes Approach for Fitting Non-Conjugate Spatial Generalized Linear Mixed Models via Basis Expansions}

\bigskip
\begin{abstract}
Large spatial datasets with non-Gaussian responses are increasingly common in environmental monitoring, ecology, and remote sensing, yet scalable Bayesian inference for such data remains challenging. Markov chain Monte Carlo methods are often prohibitive for large datasets, and existing variational Bayes methods rely on conjugacy or strong approximations that limit their applicability and can underestimate posterior variances. A scalable variational framework that incorporates semi-implicit variational inference (SIVI) with basis representations of spatial generalized linear mixed models, which may not have conjugacy, is proposed. The proposed framework accommodates gamma, negative binomial, Poisson, Bernoulli, and Gaussian responses on continuous spatial domains. Across 20 simulation scenarios with 50,000 locations, SIVI achieves predictive accuracy and posterior distributions comparable to Metropolis--Hastings and Hamiltonian Monte Carlo while providing notable computational speedups. Applications to remotely-sensed land surface temperature and blue jay abundance further demonstrate the utility of the approach for large non-Gaussian spatial datasets.
\end{abstract}

\begin{keyword}
Variational Bayes \sep Spatial Statistics \sep Semi-Implicit Variational Inference \sep Basis Representation \sep Non-Gaussian Spatial Data


\end{keyword}

\end{frontmatter}

\section{Introduction}

Large Gaussian and non-Gaussian spatial datasets with inherent spatial dependencies arise in numerous disciplines, including economics \citep{redding2017quantitative}, hydrology \citep{zhou2020geo}, public health \citep{rushton2003public}, and genetics \citep{wagner2013conceptual}. Advances in data collection technologies have enabled the acquisition of spatially indexed datasets comprising millions of observation locations, often over non-stationary and heterogeneous spatial domains.

To analyze such spatial data, a widely used approach is the spatial generalized linear mixed model (SGLMM) \citep{diggle1998model}. SGLMMs provide a flexible framework for modeling data with spatial random effects and have been extensively applied to both Gaussian and non-Gaussian spatially correlated datasets \citep{bonat2016practical, lee2023scalable, zilber2021vecchia}. The spatial random effects are typically modeled as latent Gaussian processes (GPs) with a specified spatial covariance function. Within a Bayesian hierarchical modeling framework \citep{wikle1998hierarchical}, posterior inference is commonly performed using Markov chain Monte Carlo (MCMC) methods. However, when the latent variables are high-dimensional and exhibit strong spatial dependence, the Markov chains tend to mix slowly \citep{haran2003accelerating}, and the associated computational costs scale cubically with the number of locations.

Scalable approaches for spatial models have been developed including low-rank and basis representations \citep{cressie2008fixed, banerjee2008gaussian, katzfuss2017multi, higdon1998process, cressie2015statistics} and methods that exploit sparsity in covariance or precision matrices, such as the nearest neighbor Gaussian process (NNGP) \citep{datta2016hierarchical}, Vecchia approximations \citep{katzfuss2017multi}, and covariance tapering \citep{furrer2006covariance}. Other strategies include integrated nested Laplace approximations (INLA) \citep{rue2009approximate}, as well as conjugate spatial models such as P\'olya–Gamma mixtures \citep{polson2013bayesian} and latent conjugate models \citep{bradley2020bayesian}. While these approaches offer substantial computational gains, many still require MCMC-based posterior sampling, which becomes prohibitive for very large datasets, or depend on normality and Laplace approximations that are known to underestimate uncertainty \citep{ferkingstad2015improving}.

Variational Bayes (VB) \citep{blei2006variational, fox2012tutorial, jordan1999introduction} is an optimization approach that approximates the target distribution (e.g., posterior distribution) by specifying a variational distribution that minimizes the Kullback–Leibler (KL) divergence from the target. Variants such as mean-field variational Bayes (MFVB) \citep{mohammad2009variational, han2019statistical, blei2006variational}, hybrid MFVB methods \citep{wang2013variational, tran2021practical}, and integrated non-factorized variational Bayes (INFVB) \citep{han2013integrated, lee2024scalable} have become popular among practitioners for analyzing large datasets efficiently. Semi-implicit variational inference (SIVI) \citep{yin2018semi} defines the variational mixing distribution through a neural network; thereby providing the flexibility needed to adequately model dependence in multivariate posterior distributions.

However, the application of VB methods to continuous spatial domains for non-Gaussian SGLMMs remains limited. \citet{ren2011variational} applied VB to small Gaussian spatial datasets. \citet{wu2018fast} use integrated non-factorized variational Bayes (INFVB) for Gaussian areal data, and \citet{bansal2021fast} and \citet{parker2022computationally} modeled count and binary areal data. \citet{cao2023variational} employed a variational approach with a sparse inverse Cholesky representation for the latent Gaussian process, achieving faster convergence compared to alternative methods. \citet{lee2024scalable} propose a scalable INFVB approach for modeling non-Gaussian spatial data, but their approach is limited to spatial data from Gaussian, Bernoulli, and Poisson data distributions and relies on conjugate or near-conjugate approximations. \citet{song2025fast} embed NNGP within an MFVB framework; however, their approach applies only to Gaussian responses and inherits key limitations of MFVB, including block-independence assumptions and variational functions that can underestimate posterior variances \citep{blei2006variational,han2013integrated}. \citet{garneau2025semi} incorporate NNGP within SIVI, but their framework is restricted to Gaussian and small Poisson datasets (e.g., $n=500$) and does not cover other response types such as negative binomial, gamma, or Bernoulli. Moreover, their large-scale application focuses on Gaussian spatial data, where the latent spatial random effects are readily integrated out; thus, notably reducing the number of estimable model parameters.

We propose a scalable variational framework that integrates SIVI with spatial basis representations to enable fast Bayesian inference for large-scale, continuous-domain SGLMMs across a range of non-Gaussian response types. This study focuses on spatial models defined over continuous spatial domains (i.e., point-referenced data), rather than areal data models such as Markov random fields or CAR/SAR models~\citep{cressie2015statistics}. By combining an implicit neural mixing distribution with an explicit Gaussian variational layer for the basis coefficients and fixed effects, our SIVI–basis approach flexibly captures posterior dependencies while remaining computationally efficient. The methodology accommodates a broad class of response types, including negative binomial, gamma, Poisson, Bernoulli, and Gaussian, without relying on Laplace or conjugacy-based approximations. Through extensive simulation studies considering 20 scenarios and two real-world applications, we demonstrate that the proposed SIVI–basis framework achieves predictive accuracy and posterior summaries comparable to MCMC, while reducing computation time by an order of magnitude or more.

In the context of continuous-domain SGLMMs, there is a dearth of VB methods that simultaneously: (i) accommodate gamma and negative binomial responses with dispersion, alongside Bernoulli, Poisson, and Gaussian data; (ii) exploit basis representations to scale to large spatial datasets; and (iii) avoid ad-hoc Laplace or quadratic approximations required for conditional conjugacy. Our proposed framework combines SIVI with a basis representation of latent spatial processes, which allows joint estimation of regression effects, dispersion parameters, and latent spatial effects without relying on conjugacy-based approximations. This extends scalable variational inference to a wider range of spatial models, while still remaining computationally efficient for modeling large datasets.

The remainder of the paper is organized as follows. 
Section~\ref{sec:SIVISGLMMs} provides an overview of SGLMMs and their basis representation extension (basis-SGLMM). Section~\ref{Sec:SIVI} reviews variational Bayes and SIVI methods and introduces our proposed SIVI-based inference framework for basis-SGLMMs, including algorithmic details. Section~\ref{sec:SIVIsimulations} presents an extensive simulation study with comparisons to competing methods, and Section~\ref{Sec:Applications} applies the approach to two large non-Gaussian spatial datasets: one from remote sensing and one from the North American Breeding Bird Survey. Section~\ref{Sec:Discussion} concludes with a discussion of limitations, practical guidance, and directions for future research.

\section{Spatial Generalized Linear Mixed Models (SGLMMs)} \label{sec:SIVISGLMMs} 
Spatial generalized linear mixed models (SGLMMs) \citep{diggle1998model} are a widely used framework for modeling non-Gaussian spatial data, supporting diverse response types \citep{bonat2016practical} and spatially-correlated random effects. Let $\bZ = \{\bZ(\bs_i)\}_{i=1}^N$ denote the observations collected at spatial locations $\bs_i \in \bS \subseteq \mathbb{R}^2$, and let $\bX \in \mathbb{R}^{N \times p}$ be the corresponding matrix of covariates. Spatial dependence is introduced via the random effects $\bomega = \{\bomega(\bs_i)\}_{i=1}^N \in \mathbb{R}^N$, often modeled as a zero-mean Gaussian process with covariance function $C(\Psi)$, where $\Psi$ represents the covariance parameters. For a finite set of locations, the spatial random effects follow a multivariate normal distribution $\bomega \sim \mathcal{N}(\bzero, \bSig(\Psi))$, with covariance matrix $\bSig(\Psi) \in \mathbb{R}^{N \times N}$ such that $\bSig(\Psi)_{ij} = C(\Psi)_{ij}$ for sites $\bs_i$ and $\bs_j$. To simplify notation, we set $Z_i:=Z(\bs_i)$,  and $\omega_i:=\omega(\bs_i)$. The Bayesian hierarchical formulation of the SGLMM is
\begin{align}
\text{Data model:} \quad 
& Z_i \mid \bbe, \bomega, \gamma \;\stackrel{\text{ind}}{\sim}\; F(\,\cdot \mid \eta_i,\gamma\,),
   \quad i = 1,\dots,N, \nonumber \\
& g\!\left\{ \mathbb{E}\big(Z_i \mid \bbe, \omega_i\big) \right\} = \eta_i 
  = \bx(\bs_i)^\top \bbe + \omega_i, 
   \nonumber \\[0.5em]
\text{Process model:} \quad 
& \bomega \mid \Psi \;\sim\; \mathcal{N}_N\!\left(\bzero,\; \bSig(\Psi)\right), \label{eq:bhm_sglmm} \\
\text{Parameter model:} \quad 
& \bbe \sim p(\bbe), 
  \qquad \Psi \sim p(\Psi), \qquad \gamma \sim p(\gamma), \nonumber
\end{align}
\noindent where $F(\cdot)$ denotes the probability distribution of the response (e.g., Normal for Gaussian data or negative binomial for count data), and $g(\cdot)$ is a known link function applied componentwise. 
$\bbe \in \mathbb{R}^p$ denotes the regression coefficients. $\bx(\bs_i)$ is the row of $\bX$ corresponding to the location $\bs_i$. The additional parameter $\gamma$ represents the extra distributional parameter corresponding to each response type; for example, the nugget variance $\tau^2$ for the Gaussian model and the dispersion parameters $\alpha$ and $\kappa$ for the gamma and negative binomial models. In this study, we consider $\gamma \in \{\tau^{2},\, \kappa,\, \alpha\}$. The prior distributions are $p(\bbe)$, $p(\Psi)$, and $p(\gamma)$. If not stated otherwise, we assume prior independence, so that
$p(\bbe, \Psi,\gamma) = p(\bbe)\,p(\Psi)\,p(\gamma)$. In this study, we employ the logit link function for the Bernoulli case and the log link function for the Poisson, gamma, and negative binomial cases. 

In practice, SGLMMs face substantial computational challenges when applied to large datasets.  First, matrix operations involving dense $N \times N$ covariance matrices are prohibitive, as Cholesky decompositions require $\mathcal{O}(N^3)$ floating-point operations. Next, the spatial random effects $\{\omega_i\}_{i=1}^N$ are often strongly correlated, which can lead to poor mixing in Markov chain Monte Carlo (MCMC) algorithms \citep{haran2003accelerating}. Finally, SGLMMs are overparameterized, requiring inference for the full set of latent spatial effects $\omega_i$.

\subsection{Basis Representations of SGLMMs (Basis-SGLMM)}
\label{SubSec:BasisSGLMM}

To address computational issues associated with fitting SGLMMs to large datasets, basis representations of $\bomega$ (basis-SGLMM)\citep{higdon1998process, sengupta2013hierarchical, bradley2016comparison, lee2022picar, lee2023scalable} have been employed to reduce the dimensionality of $\bomega$, bypass costly operations on large matrices, and weaken the correlations across the spatial random effects $\omega_i$.

In basis-SGLMMs, $\omega(\bs_i)$ is represented as a basis expansion of $m$ spatial basis functions. Specifically, $\bomega \approx \bgPhi\,\bdel$, where $\bgPhi \in \mathbb{R}^{N \times m}$ is the matrix of basis functions where the $j$th column $\bgPhi_j$ contains the $j$th basis function evaluated at all locations $\bs_i$ and $\bdel \in \mathbb{R}^m$ denotes the corresponding basis coefficients.

The Bayesian hierarchical model for the basis representation of SGLMMs is given by
\begin{align}
\text{Data model:} \quad 
& Z_i \mid \bbe, \bdel \;\stackrel{\text{ind}}{\sim}\; 
  F\!\left(\,\cdot \;\middle|\; \eta_i, \gamma \right),
  \qquad i = 1,\dots,N, \nonumber \\[0.25em]
& g\!\left( \mathbb{E}[Z_i \mid \bbe, \bdel] \right) 
    = \eta_i 
    = \bx(\bs_i)^\top \bbe + \bgPhi(\bs_i)^\top \bdel, \nonumber \\[0.5em]
\text{Process model:} \quad 
& \bdel \mid \zeta \;\sim\; 
  \mathcal{N}_m\!\left( \mathbf{0},\, \bSig_{\delta}(\zeta) \right), \label{eq:basisSGLMM_bhm} \\[0.5em]
\text{Parameter model:} \quad 
& \bbe \sim p(\bbe), 
  \qquad 
  \zeta \sim p(\zeta), \qquad \gamma\sim p(\gamma), \nonumber
\end{align}
where $\bSig_{\delta}(\zeta)$ denotes the prior covariance matrix for the basis coefficients $\bdel$, with covariance parameters $\zeta$. The prior distribution for $\zeta$ is $p(\zeta)$. $\bgPhi(\bs_i)$ denotes the row of $\bgPhi$ corresponding to the location $\bs_i$. We assume that the basis functions comprising $\bgPhi$ are fixed prior to model fitting. 

A key advantage of the basis representation is the substantial dimension reduction it affords, since $\bdel \in \mathbb{R}^m$ with $m \ll N$. This reduction not only lowers computational costs ($\mathcal{O}(Nm)$) for fitting the model but also weakens the dependence among the components of $\bdel$, leading to faster-mixing Markov chains \citep{haran2003accelerating}. MCMC algorithms can be prohibitive in large-$N$ since each iteration of the algorithm becomes computationally expensive, often resulting in low effective samples per second and long walltimes. The low-dimensional parameter vector $(\bbe, \bdel, \log \sigma^{2}, \gamma)$ is the focus of our SIVI approximation in Section~\ref{Sec:SIVI}.

\section{Semi-Implicit Variational Inference for Basis-SGLMMs}
\label{Sec:SIVI}
In this section, we briefly review variational inference methods, including SIVI, and then introduce our SIVI-based approach for fitting Basis-SGLMMs.
\subsection{Variational Inference}
\label{Sec:VI}
Variational Bayes (VB) methods frame Bayesian inference as an optimization problem rather than relying on sampling-based approaches such as MCMC \citep{bishop2006pattern}. Let $\bZ$ denote the observed data, $\btheta$ the collection of latent variables and parameters, and $\bpsi$ the variational parameters. VB approximates the target posterior $p(\btheta \mid \bZ)$ by introducing a variational distribution $q(\btheta \mid \bpsi)$ and choosing $\bpsi$ to minimize the Kullback-Leibler (KL) divergence
$q^{*}(\btheta \mid \bpsi)
=
\argmin_{q \in \mathcal{Q}}
\text{KL}\!\big(q(\btheta \mid \bpsi)\,\|\,p(\btheta \mid \bZ)\big)$. This is equivalent to maximizing a lower bound on the marginal log-likelihood 
$\log p(\bZ)$, commonly referred to as the Evidence Lower Bound (ELBO) \citep{bishop2013variational, blei2006variational}:

\begin{equation} \label{EQ:ELBO}
    ELBO(q):= \int q(\btheta|\bpsi) \log \frac{p(\bZ|\btheta) \cdot p(\btheta)}{q(\btheta|\bpsi)}d\btheta = \mathbb{E}_{q} \left( \log \frac{p(\bZ|\btheta) \cdot p(\btheta)}{q(\btheta|\bpsi)} \right).
\end{equation}

\paragraph{Coordinate Ascent and Implicit Variational Inference}
The mean-field variational Bayes (MFVB) approximation factorizes the joint variational density as $q({\btheta})=\prod_{k=1}^K q_k(\btheta_k)$, where $\btheta = (\btheta_1,\ldots,\btheta_K)$ denotes blocks of parameters and latent variables, and $q_k$ is the variational density for block $k$. Under this factorization, the ELBO can be optimized by coordinate-ascent variational inference (CAVI), which updates each $q_k$ in turn using closed-form expressions when conjugacy is available \citep{blei2017variational}. Though computationally efficient, CAVI often underestimates posterior variance and may not fully capture dependence among parameters \citep{blei2006variational, han2013integrated, blei2017variational, wu2018fast}. Moreover, these methods typically rely on conditional conjugacy \citep{wang2013variational} or analytic approximations \citep{jaakkola1997variational, lee2024scalable}, which do not easily extend to more complex distributions such as the gamma or negative binomial.

Implicit variational families \citep{mohamed2016learning, tran2017hierarchical, li2017gradient, shi2017implicit, huszar2017variational, mescheder2017adversarial} increase flexibility beyond MFVB by defining
\[
\btheta = g_{\lambda}(\bep), \quad \bep \sim p(\bep),
\]
where $g_{\lambda}$ is a deterministic transformation and $p(\bep)$ is a base distribution. While sampling is straightforward, the density $q_{\lambda}(\btheta)$ is not available in closed form, preventing direct evaluation of the log-density term $\log q_{\lambda}(\btheta) - \log p(\btheta,\bZ)$ in the ELBO in \eqref{EQ:ELBO}.

\subsubsection{Semi-implicit Variational Inference}
Semi-implicit variational inference (SIVI) \citep{yin2018semi} combines an explicit base variational distribution with an implicit mixing distribution, yielding a hierarchical variational model. Let $\bpsi$ denote an auxiliary variable and $\phi$ the parameters of a mixing distribution $q_{\phi}(\bpsi)$. The semi-implicit variational family is defined hierarchically as
\begin{equation}
\btheta \sim q(\btheta \mid \bpsi),
\qquad
\bpsi \sim q_{\phi}(\bpsi),
\label{eq:SIVI-hier}
\end{equation}
where $q(\btheta \mid \bpsi)$ is an explicit, reparameterizable distribution (e.g., Gaussian) and $q_{\phi}(\bpsi)$ may be either explicit or implicit (e.g., represented by a neural network). Marginalizing over $\bpsi$ yields the semi-implicit variational density
\begin{equation}
h_{\phi}(\btheta)
=
\int q(\btheta \mid \bpsi)\,q_{\phi}(\bpsi)\,d\bpsi
=
\mbE_{\bpsi \sim q_{\phi}(\bpsi)}\big[q(\btheta \mid \bpsi)\big].
\label{eq:SIVI-marg}
\end{equation}

This construction restores dependence among components of $\btheta$ while retaining the tractable (explicit) conditional density $q(\btheta \mid \bpsi)$ for ELBO optimization. Although $h_{\phi}(\btheta)$ is generally intractable when $q_{\phi}(\bpsi)$ is implicit, \citet{yin2018semi} derive an asymptotically tight lower bound,
\[
\underline{\mathcal{L}}
=
\mathbb{E}_{\bpsi \sim q_{\phi}(\bpsi)}
\mathbb{E}_{\btheta \sim q(\btheta \mid \bpsi)}
\left[
\log \frac{p(\bZ,\btheta)}{q(\btheta \mid \bpsi)}
\right],
\]
which depends only on $q(\btheta \mid \bpsi)$ and samples from $q_{\phi}(\bpsi)$. See Supplement \ref*{sec:DerivationELBOlower} for the derivation of $\underline{\mathcal{L}}$, and \ref*{sec:repar_mlp} for details on reparameterization and the neural network construction of $q_\phi(\bpsi)$.
  
\subsection{Our Approach}
We integrate SIVI into the basis-SGLMM framework to enable scalable analysis of large spatial datasets ($N\approx 50{,}000$) without MCMC. The proposed approach accommodates response types from a broad class of response distributions, including gamma, negative binomial, Poisson, Bernoulli, and Gaussian distributions. The Bayesian hierarchical formulation of the proposed model is: 

\begin{align}
\text{Data model:} \quad 
& Z_i \mid \bbe, \bdel, \gamma \;\stackrel{\text{ind}}{\sim}\; 
  F\!\left(\,\cdot \;\middle|\; \eta_i,\gamma \right),
  \qquad i = 1,\dots,N, \nonumber \\[0.25em]
& g\!\left( \mathbb{E}[Z_i \mid \bbe, \bdel] \right) 
    = \eta_i 
    = \bx(\bs_i)^\top \bbe + \bgPhi(\bs_i)^\top \bdel, \nonumber \\[0.5em]
\text{Process model:} \quad 
 &\bdel\mid \ssq \sim \mathcal{N}(\bzero,\ssq \bSig_\delta),  \label{eq:basisSGLMM_bhm2} \\[0.5em]
\text{Parameter model:} \quad 
& \bbe \sim p(\bbe), \qquad \log \ssq \sim \mbox{N}(\mu_{\sigma},\tau_{\sigma}^2), \qquad \gamma  \sim p(\gamma), \nonumber
\end{align}
where $\mu_{\sigma}$ and $\tau_{\sigma}^2$ represent the prior mean and variance of the reparameterized $\ssq$. Depending on the chosen data model, the additional parameter $\gamma$ is defined as one of $
\gamma \in \{\log \tau^{2},\, \kappa,\, \log \alpha \}$, and we assign priors accordingly: (i) $\log \tau^2\sim \mbox{N}(\mu_{\tau},\ssq_{\tau})$ for the Gaussian model; (ii)  $\kappa \sim \mbox{Gamma}(a_{\kappa},b_{\kappa})$ for the negative binomial model; and (iii) $\log \alpha \sim \mbox{N}(\mu_{\alpha},\ssq_{\alpha})$ for the gamma model. The basis functions $\bgPhi$ are precomputed prior to applying the SIVI algorithm. The estimable parameters include $\bbe$, $\bdel$, $\ssq$, and the response distribution-specific parameters $\tau^2$, $\alpha$, and $\kappa$. Note that the Bayesian hierarchical model above presents only the conditional distributions and priors of the spatial model itself, and the variational components required for SIVI, (namely the implicit mixing distribution and explicit conditional variational distribution), are not shown. Instead, they are introduced in the following subsection.

\begin{figure}[ht]
 \begin{center}
\includegraphics[width=1.1\linewidth]{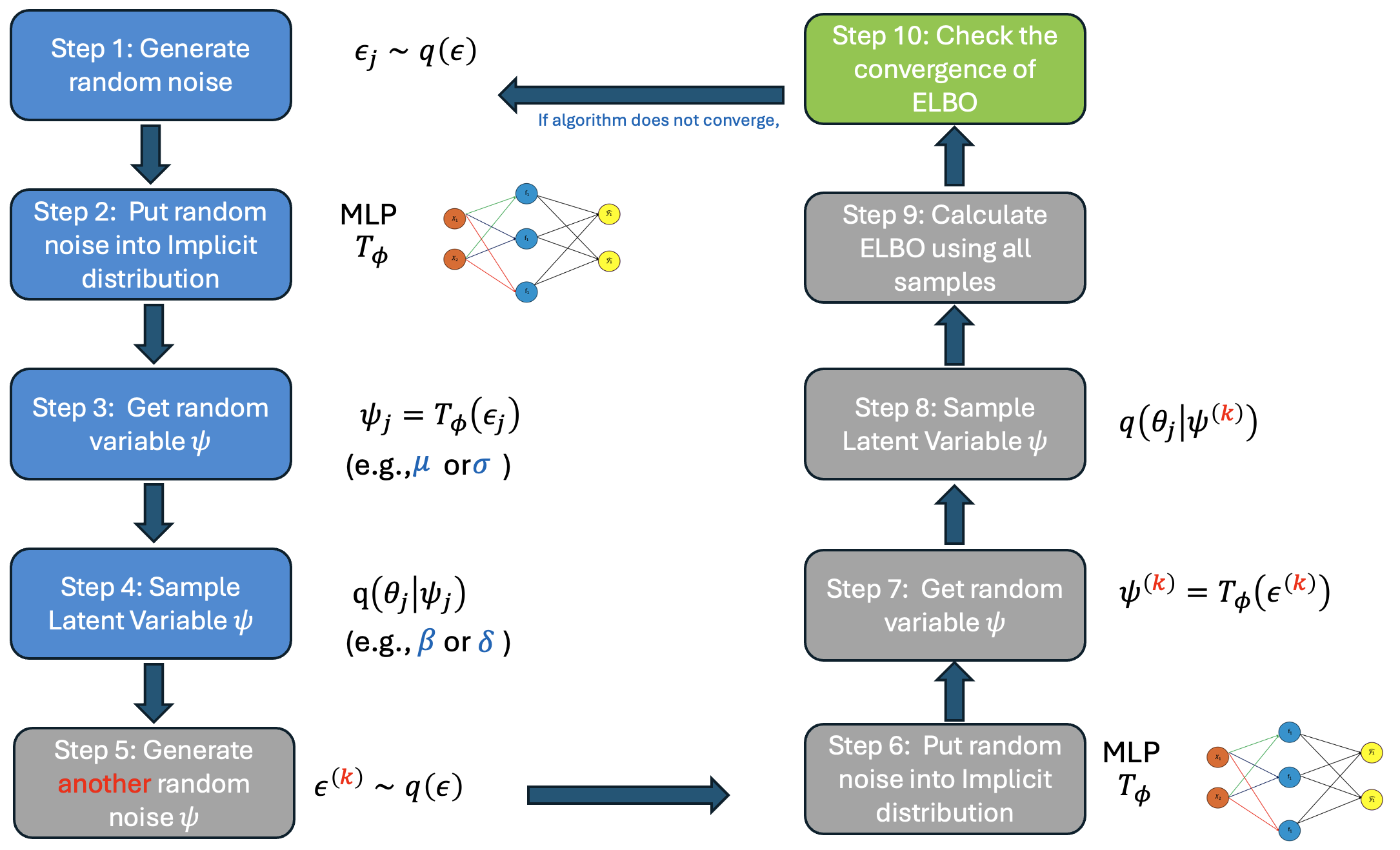}
\caption{ This illustrates an overview of the SIVI workflow of Algorithm~\ref{alg:sivi_basis} }\label{Fig:SIVIalgorithm}
\end{center}
\end{figure}

\paragraph{Variational Inference for Basis-SGLMMs}
We now describe the variational inference framework, including the associated variational distributions, for basis-SGLMMs. The SIVI workflow integrates readily with the basis-SGLMM framework by treating all model parameters and latent variables as a unified parameter vector $\btheta=(\bbe, \bdel, \log \sigma^{2}, \gamma)$. Algorithm~\ref{alg:sivi_basis} summarizes the steps, and Figure~\ref{Fig:SIVIalgorithm} illustrates the SIVI workflow within the basis-SGLMM framework.

At each iteration of the algorithm, random noise $\epsilon_j \sim q(\epsilon)$ is generated and mapped through the multilayer perceptron $T_{\phi}$ where $\phi$ denotes the MLP weights and biases to obtain an implicit mixing variable $\bpsi_j$ (Steps~1-3). This mixing variable parameterizes the explicit conditional variational distribution $q(\btheta_{j} \mid \bpsi_{j})$, which must be chosen
so that the reparameterization trick can be applied. In the context of
a basis-SGLMM, $\bpsi_{j}$ determines the variational mean or covariance for the low-dimensional basis coefficients $\bdel$ as well as the fixed effects  $\bbe$, enabling the variational family to adaptively capture posterior dependence across these parameters. Since the latent spatial process is approximated using the precomputed basis matrix $\bgPhi$ (with $\bomega \approx \bgPhi \bdel$), sampling $\btheta_{j}$ avoids the need to manipulate high-dimensional Gaussian process realizations; thereby substantially reducing computational 
cost.

To construct the semi-implicit variational density $h_{\phi}(\btheta)$, SIVI draws an additional $K$ auxiliary noise variables $\epsilon^{(k)}$ and generates corresponding mixing variables $\bpsi^{(k)} = T_{\phi}(\epsilon^{(k)})$ (Steps 5-7). The explicit conditional density $q(\btheta_{j} \mid \bpsi^{(k)})$ is then evaluated at $\btheta_{j}$ for each auxiliary mixing variable $\bpsi^{(k)}$ (Step 8), and the entire collection of samples is used to compute the surrogate lower bound of ELBO ($\underline{\mathcal{L}}$) that approximates the marginal density $h_{\phi}(\btheta)$ (Step 9). This ELBO ($\underline{\mathcal{L}}$) incorporates the non-Gaussian likelihood of the basis-SGLMM as well as the spatial structure induced by the basis expansion. The algorithm iteratively updates the neural network parameters $\phi$ until the ELBO ($\underline{\mathcal{L}}$) converges (Step 10), resulting in a flexible variational posterior that captures complex dependencies among $\btheta=(\bbe, \bdel, \log \sigma^{2}, \gamma)$ while remaining computationally scalable 
for large spatial datasets.

\begin{algorithm}
\caption{Semi-Implicit Variational Inference (SIVI) for Basis-SGLMMs}
\label{alg:sivi_basis}
\begin{algorithmic}[1]
\State \textbf{Input:} Data $\{\bZ_i\}_{1:N}$, likelihood
$p(\bZ_i \mid \bbe,\bdel,\log\sigma^{2},\gamma)$, basis matrix $\bgPhi$,
joint model $p(\bZ,\btheta)$ with $\btheta=(\bbe,\bdel,\log\sigma^{2},\gamma)$,
explicit variational distribution $q(\btheta \mid \bpsi)$ with
reparameterization $\btheta=f(\epsilon,\bpsi)$, implicit MLP $T_{\phi}(\epsilon)$,
noise source $\epsilon \sim q(\epsilon)$, number of Monte Carlo samples $J$ for the ELBO estimate
\State \textbf{Output:} Implicit variational parameter $\phi$ defining $q_{\phi}(\bpsi)$
\Statex
\State Initialize $\phi$ randomly; set $t \gets 0$
\While{not converged}
    \State Set $\underline{\mathcal{L}}_{K_t} \gets 0$, step size $\eta_t$; choose non-decreasing $K_t \ge 0$
    \For{$j=1$ to $J$}
        \State Generate random noise $\epsilon_j \sim q(\epsilon)$ \Comment{Step 1}
        \State Map $\epsilon_j$ through the implicit MLP $T_{\phi}$ \Comment{Step 2}
        \State Obtain mixing variable $\bpsi_j = T_{\phi}(\epsilon_j)$ \Comment{Step 3}
        \State Sample $\btheta_j = f(\epsilon_j,\bpsi_j) \sim q(\btheta_j \mid \bpsi_j)$ \Comment{Step 4}
        \For{$k=1$ to $K_t$}
            \State Generate another random noise $\epsilon^{(k)} \sim q(\epsilon)$ \Comment{Step 5}
            \State Map $\epsilon^{(k)}$ through the implicit MLP $T_{\phi}$ \Comment{Step 6}
            \State Obtain mixing variable $\bpsi^{(k)} = T_{\phi}(\epsilon^{(k)})$ \Comment{Step 7}
            \State Evaluate $q(\btheta_j \mid \bpsi^{(k)})$ at $\btheta_j$ \Comment{Step 8}
        \EndFor
        \State Accumulate the surrogate lower bound $\underline{\mathcal{L}}$: \Comment{Step 9}
        \Statex
        \hspace{\algorithmicindent}%
        \resizebox{0.97\linewidth}{!}{$\displaystyle
        \underline{\mathcal{L}}_{K_t} \gets \underline{\mathcal{L}}_{K_t} + \frac{1}{J}\left\{
        -\log\!\left(\frac{1}{K_t+1}\left[\sum_{k=1}^{K_t} q(\btheta_j \mid \bpsi^{(k)})
        + q(\btheta_j \mid \bpsi_j)\right]\right)
        + \log p(\bZ \mid \btheta_j) + \log p(\btheta_j)\right\}$}
    \EndFor
    \State Update $\phi \gets \phi + \eta_t \nabla_{\phi}\,\underline{\mathcal{L}}_{K_t}$;\quad $t \gets t+1$
    \State \textbf{Check ELBO convergence} \Comment{Step 10}
\EndWhile
\end{algorithmic}
\end{algorithm}

\paragraph{Implementation Details} \label{subsec:SIVIimpledetail}
The proposed SIVI algorithm requires the specification of several tuning parameters prior to implementation. These include the: (i) stopping criterion threshold $\epsilon_{*}$; (ii) maximum number of iterations for optimization; (iii) batches $J$ used when generating samples; (iv) number of auxiliary samples $K$; (v) scale parameters for the conditional (explicit) distribution; and (vi) choice of reparameterized priors used for gradient-based optimization. We provide detailed discussion of each point and additional sensitivity analyses in the Supplement.

\section{Simulation Study} \label{sec:SIVIsimulations}
The proposed SIVI framework is evaluated through an extensive simulation study based on large spatial datasets ($N = 50{,}000$) generated under robust specifications, including varying response distributions, smoothness levels of the latent spatial random field, and decay in spatial correlation. Comparative analyses are conducted against competing approaches, including the Metropolis--Hastings (MH) algorithm and Hamiltonian Monte Carlo (HMC).

\subsection{Simulation Design}
We consider $N = 50{,}000$ spatial locations $\bs_i \in \mathcal{D} = [0,1]^2 \subset \mathbb{R}^2$, of which $N_{\mathrm{train}} = 40{,}000$ are used for model fitting and $N_{\mathrm{test}} = 10{,}000$ are reserved for validation. The observation vector $\bZ = \big(Z(\bs_1), \ldots, Z(\bs_N)\big)^\top$ is generated under the basis-SGLMM framework described in subsection~\ref{SubSec:BasisSGLMM}, with covariates $\bX = [\bX_1, \bX_2]$ where $\bX_1, \bX_2 \stackrel{\mathrm{i.i.d.}}{\sim} \mathrm{Unif}(-1,1)$ and regression coefficients $\bbe = (1, 1)^\top$. Four configurations of the spatial random effects $\bomega = \{\omega(\bs_i) : \bs_i \in \mathcal{D}\}$ are generated from a zero-mean Gaussian process with a Mat\'ern covariance function with covariance parameters $\nu$, $\phi$, and $\sigma^2$. 

Table~\ref{tab:SimulationScenario} summarizes the simulation design, which considers two smoothness parameters ($\nu \in \{0.5, 1.5\}$) and two range parameters ($\phi \in \{0.1, 0.3\}$) across five data types: gamma, negative binomial, binary, count, and Gaussian. The marginal variance is fixed at $\sigma^2 = 1$. This yields 20 simulation scenarios, with 50 replicates per scenario, resulting in 1,000 datasets in total.  

The basis-SGLMM approach (Section~\ref{SubSec:BasisSGLMM}) is employed, approximating $\bomega \approx \bgPhi \bdel$, where $\bgPhi$ is an $N \times m$ matrix of spatial eigenbasis functions \citep{banerjee2008gaussian, guan2018computationally}, taken as the $m$ leading eigenvectors of a Mat\'ern covariance matrix. Throughout, we use $m = 50$ basis functions. Priors are specified as $\bbe_j \sim \mathcal{N}(0,100)$, $\log\tau^2 \sim \mathcal{N}(0,1)$, $\log\sigma^2 \sim \mathcal{N}(1,1)$, $\kappa \sim \mathrm{Gamma}(2,1)$, and $\log\alpha \sim \mathcal{N}(1,1)$.

For SIVI, the stopping criterion is set to a threshold of $\epsilon_{*} = 1 \times 10^{-2}$ (see Algorithm~\ref{alg:sivi_basis}). The maximum number of iterations is fixed at $5{,}000$, and we set $J = 20$ as the number of samples per batch for SIVI. In this study, we utilize a multilayer perceptron (MLP) architecture consisting of three hidden layers of sizes $40$, $60$, and $40$, respectively. Each pair of consecutive layers is fully connected, with the tanh activation function applied to the hidden layers for the gamma data model and the ReLU activation function for the negative binomial, binary, count, and Gaussian data models, and a linear activation function for the output layer. In our implementation of the MLP, we set the learning rate to 0.001 and use the Adam algorithm \citep{kingma2014adam}. For the MH algorithm, we obtain $100{,}000$ posterior samples, assessing convergence via batch means standard errors (BMSE)~\citep{flegal2008markov} and trace plots. For HMC, $2{,}000$ posterior samples are drawn, which yield an effective sample size (ESS)~\citep{liu2001monte} comparable to that of the MH sampler. 

Model performance is evaluated using the root mean squared prediction error,
\[
\mathrm{RMSPE} = \sqrt{\frac{1}{N_{\mathrm{test}}} \sum_{i=1}^{N_{\mathrm{test}}} \left(Z_i - \hat{Z}_i\right)^2},
\]
for the negative binomial, gamma, count, and Gaussian data settings, and the area under the receiver operating characteristic curve (AUC) for the binary case.

All simulations are executed on a high-performance computing (HPC) system, with walltimes reported based on a single 2.4 GHz Intel Xeon Gold 6240R processor. The Metropolis--Hastings (MH) MCMC algorithm is implemented in R version 4.1.2, while Hamiltonian Monte Carlo (HMC) and Semi-Implicit Variational Inference (SIVI) are implemented in PyTorch using Python (version 3.10.1).

\subsection{Results}
Results for inference and out-of-sample predictions are provided for the negative binomial and gamma cases. The Bernoulli, Poisson and Gaussian cases are available in the Supplement. 

\paragraph{Negative Binomial Responses}

Table~\ref{tab:NB_speedup} summarizes the out-of-sample predictive performance, measured by RMSPE, across all approaches. The results indicate that all three methods achieve nearly identical predictive accuracy. For instance, when the smoothness parameter is set to $\nu = 0.5$ and the spatial range parameter to $\phi = 0.1$, MH, HMC, and SIVI all yield an RMSPE of 3.473. However, the computational cost differs substantially: SIVI is approximately $44.4$ times faster than MH and $2.0$ times faster than HMC. Across the negative binomial experiments, the speedup of SIVI relative to MH ranges from 39 to 44, whereas the improvement relative to HMC is about a factor of two. Notably, the computational gains from SIVI would be even more pronounced under looser stopping criteria (see Supplements~\ref*{sec:VB2smallcriteria1e1}, \ref*{sec:VB2smallcriteria1e3}, and \ref*{sec:VB2smallcriteria1e4}).

\begin{figure}[H]
 \begin{center}
\includegraphics[width=1.0\linewidth]{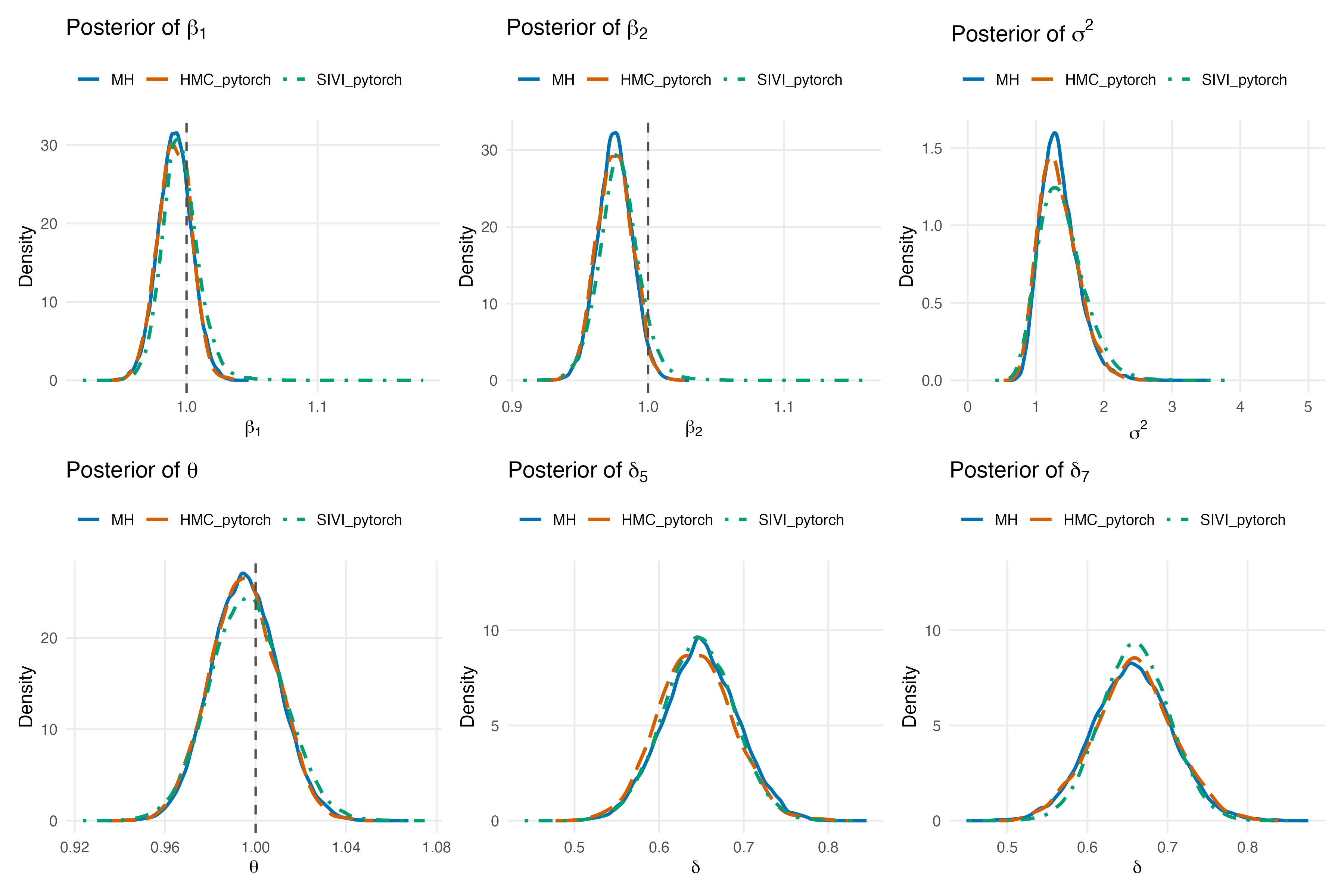}
\caption{Posterior density estimates of selected model parameters under three inference methods for the negative binomial (NB) data when $\nu = 0.5$ and $\phi = 0.1$: 
    Metropolis--Hastings (MH), Hamiltonian Monte Carlo (HMC), and Semi-Implicit Variational 
    Inference (SIVI). Panels show results for regression coefficients 
    ($\beta_{1}, \beta_{2}$), variance components ($\sigma^{2}, \kappa$), and selected spatial 
    random effects ($\delta_{5}, \delta_{7}$). Dashed vertical lines indicate the corresponding 
    true parameter values where available. Overall, all three methods yield nearly indistinguishable 
    posterior distributions, highlighting the accuracy of SIVI relative to MCMC-based approaches.}\label{Fig:SIVINBAll}
\end{center}
\end{figure}

For all negative binomial datasets, the posterior distributions of the model parameters obtained from MH, HMC, and SIVI are largely comparable. Figure~\ref{Fig:SIVINBAll} illustrates this comparison for the regression coefficients ($\beta_{1}, \beta_{2}$), variance components ($\sigma^{2}, \kappa$), and selected spatial random effects ($\delta_{5}, \delta_{7}$) under the setting $\nu = 0.5$ and $\phi = 0.1$. Although variational methods are often noted for underestimating posterior variance \citep{blei2006variational, wu2018fast}, our proposed SIVI approach not only achieves strong predictive accuracy but also produces posterior distributions that closely resemble those from MCMC-based methods, at least in our simulation study. We compare $95\%$ posterior credible intervals under the setting with smoothness parameter $\nu = 1.5$ and range parameter $\phi = 0.3$ for the negative binomial model across $50$ simulation replicates. The average empirical coverage probabilities are $0.944$ for MH-MCMC, $0.951$ for HMC, and $0.925$ for SIVI. These results indicate that SIVI does not exhibit substantial underestimation of posterior variance. Moreover, Figure~\ref{Fig:SIVINBAll} in the main manuscript shows that all three methods produce posterior distributions with similar variances.

\begin{table}[ht]
\caption{Comparison of RMSPE, walltime (in seconds), and speedup for MH, HMC, and SIVI under different parameter settings for the negative binomial (NB) case.}
\label{tab:NB_speedup}
\centering
\resizebox{\textwidth}{!}{
\begin{tabular}{lcccccccc}
\toprule
 & \multicolumn{2}{c}{MH} & \multicolumn{2}{c}{HMC} & \multicolumn{2}{c}{SIVI} & \multicolumn{2}{c}{Speedup} \\
\cmidrule(lr){2-3}\cmidrule(lr){4-5}\cmidrule(lr){6-7}\cmidrule(lr){8-9}
 & RMSPE & Walltime & RMSPE & Walltime & RMSPE & Walltime & MH/SIVI & HMC/SIVI \\
\midrule
\rowcolor{yellow!30} \textbf{NB, $\nu=0.5$} & & & & & & & & \\
$\phi=0.1$ & 3.473 & 3194.339 & 3.473 & 141.085 & 3.473 & 71.909 & \textcolor{red}{44.422} & \textcolor{red}{1.962} \\
$\phi=0.3$ & 3.930 & 3308.680 & 3.930 & 143.947 & 3.931 & 81.522 & \textcolor{red}{40.586} & \textcolor{red}{1.766} \\
\midrule
\rowcolor{yellow!30} \textbf{NB, $\nu=1.5$} & & & & & & & & \\
$\phi=0.1$ & 3.971 & 3162.185 & 3.971 & 126.351 & 3.971 & 79.344 & \textcolor{red}{39.854} & \textcolor{red}{1.592} \\
$\phi=0.3$ & 3.887 & 3334.172 & 3.887 & 151.783 & 3.888 & 78.383 & \textcolor{red}{42.537} & \textcolor{red}{1.936} \\
\bottomrule
\end{tabular}
}
\end{table}

\paragraph{Gamma Responses}

Table~\ref{tab:Gamma_results} reports the RMSPE and walltime for Metropolis--Hastings (MH), Hamiltonian Monte Carlo (HMC), and Semi-Implicit Variational Inference (SIVI) applied to the gamma data using the $50$ leading eigenvectors. Similar to the negative binomial results, all three methods achieve nearly identical predictive accuracy; however, the computational gains from SIVI are substantially greater. For example, when $\nu = 0.5$ and $\phi = 0.1$, MH, HMC, and SIVI all yield an RMSPE of 3.945, yet SIVI is approximately $145$ times faster than MH and about $4$ times faster than HMC. Across the gamma simulation studies, SIVI demonstrates speedups of 117–145 relative to MH and 3–4 relative to HMC. Notably, the gamma data provide the most pronounced computational advantage for SIVI compared with other models such as negative binomial, count, binary, and Gaussian.

\begin{figure}[H]
 \begin{center}
\includegraphics[width=1.0\linewidth]{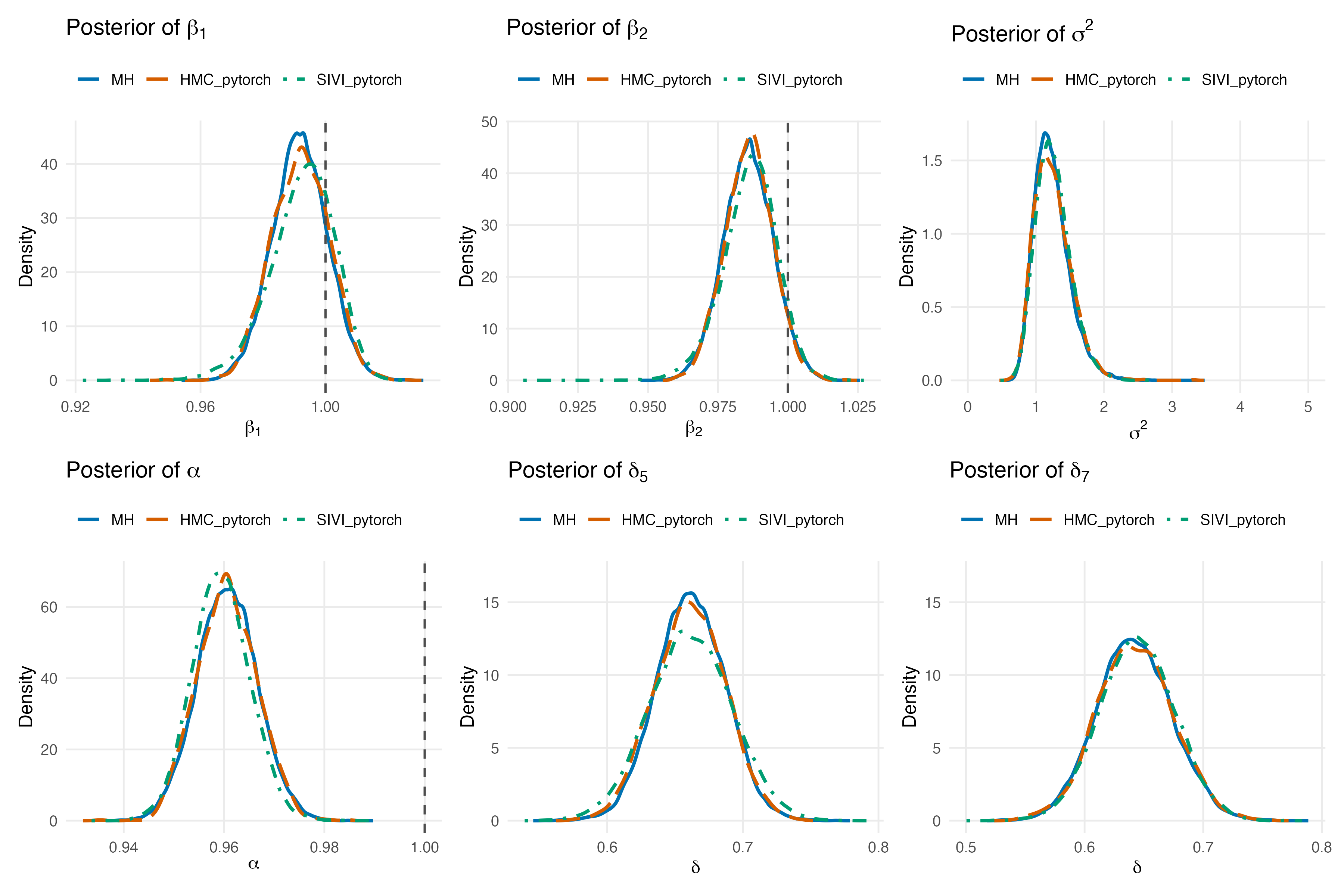}
\caption{Posterior density estimates of selected model parameters under three inference methods for the gamma data when $\nu = 0.5$ and $\phi = 0.3$:
    Metropolis--Hastings (MH), Hamiltonian Monte Carlo (HMC), and Semi-Implicit Variational 
    Inference (SIVI). Panels show results for regression coefficients 
    ($\beta_{1}, \beta_{2}$), variance components ($\sigma^{2}, \alpha$), and selected spatial 
    random effects ($\delta_{5}, \delta_{7}$). Dashed vertical lines indicate the corresponding 
    true parameter values where available. Overall, all three methods yield nearly indistinguishable 
    posterior distributions, highlighting the accuracy of SIVI relative to MCMC-based approaches.}\label{Fig:SIVIGamAll}
\end{center}
\end{figure}

As with the other data models, the posterior distributions of the model parameters obtained from MH, HMC, and SIVI remain largely consistent. Figure~\ref{Fig:SIVIGamAll} illustrates this comparison for the setting $\nu = 0.5$ and $\phi = 0.3$.

\begin{table}[ht]
\caption{Comparison of RMSPE, walltime (in seconds), and speedup for MH, HMC, and SIVI under different parameter settings for the gamma case.}
\label{tab:Gamma_results}
\centering
\resizebox{\textwidth}{!}{
\begin{tabular}{lcccccccc}
\toprule
 & \multicolumn{2}{c}{MH} & \multicolumn{2}{c}{HMC} & \multicolumn{2}{c}{SIVI} & \multicolumn{2}{c}{Speedup} \\
\cmidrule(lr){2-3}\cmidrule(lr){4-5}\cmidrule(lr){6-7}\cmidrule(lr){8-9}
 & RMSPE & Walltime & RMSPE & Walltime & RMSPE & Walltime & MH/SIVI & HMC/SIVI \\
\midrule
\rowcolor{yellow!30} \textbf{Gamma, $\nu=0.5$} & & & & & & & & \\
$\phi=0.1$ & 3.945 & 4291.942 & 3.945 & 121.865 & 3.945 & 29.560 & \textcolor{red}{145.193} & \textcolor{red}{4.123} \\
$\phi=0.3$ & 3.855 & 4071.220 & 3.855 & 129.565 & 3.855 & 32.061 & \textcolor{red}{126.985} & \textcolor{red}{4.041} \\
\midrule
\rowcolor{yellow!30} \textbf{Gamma, $\nu=1.5$} & & & & & & & & \\
$\phi=0.1$ & 4.614 & 4204.997 & 4.614 & 113.861 & 4.614 & 35.745 & \textcolor{red}{117.638} & \textcolor{red}{3.185} \\
$\phi=0.3$ & 3.559 & 4577.599 & 3.559 & 129.269 & 3.559 & 38.329 & \textcolor{red}{119.429} & \textcolor{red}{3.373} \\
\bottomrule
\end{tabular}
}
\end{table}

\subsubsection{Computational Costs}

Table~\ref{tab:quantile_speedups_all} summarizes the distribution of walltimes for MH, HMC, and SIVI across different data types, smoothness parameters, and spatial ranges. While the previous comparisons were based on mean walltimes, here we report the 25th, 50th (median), and 75th quantiles to capture the variability in computational costs. This quantile-based summary provides a more robust characterization, particularly since walltimes for SIVI methods can fluctuate across replications due to the seed of the samples and convergence behavior. 

The results demonstrate that SIVI exhibits notable speedups across all data types and settings. For example, in the negative binomial and gamma cases, SIVI reduces computation time by roughly 50 to 100 times compared to MH, and by factors of two to five relative to HMC. These speedup ratios are largely preserved across the 25th, 50th, and 75th percentiles. Binary and count data scenarios maintain speedups ranging from 20- to 75-fold relative to MH, and from 3- to 12-fold relative to HMC. Gaussian data show the smallest but still meaningful gains, with speedups of around 21- to 38-fold relative to MH and 2- to 4-fold relative to HMC.
Based on these results, SIVI consistently delivers computational speedups over MCMC methods across data types and settings, even after accounting for variability in walltimes.

\begin{table}[ht]
\caption{Walltime quantiles (25\%, 50\%, 75\%) for MH, HMC, and SIVI, and corresponding speedups (MH/SIVI, HMC/SIVI) across data types and smoothness settings.}
\label{tab:quantile_speedups_all}
\centering
\setlength{\tabcolsep}{5pt}    
\renewcommand{\arraystretch}{1.1}
\resizebox{\textwidth}{!}{
\begin{tabular}{l ccc ccc ccc ccc ccc}
\toprule
 & \multicolumn{3}{c}{\textbf{MH}} & \multicolumn{3}{c}{\textbf{HMC}} & \multicolumn{3}{c}{\textbf{SIVI}}
 & \multicolumn{3}{c}{\textbf{Speedup (MH/SIVI)}} & \multicolumn{3}{c}{\textbf{Speedup (HMC/SIVI)}}\\
\cmidrule(lr){2-4}\cmidrule(lr){5-7}\cmidrule(lr){8-10}\cmidrule(lr){11-13}\cmidrule(lr){14-16}
 & 25\% & 50\% & 75\% & 25\% & 50\% & 75\% & 25\% & 50\% & 75\% & 25\% & 50\% & 75\% & 25\% & 50\% & 75\% \\
\midrule

\rowcolor{yellow!30}\multicolumn{16}{l}{\textbf{NB}}\\
\rowcolor{yellow!15}\multicolumn{16}{l}{\textbf{$\nu=0.5$}}\\
$\phi=0.1$ &
3038.053 & 3205.406 & 3332.122 &
121.477  & 134.756  & 159.281  &
58.569   & 68.310   & 86.513   &
51.871   & \textcolor{red}{46.924} & 38.516 &
2.074    & \textcolor{red}{1.973}  & 1.841 \\
$\phi=0.3$ &
3205.791 & 3337.000 & 3431.428 &
124.543  & 133.702  & 160.183  &
51.481   & 79.814   & 98.385   &
62.272   & \textcolor{red}{41.809} & 34.878 &
2.419    & \textcolor{red}{1.675}  & 1.628 \\
\rowcolor{yellow!15}\multicolumn{16}{l}{\textbf{$\nu=1.5$}}\\
$\phi=0.1$ &
3041.410 & 3165.704 & 3265.998 &
106.224  & 123.283  & 142.447  &
61.417   & 78.272   & 92.491   &
49.521   & \textcolor{red}{40.445} & 35.312 &
1.730    & \textcolor{red}{1.575}  & 1.540 \\
$\phi=0.3$ &
3212.261 & 3385.727 & 3450.251 &
136.075  & 146.063  & 166.988  &
57.535   & 72.337   & 89.648   &
55.831   & \textcolor{red}{46.805} & 38.487 &
2.365    & \textcolor{red}{2.019}  & 1.863 \\
\midrule

\rowcolor{yellow!30}\multicolumn{16}{l}{\textbf{Gamma}}\\
\rowcolor{yellow!15}\multicolumn{16}{l}{\textbf{$\nu=0.5$}}\\
$\phi=0.1$ &
3435.886 & 3540.562 & 3684.057 &
105.796  & 122.750  & 139.434 &
18.115   & 27.315   & 41.319  &
189.667  & \textcolor{red}{129.621} & 89.161 &
5.840    & \textcolor{red}{4.494}    & 3.375 \\
$\phi=0.3$ &
3216.671 & 3337.640 & 3478.441 &
110.334  & 131.261  & 147.809 &
18.892   & 33.153   & 47.371  &
170.267  & \textcolor{red}{100.675} & 73.430 &
5.840    & \textcolor{red}{3.959}    & 3.120 \\
\rowcolor{yellow!15}\multicolumn{16}{l}{\textbf{$\nu=1.5$}}\\
$\phi=0.1$ &
3287.160 & 3479.287 & 3607.622 &
93.948   & 109.456  & 129.891 &
22.993   & 36.281   & 44.389  &
142.963  & \textcolor{red}{95.899} & 81.272 &
4.086    & \textcolor{red}{3.017}  & 2.926 \\
$\phi=0.3$ &
2975.072 & 3150.037 & 3357.022 &
112.922  & 131.158  & 144.769 &
27.034   & 34.083   & 50.282  &
110.049  & \textcolor{red}{92.424} & 66.764 &
4.177    & \textcolor{red}{3.848}  & 2.879 \\
\midrule

\rowcolor{yellow!30}\multicolumn{16}{l}{\textbf{Binary}}\\
\rowcolor{yellow!15}\multicolumn{16}{l}{\textbf{$\nu=0.5$}}\\
$\phi=0.1$ &
1102.228 & 1361.181 & 1534.596 &
178.892  & 184.339  & 195.206 &
14.728   & 22.249   & 29.604 &
74.840   & \textcolor{red}{61.180} & 51.837 &
12.147   & \textcolor{red}{8.285}  & 6.594 \\
$\phi=0.3$ &
1100.464 & 1371.999 & 1527.190 &
179.067  & 184.389  & 191.079 &
17.481   & 26.021   & 33.474 &
62.952   & \textcolor{red}{52.727} & 45.623 &
10.243   & \textcolor{red}{7.086}  & 5.708 \\
\rowcolor{yellow!15}\multicolumn{16}{l}{\textbf{$\nu=1.5$}}\\
$\phi=0.1$ &
1060.812 & 1158.577 & 1504.216 &
176.950  & 179.799  & 184.318 &
15.382   & 27.420   & 31.645 &
68.962   & \textcolor{red}{42.253} & 47.535 &
11.503   & \textcolor{red}{6.557}  & 5.825 \\
$\phi=0.3$ &
1048.858 & 1260.038 & 1503.935 &
179.101  & 182.640  & 189.276 &
18.710   & 26.934   & 34.497 &
56.058   & \textcolor{red}{46.782} & 43.596 &
9.572    & \textcolor{red}{6.781}  & 5.487 \\
\midrule

\rowcolor{yellow!30}\multicolumn{16}{l}{\textbf{Count}}\\
\rowcolor{yellow!15}\multicolumn{16}{l}{\textbf{$\nu=0.5$}}\\
$\phi=0.1$ &
1043.119 & 1199.590 & 1228.619 &
183.496  & 200.628  & 216.618 &
39.466   & 50.036   & 64.191 &
26.431   & \textcolor{red}{23.975} & 19.140 &
4.649    & \textcolor{red}{4.010}  & 3.375 \\
$\phi=0.3$ &
1045.639 & 1185.762 & 1239.167 &
169.726  & 184.288  & 201.212 &
42.037   & 54.113   & 69.422 &
24.874   & \textcolor{red}{21.913} & 17.850 &
4.038    & \textcolor{red}{3.406}  & 2.898 \\
\rowcolor{yellow!15}\multicolumn{16}{l}{\textbf{$\nu=1.5$}}\\
$\phi=0.1$ &
1089.074 & 1367.412 & 1515.595 &
158.783  & 182.745  & 202.656 &
41.847   & 56.343   & 63.753 &
26.025   & \textcolor{red}{24.270} & 23.773 &
3.794    & \textcolor{red}{3.243}  & 3.179 \\
$\phi=0.3$ &
1100.422 & 1461.657 & 1525.157 &
157.560  & 182.322  & 205.267 &
46.807   & 57.893   & 65.707 &
23.510   & \textcolor{red}{25.247} & 23.211 &
3.366    & \textcolor{red}{3.149}  & 3.124 \\
\midrule

\rowcolor{yellow!30}\multicolumn{16}{l}{\textbf{Gaussian}}\\
\rowcolor{yellow!15}\multicolumn{16}{l}{\textbf{$\nu=0.5$}}\\
$\phi=0.1$ &
560.900 & 678.775 & 692.222 &
53.365  & 60.349  & 72.330  &
18.168  & 23.028  & 27.417  &
30.873  & \textcolor{red}{29.476} & 25.248 &
2.937   & \textcolor{red}{2.621}  & 2.638 \\
$\phi=0.3$ &
559.822 & 679.687 & 692.743 &
57.518  & 61.721  & 70.991  &
17.262  & 26.955  & 33.265  &
32.431  & \textcolor{red}{25.216} & 20.825 &
3.332   & \textcolor{red}{2.290}  & 2.134 \\
\rowcolor{yellow!15}\multicolumn{16}{l}{\textbf{$\nu=1.5$}}\\
$\phi=0.1$ &
672.899 & 684.509 & 688.272 &
51.098  & 57.222  & 67.887  &
17.947  & 27.261  & 31.509  &
37.494  & \textcolor{red}{25.109} & 21.844 &
2.847   & \textcolor{red}{2.099}  & 2.155 \\
$\phi=0.3$ &
645.736 & 681.685 & 688.288 &
66.607  & 78.775  & 85.513  &
17.174  & 20.956  & 30.334  &
37.600  & \textcolor{red}{32.529} & 22.690 &
3.878   & \textcolor{red}{3.759}  & 2.819 \\
\bottomrule
\end{tabular}
}
\end{table}

\section{Applications}
\label{Sec:Applications}
The proposed Semi-Implicit Variational Inference (SIVI) method is demonstrated using two real-world spatial datasets: (i) land surface temperature from the NASA Aqua satellite \citep{zilber2021vecchia}, and (ii) Blue Jay bird-count data \citep{BBS_2022}. For benchmarking, we also compare SIVI with MH and HMC as in the simulation study.

\subsection{MODIS Land Surface Temperature Data}

NASA’s Aqua Satellite Mission, part of the Earth Observing System Afternoon Constellation (EOS-PM), was launched on May 4, 2002, and began routine data acquisition in June 2002, providing continuous global observations of the Earth’s atmosphere, oceans, and cryosphere. We analyze land surface temperature (LST) obtained from NASA’s Moderate Resolution Imaging Spectroradiometer (MODIS) aboard the Aqua satellite. LST data are contained in the \texttt{MYD11\_L2} product acquired on July 1, 2025, at 1:00 PM local time. The study region spans longitudes $16.5^{\circ}$–$24.0^{\circ}$ E and latitudes $36.0^{\circ}$–$41.0^{\circ}$ N, comprising $215{,}941$ grid cells at a 1 km spatial resolution. For this demonstration, we randomly selected $N = 50{,}000$ locations for model fitting, reserving an additional $20\%$ of the locations for validation. 

Given that LST, in $^\circ C$, is strictly positive, the data are modeled using a basis-SGLMM with gamma-distributed responses and a log link function. The corresponding Bayesian hierarchical model within the SIVI framework is as follows:

\begin{align*}
\label{EQ:SIVIGammaData}
\tb{Data Model:} \quad
    & \bZ \mid \bbe, \bdel, \alpha
      \sim \mbox{Gamma}\left(\alpha\mathbf{1}, \alpha\boldsymbol{\mu}^{-1}\right), \\
    & \text{where} \quad
      \boldsymbol{\mu}=\exp(\bX\bbe+\bgPhi\bdel), \\
\tb{Process Model:} \quad
    & \bdel \mid \ssq \sim \mathcal{N}(\bzero,\ssq \bSig_\delta), \\
\tb{Parameter Model:} \quad
    & \boldsymbol{\beta} \sim \mathcal{N}(\mu_{\beta},\Sigma_{\beta}), \\
    & \log \ssq \sim \mathcal{N}(\mu_{\sigma},\ssq_{\sigma}), \\
    & \log \alpha \sim \mathcal{N}(\mu_{\alpha},\ssq_{\alpha}),
\end{align*}
where $\mu$ is the conditional mean,  $\alpha$ denotes the shape parameter, $\bX$ is the matrix of covariates, $\bbe$ the regression coefficient vector, $\bgPhi$ the basis function matrix, and $\bdel$ are the basis coefficients. The matrix $\bSig_{\delta}$ characterizes the spatial covariance structure, thereby inducing correlation across spatial locations and capturing spatial dependence in the data. We represent the latent process using discretized Moran’s basis functions following \citet{lee2022picar}. Specifically, the leading 100 eigenvectors were selected, as internal sensitivity checks indicated this number provides a balance between computational efficiency and predictive accuracy. The prior distributions were specified as $\bbe \sim \mathcal{N}(0, 10^2)$, $\log \sigma^2 \sim \mathcal{N}(-8.84, 1.1)$, $\log \alpha \sim \mathcal{N}(1.270, 4.934)$, and $\bdel \sim \mathcal{N}(0, \sigma^2 I)$. MH sampler is run for $100{,}000$ iterations, while Hamiltonian Monte Carlo (HMC) is run for $2{,}000$ iterations, which is consistent with the simulation experiments. We use $\epsilon_{*}=1\times10^{-3}$ as the stopping criterion for SIVI. 

\begin{figure}[H]
 \begin{center}
\includegraphics[width=1.0\linewidth]{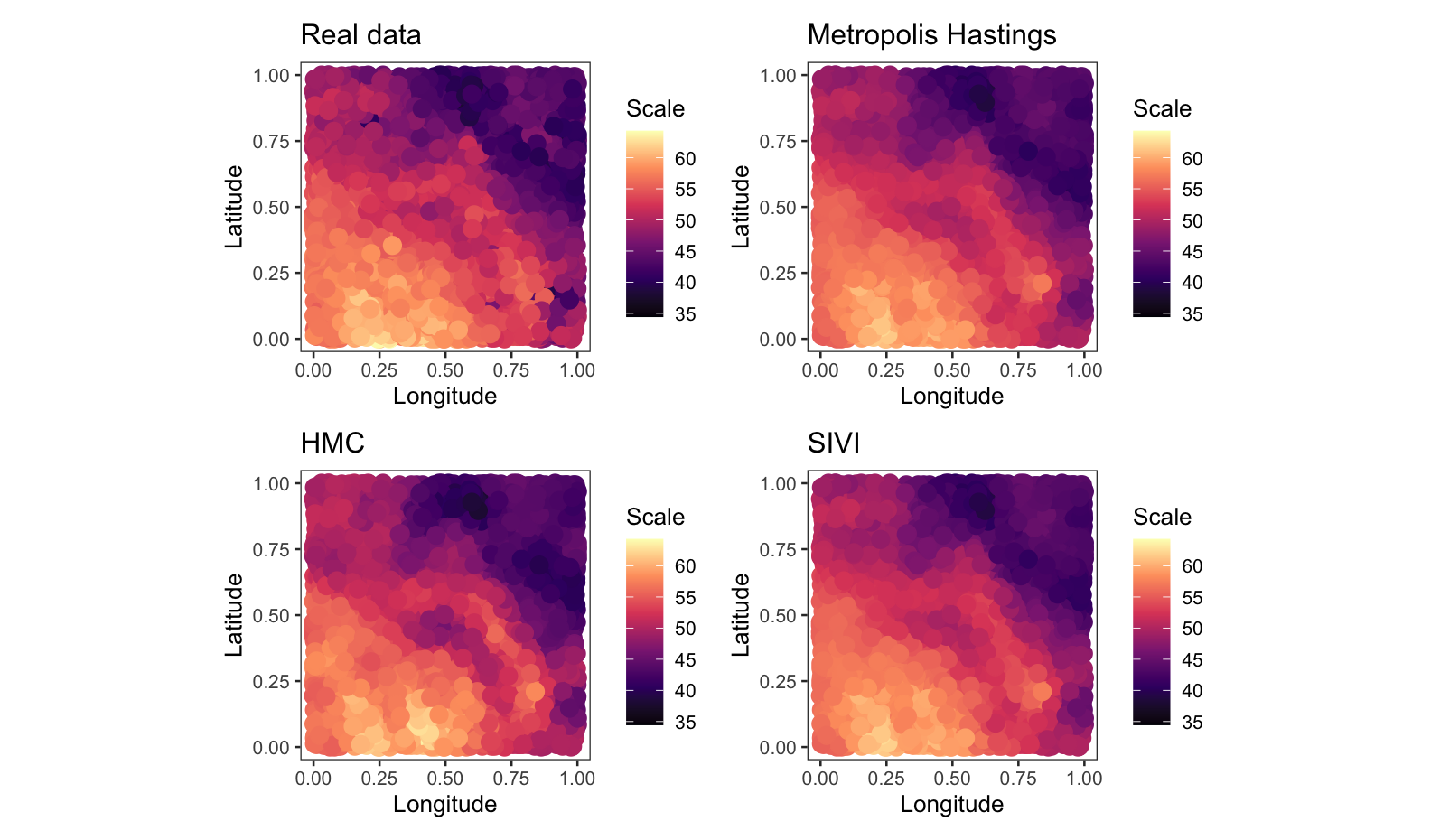}
\caption{Interpolation results from modeling the MODIS land surface temperature data. Implementations of Metropolis--Hastings (MH) MCMC, Hamiltonian Monte Carlo (HMC), and Semi-Implicit Variational Inference (SIVI) with basis function representations are shown. These panels show that the interpolated response surfaces from SIVI closely match those obtained using MCMC-based methods.}\label{Fig:SIVILST50k}
\end{center}
\end{figure}

Although MH achieved the lowest RMSPE (1.42), it required 45 minutes (2,741 seconds), as shown in Table~\ref{tab:combined_realworld_results} (a). In contrast, SIVI achieved a comparable RMSPE in only 262 seconds, representing a 10.46-fold speedup, while HMC attained a higher RMSPE of 1.85 with a runtime of 3.6 hours (12,826.73 seconds); thus making it 48.95 times slower than SIVI. Overall, these results demonstrate that SIVI provides predictive accuracy comparable to MH and HMC while being approximately 10–to-49 times faster. Figure~\ref{Fig:SIVILST50k} illustrates the land surface temperature results, showing that all three methods produce visually similar outputs that effectively capture the hottest regions. In addition to the 50,000-location setting, we consider larger datasets, which demonstrate greater computational gains as the data scale increases (see Supplement~\ref*{sec:largeMODIS} for details).

\subsection{North American Breeding Bird Survey: Blue Jay Abundance}
\begin{figure}[H]
 \begin{center}
\includegraphics[width=1.0\linewidth]{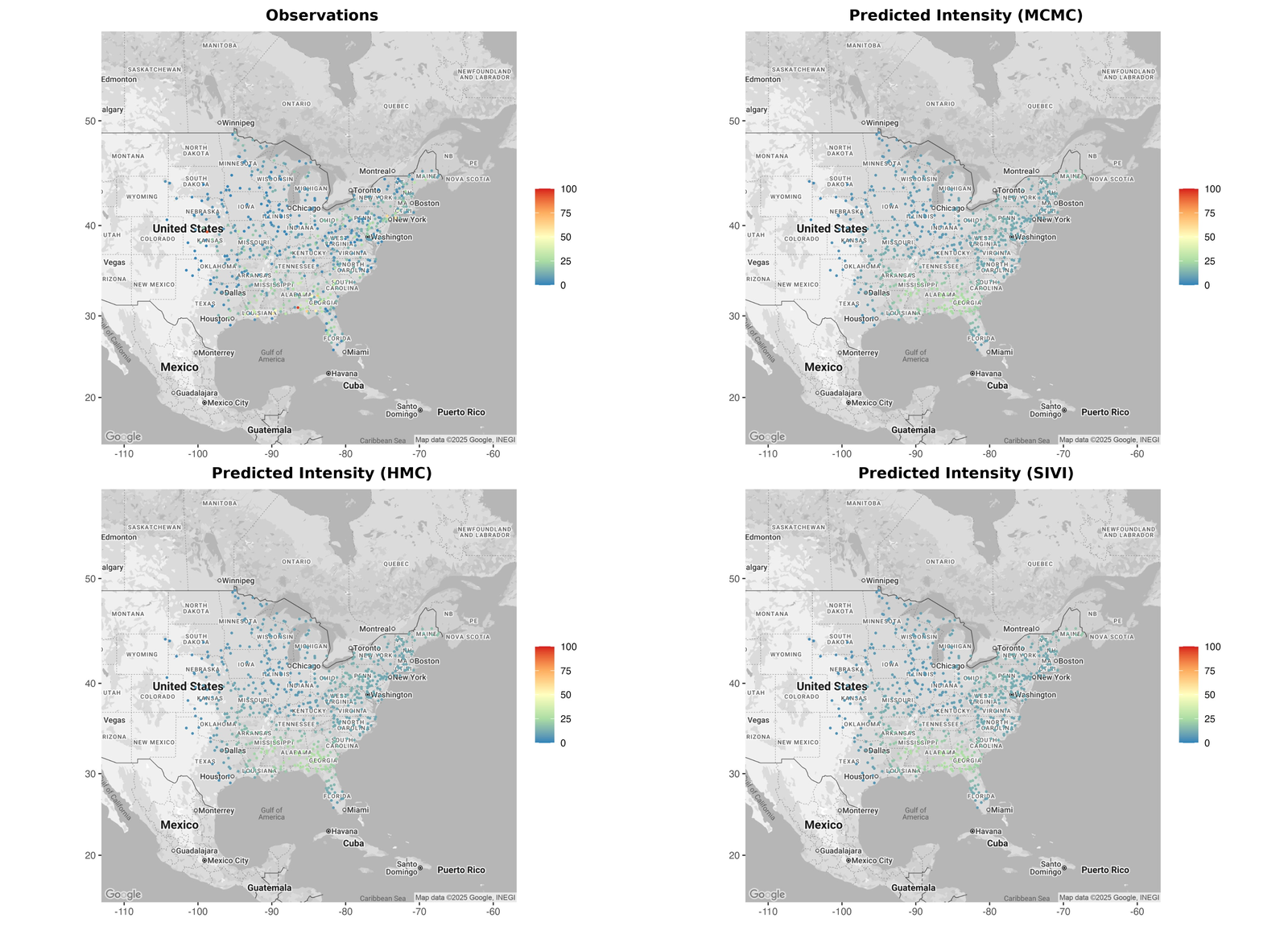}
\caption{True observations (top-left) and predicted intensity surfaces for the North American Blue Jay abundance dataset. The intensity surface is estimated using the basis-SGLMM model fit via Metropolis--Hastings MCMC (top-right), Hamiltonian Monte Carlo (HMC) (bottom-left), and Semi-Implicit Variational Inference (SIVI) (bottom-right). The predicted intensity surfaces under SIVI closely match those from MCMC-based methods.}\label{Fig:BlueJayFigureNB}
\end{center}
\end{figure}

Jointly administered by the U.S. Geological Survey’s Eastern Ecological Science Center and the Canadian Wildlife Service of Environment Canada, the North American Breeding Bird Survey (BBS) \citep{BBS_2022} is a long-term monitoring program that provides annual data on population trends and abundance for over 400 bird species across North America. We analyze observations of the Blue Jay (Cyanocitta cristata) collected in 2018 across 1,593 roadside survey sites (Figure~\ref{Fig:BlueJayFigureNB}). A total of 1,000 sites were used for model training, and the remaining 593 sites were held out for validation.

The analysis is conducted using the basis-SGLMM framework with a negative binomial response and the canonical log link function with the following hierarchical model: 
\begin{align*} 
\tb{Data Model:} \quad &\quad  \bZ|\bbe, \bdel, \kappa \sim \mbox{NB}(\boldsymbol{\mu},\kappa), \\
    \quad&\quad \text{ where} \hspace{0.1cm} \boldsymbol{\mu}=\exp(\bX\bbe+\bgPhi\bdel),\\
\tb{Process Model:} \quad
    &\quad  \bdel| \ssq \sim \mathcal{N}(\bzero,\ssq \bSig_\delta), \\
\tb{Parameter Model:} \quad
     & \quad \boldsymbol{\beta} \sim \mathcal{N}(\mu_{\beta},\Sigma_{\beta}),\\
     \quad& \quad \log \ssq \sim \mathcal{N}(\mu_{\sigma},\ssq_{\sigma}), \\
     \quad & \quad \kappa \sim \mbox{Gamma}(a,b),
\end{align*}
where $\bbe$ and $\bdel$ are the vectors of fixed-effect and basis coefficients, respectively. $\ssq$ denotes the prior marginal variance of $\bdel$, $\kappa$ is the dispersion parameter, and $\bX$ is the design matrix containing geographic coordinates (latitude and longitude) and an intercept term. The basis function matrix $\bgPhi$ is constructed from the leading 10 eigenvectors of a positive-definite covariance matrix defined using a Mat\'ern covariance function with smoothness parameter $\nu = 0.5$ and range $\phi = 0.5$, evaluated at all sampled locations. Prior distributions are specified as $\bbe \sim \mcN([0,0]',100 \bI_2)$, $\log \ssq \sim \mathcal{N}(1,1)$, and $\kappa \sim \mathrm{Gamma}(2,1)$. We compare the scalable SIVI approach with MH and HMC using the same implementation settings as the LST case.

The SIVI approach yields predictive accuracy on par with the MCMC-based approaches, while achieving substantially lower runtimes compared to both MH and HMC. Specifically, the SIVI implementation requires 13.968 seconds, corresponding to a computational speedup of 7.837 and 16.169 relative to MH and HMC, respectively in Table~\ref{tab:combined_realworld_results} (b). Although all three methods produce similar estimates of the latent intensity surface for Blue Jay abundance, as shown in Figure~\ref{Fig:BlueJayFigureNB}, SIVI achieves these results with considerably lower computational costs.

\begin{table}[!ht]
\caption{Predictive performance, computational cost, and speedup comparisons for the MODIS LST (50k locations) and Blue Jay abundance applications under Metropolis--Hastings (MH) MCMC, Hamiltonian Monte Carlo (HMC), and Semi-Implicit Variational Inference (SIVI). Wall times are reported in seconds, and computational speedup is defined as (MCMC walltime)/(SIVI walltime).}
\label{tab:combined_realworld_results}
\begin{center}

\subfloat[MODIS Land Surface Temperature (LST)]{
\begin{tabular}{c c c c}
\hline
& MH MCMC & HMC & SIVI \\
\hline
RMSPE & 1.42 & 1.85 & 1.43 \\
Walltime (sec) & (2741.00) & (12826.73) & (262.05) \\
Computational Speedup & 10.46 & 48.95 & --- \\
\hline
\end{tabular}
}

\vspace{1em}

\subfloat[Blue Jay abundance data]{
\begin{tabular}{c c c c}
\hline
& MH MCMC & HMC & SIVI \\
\hline
RMSPE & 9.848 & 9.846 & 9.857 \\
Walltime (sec) & (109.483) & (225.849) & (13.968) \\
Computational Speedup & 7.837 & 16.169 & --- \\
\hline
\end{tabular}
}

\end{center}
\end{table}

\section{Discussion} 
\label{Sec:Discussion}
We develop an SIVI approach for modeling a wide range of spatially-correlated data types in the continuous spatial domain, including negative binomial, gamma, binary, count, and Gaussian. By integrating SIVI with basis-SGLMMs, our framework extends the applicability of spatial generalized linear mixed models to large-scale settings with spatial random effects, accommodating datasets of size $N=50{,}000$. To the best of our knowledge, this is the first variational framework for continuous-domain non-Gaussian spatial data that accommodates gamma and negative binomial responses, alongside Bernoulli, Poisson, and Gaussian data, within a scalable basis-SGLMM setting. Existing VB methods have largely relied on conjugacy or strong approximations; by combining SIVI with basis-SGLMMs, we enable scalable inference without such restrictions. Through extensive simulation studies, we demonstrate that SIVI achieves results comparable to MCMC-based methods while delivering substantial computational advantages, with speedups ranging from 2-fold to 145-fold. More importantly, in our simulation settings, the SIVI approach does not severely underestimate posterior variance, which is a common limitation of VB methods. We apply our method to land surface temperature and Blue Jay count data, achieving similar computational gains while preserving predictive accuracy and posterior uncertainty. These results demonstrate that SIVI enables scalable inference for large, non-Gaussian spatial datasets where traditional methods are impractical.

For the proposed approach, both the explicit conditional distribution $q(\btheta \mid \bpsi)$ and the implicit mixing distribution $q_\phi(\bpsi)$ must be from reparameterizable families, which limits modeling flexibility \citep{yin2018semi}. Particle SIVI \citep{lim2024particle} relaxes this requirement for the mixing distribution, but the explicit conditional distribution must still be reparameterizable. More flexible, non-reparameterization-based approaches, such as Normalizing Flows \citep{rezende2015variational}, Boosting Variational Inference \citep{guo2016boosting}, and Operator Variational Inference \citep{ranganath2016operator}, could provide alternative variational distributions, particularly for more complex models (e.g., spatial extremes or spatio-temporal). Partitioned models \citep{lee2023scalable} may also improve posterior inference by fitting locally non-stationary models to the spatial domain. In addition, SIVI currently requires fixing scale parameters for the explicit variational distribution, and selecting these hyperparameters becomes increasingly difficult in high-dimensional settings. Allowing these scales to be learned automatically may improve model flexibility but would increase computational costs and introduce additional issues related to model convergence.

 Extending SIVI to support subsampling could therefore enable inference on substantially larger spatial domains. Further opportunities include adapting the method to spatio-temporal and multivariate spatial processes \citep{wikle2019spatio, hamelijnck2021spatio, gneiting2010matern, yarger2023multivariate}, where computational demands remain a major bottleneck. While we focused on eigenvector and bisquare basis functions, many other basis representations, such as wavelets \citep{nychka2002multiresolution} and empirical orthogonal functions \citep{cressie2015statistics}, could be readily embedded into the SIVI basis-SGLMM framework. Subsampling the datasets could help mitigate the computational complexity of evaluating the Evidence Lower Bound (ELBO). Moreover, subsampling naturally introduces gradient noise, which can act as a form of regularization by reducing the risk of overfitting and potentially improving convergence \citep{hoffman2013stochastic, ranganath2013adaptive, ranganath2014black}.

\section*{Data availability statement}
The authors confirm that the data supporting the findings of this study are available within the article and its supplementary materials. The data and code used in this study will be made publicly available as a GitHub repository upon publication.

\section*{Funding}

This research was not supported by any grants or funding. 

\section*{Disclosure statement}
No potential conflict of interest was reported by the authors.

\bibliographystyle{elsarticle-harv} 
\bibliography{references}

\newpage

\def\spacingset#1{\renewcommand{\baselinestretch}%
{#1}\small\normalsize} \spacingset{1}

\begin{center}
\Large \textbf{Supplemental Information for\\
``A Scalable Variational Bayes Approach for Fitting Non-Conjugate Spatial Generalized Linear Mixed Models via Basis Expansions''}
\end{center}
\vspace{1em}

\setcounter{page}{1}
\renewcommand{\thepage}{S.\arabic{page}}
\setcounter{section}{0}
\renewcommand{\thesection}{S.\arabic{section}}
\setcounter{equation}{0}
\renewcommand{\theequation}{S.\arabic{equation}}
\setcounter{table}{0}
\renewcommand{\thetable}{S.\arabic{table}}
\setcounter{figure}{0}
\renewcommand{\thefigure}{S.\arabic{figure}}
\setcounter{algorithm}{0}
\renewcommand{\thealgorithm}{S.\arabic{algorithm}}

\section{SIVI: Reparameterization and Multilayer Perceptron}\label{sec:repar_mlp}
SIVI optimizes the ELBO using stochastic gradients \citep{hoffman2013stochastic}, but the ELBO contains expectations whose gradients cannot be directly evaluated because $q(\btheta \mid \bpsi)$ depends on $\bpsi$. The reparameterization trick \citep{kucukelbir2017automatic} addresses this by expressing samples from $q_{\psi}$ as deterministic transformations of auxiliary noise variables $g_{\psi}(\epsilon)$ where $\epsilon \sim p_0(\epsilon)$ for some distribution $p_0(\cdot)$. This allows gradients to pass through expectations:
\[
\nabla_{\psi}\,\mathbb{E}_{q_{\psi}(g_{\psi}(\epsilon))}[f(g_{\psi}(\epsilon))]
=
\mathbb{E}_{\epsilon \sim p_0}\!\left[
\nabla_{\psi} f\big(g_{\psi}(\epsilon)\big)
\right].
\]
This yields low-variance gradient estimators and enables the use of automatic differentiation for variational optimization.

In this study, samples of $\bpsi$ are obtained by passing random noise $\epsilon$ through a neural network, which  provides a reparameterized representation of $q_\phi(\bpsi)$. Specifically, we employ a multilayer perceptron (MLP) \citep{almeida2020multilayer,popescu2009multilayer} with the output layer defined as $\hat{y}=f (\bW X+\bB)$ where $\hat{y}$ denotes the output layer, $\bW$ the weight matrix, $X$ the input layer, $\bB$ the bias vector, and $f(\cdot)$ a nonlinear activation function (e.g., ReLU, sigmoid, or tanh). $\phi$ corresponds to the weights and biases of an MLP, which are updated at each iteration to increase the flexibility of the posterior approximation. When a deep neural network is used to represent $q_{\phi}(\bpsi)$, the resulting distribution is implicit and the transformation is generally non-invertible. Nevertheless, $q_{\phi}(\bpsi)$ can be highly expressive, allowing complex dependencies among parameters to be captured. Once $\phi$ is updated, the MLP generates samples of $\bpsi$, which are then used to estimate both the latent variables and the observed data. Conceptually, this construction can be viewed hierarchically as $\epsilon \;\;\rightarrow\;\; \bpsi \;\;\rightarrow\;\; \btheta$
where random noise $\epsilon$ is transformed into $\bpsi$ through the MLP, and $\bpsi$ in turn defines the variational distribution of $\btheta$.

\section{Implementation Details for Our Approach}\label{sec:SIVI_implementations}
First, the stopping criterion threshold $\epsilon_{*}$ must be specified. To examine the effect of different stopping criteria on SIVI performance, we conduct a sensitivity analysis using different stopping criteria ($10^{-1}$, $10^{-3}$, and $10^{-4}$) and compare metrics such as RMSPE and walltime across different methods. Using a smaller threshold (e.g., $\epsilon_{*}=10^{-3}$ or $10^{-4}$) increases walltime but yields only modest gains in predictive accuracy (see Section~\ref*{sec:VB2smallcriteria1e1}). Larger thresholds reduce computational times but risk underestimating posterior uncertainty. In practice, the stopping criterion should be set by balancing walltime against the desired predictive accuracy (e.g., RMSPE or AUC). In our simulation study (Section~\ref{sec:SIVIsimulations}), we set the threshold to $\epsilon_{*} = 1 \times 10^{-2}$ (see Algorithm~\ref{alg:sivi_basis}). 

Second, the maximum number of optimization iterations must be specified; we set this to 5,000. Although a larger limit increases walltime, 5,000 iterations were sufficient in our experiments, with the loss function (negative ELBO) typically converging well before reaching this cap. Convergence is assessed using the change in the ELBO between successive iterations, and the maximum number of iterations is used only as a safeguard. As shown in Table~\ref{tab:NB_max_iterations}, the walltime is largely unaffected by the specified maximum number of iterations, suggesting that the algorithm typically converges before reaching the iteration cap (see Supplement S.7 for details).

Third, $K$ is the number of auxiliary samples used to approximate the mixing distribution in Algorithm~\ref{alg:sivi_basis}. The gap between the surrogate ELBO $\underline{\mathcal{L}}_K$ and the true ELBO $\mathcal{L}$ vanishes as $K \to \infty$ (Proposition~2 in \citet{yin2018semi}), and empirically, our results demonstrate that $K = 1{,}000$ is sufficient to achieve MCMC-comparable inference across all settings considered in this study, with predictive accuracy insensitive to the choice of $K$ (see Supplement S.7 for details).

Fourth, the conditional explicit distribution is specified as
\[
q(\btheta \mid \boldsymbol{\psi}) 
= \mathcal{N}\!\big(\btheta;\,\boldsymbol{\psi},\,\mathrm{diag}(\boldsymbol{\tau}^2)\big),
\]
where $\boldsymbol{\psi}$ denotes the output of a neural network transformation of an auxiliary noise variable, i.e., $\boldsymbol{\psi} = T_\phi(\boldsymbol{\epsilon})$ with $\boldsymbol{\epsilon}$ drawn from a base distribution. Specifically, $\tau_{\beta}, \tau_{\delta}, \tau_{\log\sigma^2}, \tau_{\alpha}$ represent the fixed scale parameters (standard deviations) corresponding to the $\beta$, $\delta$, $\log\sigma^2$, and $\log\alpha$ coordinates, respectively. These scale parameters control the amount of smoothing in the SIVI family. Larger scales produce smoother and more stable gradient estimates but reduce flexibility, whereas smaller scales increase flexibility but may lead to higher gradient variance and potential instability during optimization. Although these scales are fixed in our implementation, learning them is a possible extension, albeit with additional computational cost. Although the conditional distribution $q(\btheta|\boldsymbol{\psi})$ has a diagonal covariance matrix for computational stability and efficient optimization, the overall variational distribution $h_\phi(\btheta)$ remains highly expressive due to the implicit mixing distribution $q_\phi(\boldsymbol{\psi})$.

Fifth, the explicit conditional distribution $q(\btheta \mid \bpsi)$ must either be reparameterizable or have a tractable analytic density. In contrast, the implicit mixing distribution $q_\phi(\bpsi)$ is required to be reparameterizable (through a noise transformation $\bpsi = T_\phi(\epsilon)$), since its density is intractable \citep{yin2018semi}. Consequently, the choice of variational distributions is restricted to those that admit a reparameterization, which ensures the feasibility of gradient-based optimization within the SIVI framework.

 Sixth, the dominant computational cost arises from evaluating the linear predictor $\eta = \bX\bbe + \bgPhi\bdel$ across all $N$ observations, which requires matrix–vector multiplications involving the $N\times m$ basis matrix $\bgPhi$. In the SIVI algorithm, the evidence lower bound (ELBO) is estimated using $J$ Monte Carlo samples of the latent parameters and $K$ auxiliary samples for the mixing distribution. This leads to a per-iteration complexity of $O(JNm)$, which simplifies to $O(Nm)$ when $J$ and $K$ are treated as fixed constants. We also compare this complexity with the Metropolis--Hastings and Hamiltonian Monte Carlo algorithms used in our experiments, which operate under the same basis representation and therefore also require $O(Nm)$ operations per iteration. Additionally, SIVI can be combined with subsampling or mini-batching, enabling further scalability to large-scale datasets (see Supplement~S.9 for details).

\begin{figure}[ht]
 \begin{center}
\includegraphics[width=0.6\linewidth]{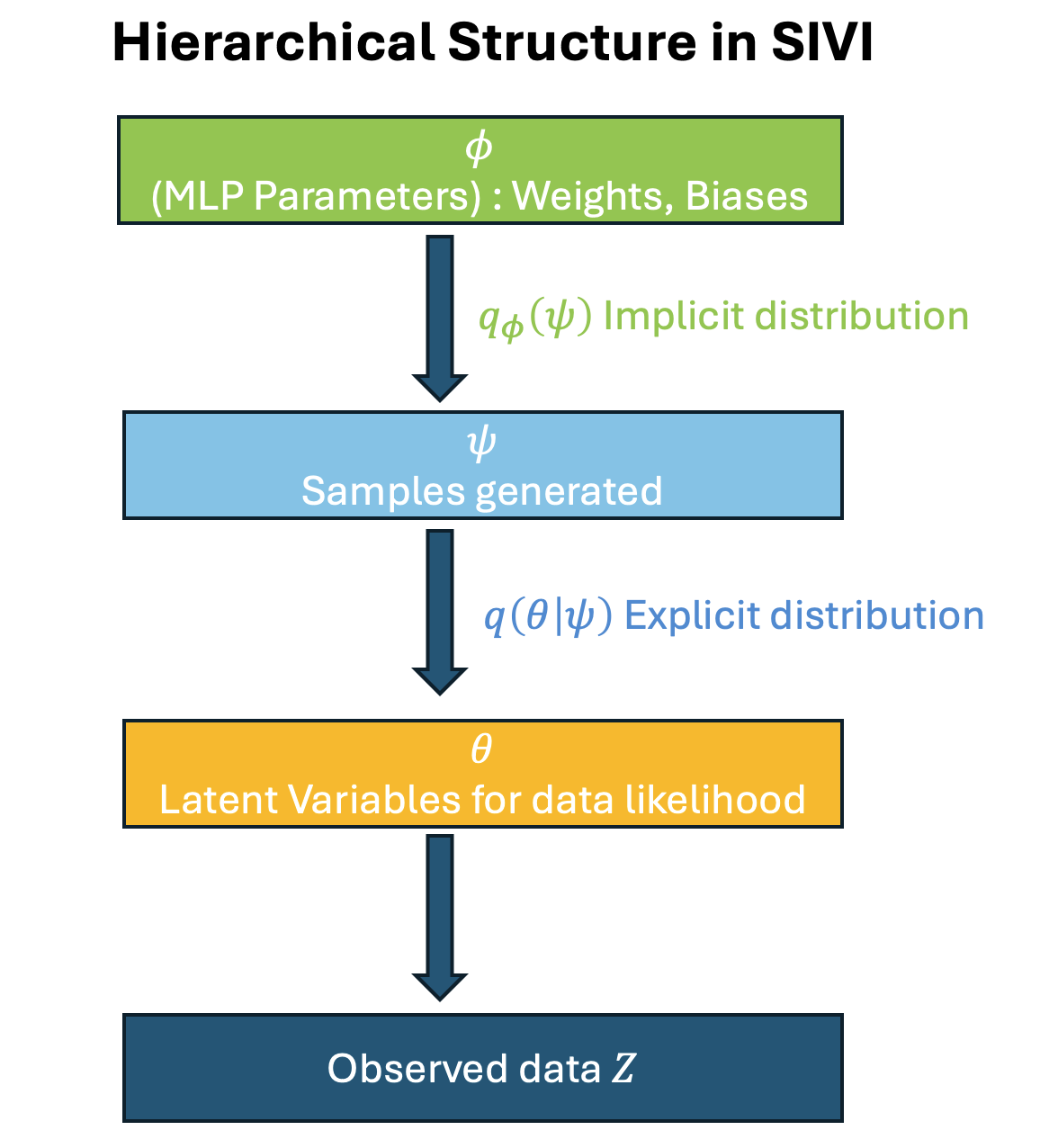}
\caption{$\phi$ denotes the weights and biases of the MLP, which define the implicit distribution, while $\psi$ represents samples generated by the Multilayer Perceptron (MLP). These samples $\psi$ are then used to estimate the latent variables and the observed data.  
}\label{Fig:SIVIHierarchicalStructure}
\end{center}
\end{figure}

\clearpage
\section{Simulation Study Results}
We provide the simulation study results for the Bernoulli, Poisson, and Gaussian cases. Table~\ref{tab:SimulationScenario} provides an overview of the settings for $\nu$ and $\phi$ used in the main simulation study. 
\begin{table}[ht]
\caption{Simulation settings for smoothness parameter $\nu$ and spatial range parameter $\phi$ across different data types including gamma, negative binomial, binary, count and Gaussian.}
\label{tab:SimulationScenario}
\centering
\begin{tabular}{ccc}
\hline
\textbf{Smoothness} & \textbf{Range} & \textbf{Data types} \\
\hline
\multirow{2}{*}{$\nu$=0.5} 
& $\phi$=0.1 & gamma, negative binomial, binary, count, and Gaussian \\
\cline{2-3}
& $\phi$=0.3 & gamma, negative binomial, binary, count, and Gaussian \\
\hline
\multirow{2}{*}{$\nu$=1.5} 
& $\phi$=0.1 & gamma, negative binomial, binary, count, and Gaussian \\
\cline{2-3}
& $\phi$=0.3& gamma, negative binomial, binary, count, and Gaussian \\
\hline
\end{tabular}
\end{table}

\subsection{Data Model: Bernoulli Distribution}

Table~\ref{tab:Binary_results} presents the AUC and walltime for Metropolis--Hastings (MH), Hamiltonian Monte Carlo (HMC), and Semi-Implicit Variational Inference (SIVI) applied to the binary data using the $50$ leading eigenvectors. With the exception of the case $\nu = 1.5$ and $\phi = 0.3$, where all three methods produce identical results, the AUC from SIVI is slightly lower than that of MH and HMC. For example, when $\nu = 0.5$ and $\phi = 0.1$, both MH and HMC achieve an AUC of 0.756, requiring 1305.052 and 195.441 seconds of computation, respectively. In contrast, SIVI attains a comparable AUC of 0.752 while completing in only 21.840 seconds—representing speedups of approximately 59-fold relative to MH and 9-fold relative to HMC. Across the binary simulation studies, SIVI achieves computational gains of 45–59 times over MH and 6–9 times over HMC.

\begin{figure}[ht]
 \begin{center}
\includegraphics[width=1.0\linewidth]{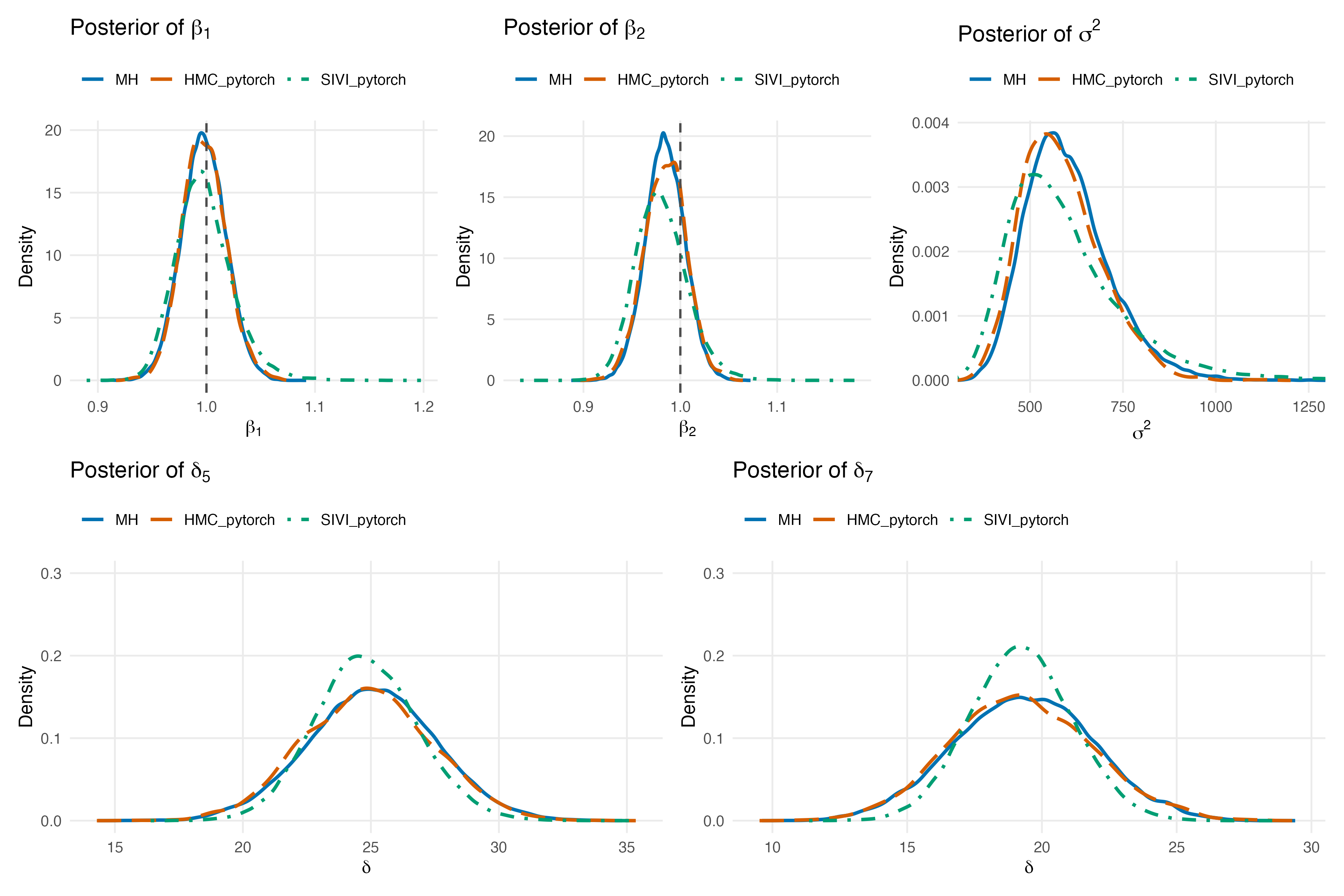}
\caption{Posterior density estimates of selected model parameters under three inference methods for the binary data when $\nu = 1.5$ and $\phi = 0.1$: 
    Metropolis--Hastings (MH), Hamiltonian Monte Carlo (HMC), and Semi-Implicit Variational 
    Inference (SIVI). Panels show results for regression coefficients 
    ($\beta_{1}, \beta_{2}$), variance components ($\sigma^{2}$), and selected spatial 
    random effects ($\delta_{5}, \delta_{7}$). Dashed vertical lines indicate the corresponding 
    true parameter values where available. Overall, all three methods yield nearly indistinguishable 
    posterior distributions, highlighting the accuracy of SIVI relative to MCMC-based approaches.}\label{Fig:SIVIBinaryAll}
\end{center}
\end{figure}

Figure~\ref{Fig:SIVIBinaryAll} further compares the posterior distributions of MH, HMC, and SIVI under the setting $\nu = 1.5$ and $\phi = 0.1$ for regression coefficients ($\beta_{1}, \beta_{2}$), the variance component ($\sigma^{2}$), and selected spatial random effects ($\delta_{5}, \delta_{7}$). The results indicate that the three methods yield broadly similar posterior distributions.

\begin{table}[ht]
\caption{Comparison of AUC, walltime (in seconds), and speedup for MH, HMC, and SIVI under different parameter settings for the Binary case.}
\label{tab:Binary_results}
\centering
\resizebox{\textwidth}{!}{
\begin{tabular}{lcccccccc}
\toprule
 & \multicolumn{2}{c}{MH} & \multicolumn{2}{c}{HMC} & \multicolumn{2}{c}{SIVI} & \multicolumn{2}{c}{Speedup} \\
\cmidrule(lr){2-3}\cmidrule(lr){4-5}\cmidrule(lr){6-7}\cmidrule(lr){8-9}
 & AUC & Walltime & AUC & Walltime & AUC & Walltime & MH/SIVI & HMC/SIVI \\
\midrule
\rowcolor{yellow!30} \textbf{Binary, $\nu=0.5$} & & & & & & & & \\
$\phi=0.1$ & 0.756 & 1305.052 & 0.756 & 195.441 & 0.752 & 21.840 & \textcolor{red}{59.755} & \textcolor{red}{8.949} \\
$\phi=0.3$ & 0.750 & 1308.432 & 0.750 & 186.651 & 0.749 & 24.719 & \textcolor{red}{52.933} & \textcolor{red}{7.551} \\
\midrule
\rowcolor{yellow!30} \textbf{Binary, $\nu=1.5$} & & & & & & & & \\
$\phi=0.1$ & 0.768 & 1242.841 & 0.768 & 181.665 & 0.763 & 23.768 & \textcolor{red}{52.290} & \textcolor{red}{7.643} \\
$\phi=0.3$ & 0.751 & 1256.131 & 0.751 & 184.890 & 0.751 & 27.401 & \textcolor{red}{45.843} & \textcolor{red}{6.748} \\
\bottomrule
\end{tabular}
}
\end{table}

\clearpage
\subsection{Data Model: Poisson Distribution}

Table~\ref{tab:Count_results} reports the RMSPE and walltime for Metropolis--Hastings (MH), Hamiltonian Monte Carlo (HMC), and Semi-Implicit Variational Inference (SIVI) applied to the count data using the $50$ leading eigenvectors. In contrast to the binary case, SIVI achieves results that are virtually identical to those of MH and HMC. For example, when $\nu = 0.5$ and $\phi = 0.1$, all three methods yield an RMSPE of 1.374. However, the computational costs differ considerably: MH requires 1187.895 seconds, HMC requires 198.707 seconds, while SIVI completes in only 50.491 seconds, corresponding to speedups of approximately 23-fold relative to MH and 4-fold relative to HMC.

\begin{figure}[ht]
 \begin{center}
\includegraphics[width=1.0\linewidth]{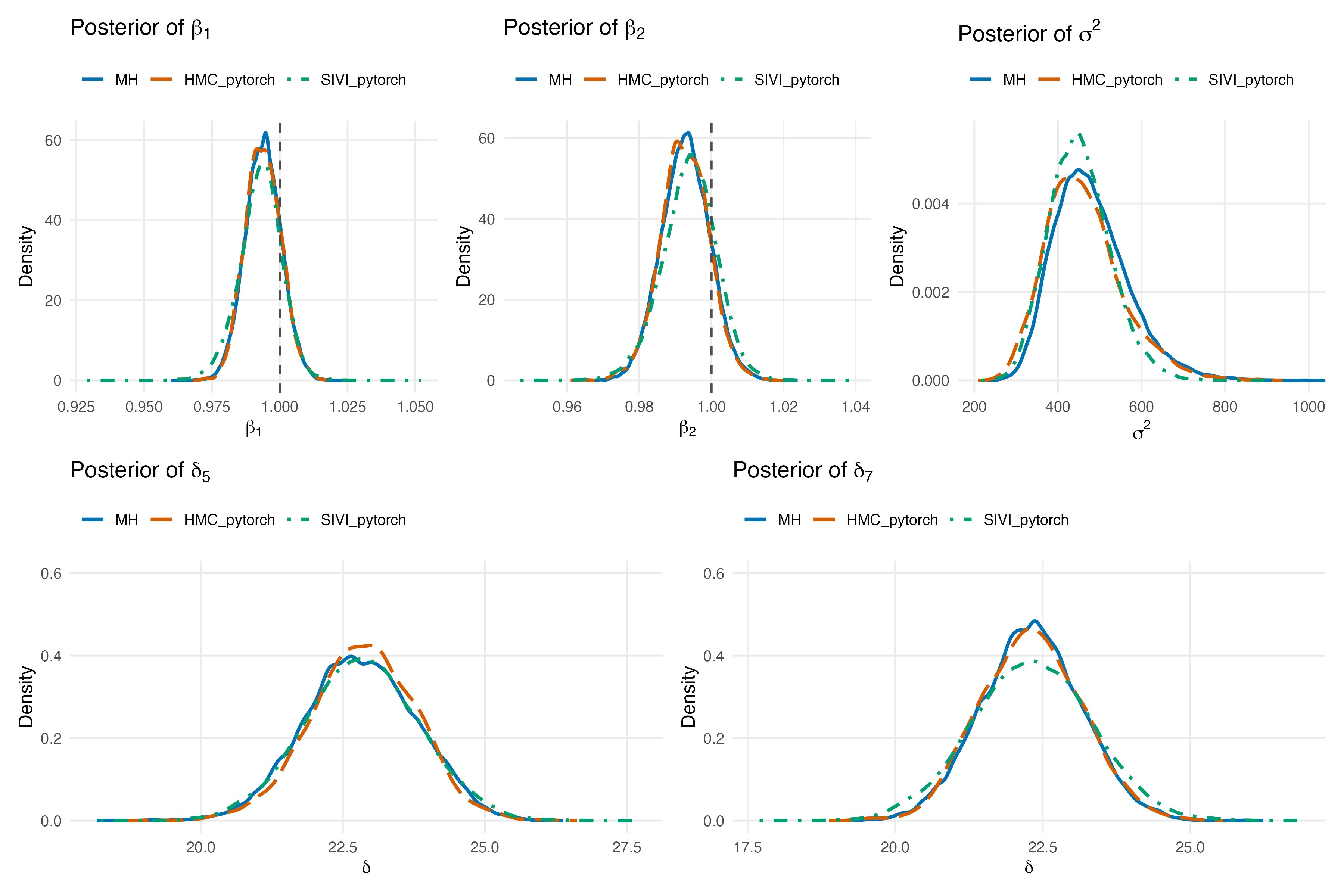}
\caption{Posterior density estimates of selected model parameters under three inference methods for the count data when $\nu = 0.5$ and $\phi = 0.1$: 
    Metropolis--Hastings (MH), Hamiltonian Monte Carlo (HMC), and Semi-Implicit Variational 
    Inference (SIVI). Panels show results for regression coefficients 
    ($\beta_{1}, \beta_{2}$), variance components ($\sigma^{2}$), and selected spatial 
    random effects ($\delta_{5}, \delta_{7}$). Dashed vertical lines indicate the corresponding 
    true parameter values where available. Overall, all three methods yield nearly indistinguishable 
    posterior distributions, highlighting the accuracy of SIVI relative to MCMC-based approaches.}\label{Fig:SIVICountAll}
\end{center}
\end{figure}

Figure~\ref{Fig:SIVICountAll} further examines the posterior distributions of model parameters under the setting ($\nu = 0.5$, $\phi = 0.1$). The results show that regression coefficients ($\beta_{1}, \beta_{2}$), the variance component ($\sigma^2$), and spatial random effects ($\delta$) are highly consistent across all three methods. Together, Figure~\ref{Fig:SIVICountAll} and Table~\ref{tab:Count_results} demonstrate that SIVI produces posterior distributions comparable to those from MH and HMC, while delivering substantial reductions in computation time.

\begin{table}[ht]
\caption{Comparison of RMSPE, walltime (in seconds), and speedup for MH, HMC, and SIVI under different parameter settings for the count case.}
\label{tab:Count_results}
\centering
\resizebox{\textwidth}{!}{
\begin{tabular}{lcccccccc}
\toprule
 & \multicolumn{2}{c}{MH} & \multicolumn{2}{c}{HMC} & \multicolumn{2}{c}{SIVI} & \multicolumn{2}{c}{Speedup} \\
\cmidrule(lr){2-3}\cmidrule(lr){4-5}\cmidrule(lr){6-7}\cmidrule(lr){8-9}
 & RMSPE & Walltime & RMSPE & Walltime & RMSPE & Walltime & MH/SIVI & HMC/SIVI \\
\midrule
\rowcolor{yellow!30} \textbf{Count, $\nu=0.5$} & & & & & & & & \\
$\phi=0.1$ & 1.374 & 1187.895 & 1.374 & 198.707 & 1.374 & 50.491 & \textcolor{red}{23.527} & \textcolor{red}{3.935} \\
$\phi=0.3$ & 1.489 & 1188.624 & 1.489 & 183.725 & 1.489 & 55.633 & \textcolor{red}{21.365} & \textcolor{red}{3.302} \\
\midrule
\rowcolor{yellow!30} \textbf{Count, $\nu=1.5$} & & & & & & & & \\
$\phi=0.1$ & 1.435 & 1304.939 & 1.435 & 184.019 & 1.450 & 53.754 & \textcolor{red}{24.276} & \textcolor{red}{3.423} \\
$\phi=0.3$ & 1.479 & 1341.094 & 1.479 & 184.804 & 1.480 & 57.300 & \textcolor{red}{23.405} & \textcolor{red}{3.225} \\
\bottomrule
\end{tabular}
}
\end{table}

\clearpage

\clearpage
\subsection{Data Model: Gaussian Distribution}

The Gaussian data case exhibits results consistent with those observed for the negative binomial, gamma, binary, and count models. Table~\ref{tab:Gaussian_results} reports the RMSPE and walltime for Metropolis--Hastings (MH), Hamiltonian Monte Carlo (HMC), and Semi-Implicit Variational Inference (SIVI). All three methods achieve an RMSPE of 1, yet their computational costs differ markedly: MH requires 650.205 seconds, HMC requires 62.720 seconds, while SIVI completes in only 23 seconds. This corresponds to speedups of approximately 28-fold relative to MH and 3-fold relative to HMC.

\begin{figure}[ht]
 \begin{center}
\includegraphics[width=1.0\linewidth]{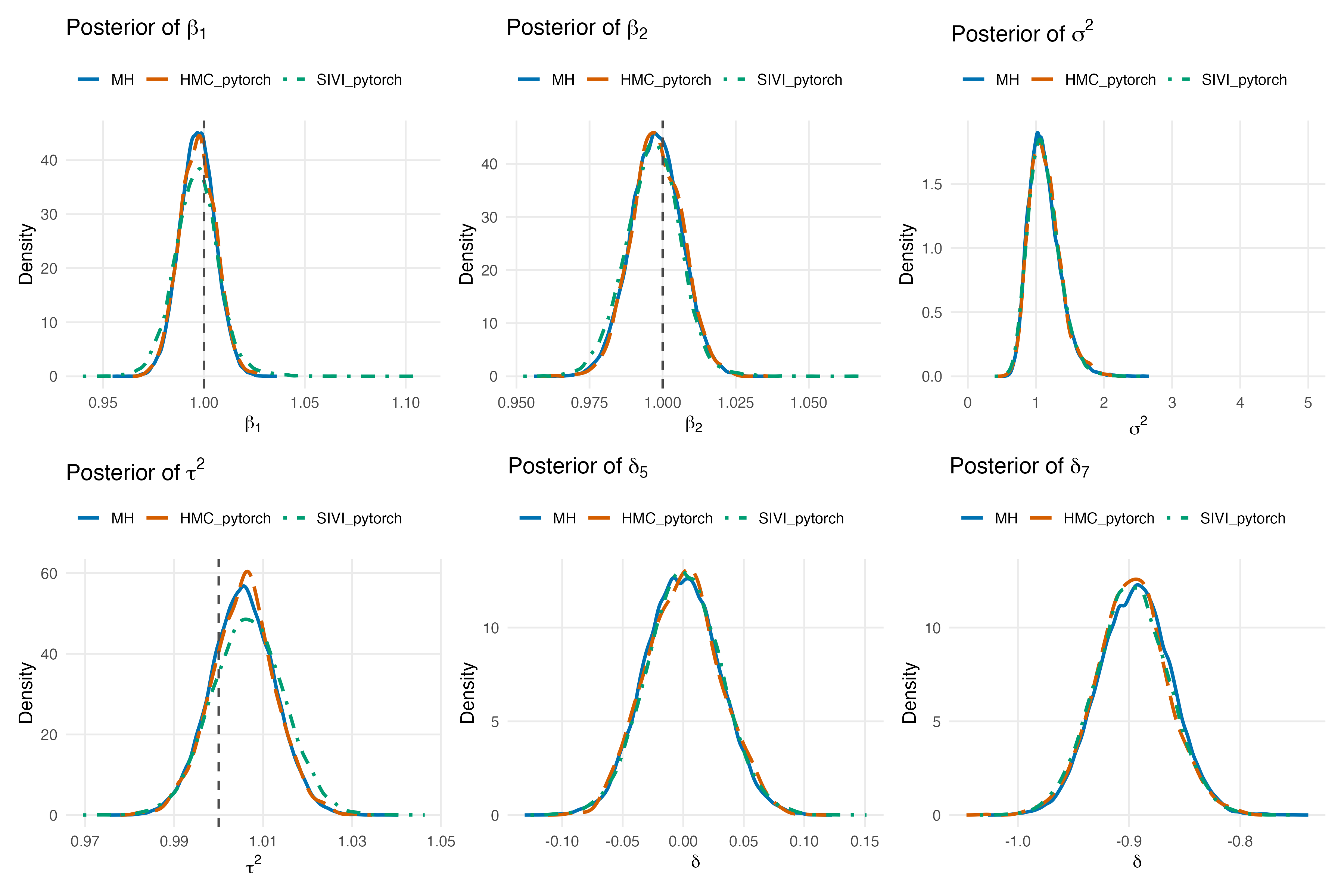}
\caption{Posterior density estimates of selected model parameters under three inference methods for the Gaussian data when $\nu = 0.5$ and $\phi = 0.1$: 
    Metropolis--Hastings (MH), Hamiltonian Monte Carlo (HMC), and Semi-Implicit Variational 
    Inference (SIVI). Panels show results for regression coefficients 
    ($\beta_{1}, \beta_{2}$), variance components ($\sigma^{2}, \tau^{2}$), and selected spatial 
    random effects ($\delta_{5}, \delta_{7}$). Dashed vertical lines indicate the corresponding 
    true parameter values where available. Overall, all three methods yield nearly indistinguishable 
    posterior distributions, highlighting the accuracy of SIVI relative to MCMC-based approaches.}\label{Fig:SIVIGaussianAll}
\end{center}
\end{figure}

Figure~\ref{Fig:SIVIGaussianAll} compares the posterior distributions of model parameters under the setting $\nu = 0.5$ and $\phi = 0.1$. The results indicate that regression coefficients ($\beta_{1}, \beta_{2}$), the variance component ($\sigma^2$), and spatial random effects ($\delta$) are nearly indistinguishable across MH, HMC, and SIVI.

Although variational inference methods, including SIVI, are often criticized for underestimating posterior variance \citep{blei2006variational, hoffman2013stochastic, wu2018fast}, our findings suggest that the proposed SIVI approach not only maintains predictive accuracy but also recovers posterior distributions that closely align with those from MCMC-based methods. This consistency holds across all data models considered—negative binomial, gamma, binary, count, and Gaussian—while providing substantial computational efficiency gains.

\begin{table}[ht]
\caption{Comparison of RMSPE, walltime (in seconds), and speedup for MH, HMC, and SIVI under different parameter settings for the Gaussian case.}
\label{tab:Gaussian_results}
\centering
\resizebox{\textwidth}{!}{
\begin{tabular}{lcccccccc}
\toprule
 & \multicolumn{2}{c}{MH} & \multicolumn{2}{c}{HMC} & \multicolumn{2}{c}{SIVI} & \multicolumn{2}{c}{Speedup} \\
\cmidrule(lr){2-3}\cmidrule(lr){4-5}\cmidrule(lr){6-7}\cmidrule(lr){8-9}
 & RMSPE & Walltime & RMSPE & Walltime & RMSPE & Walltime & MH/SIVI & HMC/SIVI \\
\midrule
\rowcolor{yellow!30} \textbf{Gaussian, $\nu=0.5$} & & & & & & & & \\
$\phi=0.1$ & 1.000 & 650.205 & 1.000 & 62.720 & 1.000 & 23.143 & \textcolor{red}{28.095} & \textcolor{red}{2.710} \\
$\phi=0.3$ & 1.001 & 650.551 & 1.001 & 65.297 & 1.001 & 27.141 & \textcolor{red}{23.970} & \textcolor{red}{2.406} \\
\midrule
\rowcolor{yellow!30} \textbf{Gaussian, $\nu=1.5$} & & & & & & & & \\
$\phi=0.1$ & 1.000 & 684.511 & 1.000 & 60.818 & 1.000 & 26.325 & \textcolor{red}{26.003} & \textcolor{red}{2.310} \\
$\phi=0.3$ & 1.001 & 660.625 & 1.001 & 77.063 & 1.015 & 25.366 & \textcolor{red}{26.044} & \textcolor{red}{3.038} \\
\bottomrule
\end{tabular}
}
\end{table}

\clearpage

\clearpage
\subsection{Smaller datasets Results}

Table~\ref{tab:NB_sample_size} compares the RMSPE, walltime (in seconds), and speedup for MH, HMC, and SIVI under different sample sizes $N$ for the negative binomial model with smoothness parameter $\nu = 0.5$ and spatial range parameter $\phi = 0.1$. When the sample size is $N = 10{,}000$, SIVI yields an RMSPE of 3.953, slightly better than MH and HMC (3.958), with a 55.8-fold speedup over MH. For larger sample sizes such as $N = 25{,}000$ and $50{,}000$, SIVI continues to achieve comparable RMSPE to MH and HMC while maintaining substantial computational advantages. These results demonstrate that SIVI performs well across a range of sample sizes, with the speedup becoming more pronounced as the dataset size increases. 

    \begin{table}[ht]
    \caption{Comparison of RMSPE, walltime (in seconds), and speedup for MH, HMC, and SIVI under different sample sizes $N$ for the negative binomial model when smoothness parameter $\nu = 0.5$ and spatial range parameter $\phi = 0.1$.}
    \label{tab:NB_sample_size}
    \centering
    \resizebox{\textwidth}{!}{
    \begin{tabular}{lcccccccc}
    \toprule
     & \multicolumn{2}{c}{MH} & \multicolumn{2}{c}{HMC} & \multicolumn{2}{c}{SIVI} & \multicolumn{2}{c}{Speedup} \\
    \cmidrule(lr){2-3}\cmidrule(lr){4-5}\cmidrule(lr){6-7}\cmidrule(lr){8-9}
    $N$ & RMSPE & Walltime & RMSPE & Walltime & RMSPE & Walltime & MH/SIVI & HMC/SIVI \\
    \midrule
    \rowcolor{yellow!30} \textbf{negative binomial} & & & & & & & & \\
    10{,}000  & 3.958 & 690.349    & 3.958 & 66.689  & \textcolor{red}{\textbf{3.953}} & 12.365 & \textcolor{red}{\textbf{55.832}} & \textcolor{red}{\textbf{5.393}} \\
    25{,}000  & 3.924 & 1667.936   & 3.924 & 86.408  & \textcolor{red}{\textbf{3.924}} & 29.230 & \textcolor{red}{\textbf{57.063}} & \textcolor{red}{\textbf{2.956}} \\
    50{,}000  & 3.401 & 3225.591   & 3.401 & 128.912 & \textcolor{red}{\textbf{3.401}} & 51.198 & \textcolor{red}{\textbf{63.002}} & \textcolor{red}{\textbf{2.518}} \\
    \bottomrule
    \end{tabular}
    }
\end{table}

\clearpage
\section{Derivation of ELBO Lower Bound in SIVI}
\label{sec:DerivationELBOlower}
SIVI incorporates an implicit mixing distribution $q_{\phi}(\bpsi)$, which distinguishes it from the existing literature that uses explicit variational distributions. In this study, we focus on the lower bound of the ELBO that arises from combining implicit and explicit distributions \citep{yin2018semi}.   When the variational family includes an implicit distribution, the ELBO cannot be evaluated directly since the marginal variational density is intractable \citep{huszar2017variational, mohamed2016learning}. SIVI overcomes this by combining an explicit conditional distribution $q(\btheta \mid \bpsi)$ with an
implicit mixing distribution $q_{\phi}(\bpsi)$. The resulting marginal
variational density is
\[
h_{\phi}(\btheta)
= \mathbb{E}_{\bpsi \sim q_{\phi}(\bpsi)}\big[q(\btheta \mid \bpsi)\big].
\]

Using Jensen's inequality and the convexity of the KL divergence
\citep{cover1999elements}, \citet{yin2018semi} show that
\[
\mathbb{E}_{\bpsi \sim q_{\phi}(\bpsi)}
\text{KL}\!\left(q(\btheta \mid \bpsi)\,\|\,p(\btheta)\right)
\;\ge\;
\text{KL}\!\left( h_{\phi}(\btheta)\,\|\,p(\btheta)\right).
\]

Applying this inequality to the ELBO yields the SIVI lower bound
\[
\underline{\mathcal{L}}
=
\mathbb{E}_{\bpsi \sim q_{\phi}(\bpsi)}
\mathbb{E}_{\btheta \sim q(\btheta \mid \bpsi)}
\bigg[
\log \frac{p(\bZ,\btheta)}{q(\btheta \mid \bpsi)}
\bigg],
\]
which satisfies
\[
\underline{\mathcal{L}}
\;\le\;
\mathcal{L}
=
\mathbb{E}_{\btheta \sim h_{\phi}}
\bigg[
\log \frac{p(\bZ,\btheta)}{h_{\phi}(\btheta)}
\bigg].
\]

The bound $\underline{\mathcal{L}}$ is computable because it depends only on
the explicit density $q(\btheta \mid \bpsi)$, while the implicit mixing
distribution $q_{\phi}(\bpsi)$ enters only through Monte Carlo sampling.
Importantly, \citet{yin2018semi} show that $\underline{\mathcal{L}}$ is
asymptotically exact to the true ELBO; thereby allowing SIVI to retain the flexibility of implicit
variational families while keeping ELBO optimization tractable (e.g, straightforward Monte Carlo estimation) of $\underline{\mathcal{L}}$.

\section{General overview of SIVI Algorithm}
\label{sec:OverviewSIVI}

Algorithm~\ref*{alg:sivi} outlines SIVI when the explicit variational distribution is reparameterizable and the implicit mixing distribution is parameterized by a neural network. The algorithm requires the following inputs: the observed data $\{\bZ_i\}_{i=1}^N$, the joint likelihood $p(\bZ,\btheta)$, an explicit variational distribution $q(\btheta \mid \bpsi)$ with reparameterization $\btheta = f(\epsilon, \bpsi)$ where $\epsilon \sim p(\epsilon)$, and an implicit mixing distribution $q_\phi(\bpsi)$ defined through the neural network transformation $T_\phi(\epsilon)$ with randomness $\epsilon \sim q(\epsilon)$. The goal of the algorithm is to optimize the variational parameter $\phi$, which corresponds to the weights and biases of the implicit neural network.  

First, $\phi$ is initialized randomly. At each iteration, the surrogate lower bound $\underline{L}_K$ is initialized to zero, and the step size $\eta_t$ and the number of auxiliary samples $K$ are specified. The implicit distribution $q_\phi(\bpsi)$ is then approximated by drawing random noise $\epsilon^{(k)} \sim q(\epsilon)$ and mapping it through the implicit neural network, yielding auxiliary samples $\bpsi^{(k)} = T_\phi(\epsilon^{(k)})$.  

To address the intractability of the implicit marginal distribution 
\[
h_\phi(\btheta) = \int q(\btheta \mid \bpsi) q_\phi(\bpsi) \, d\bpsi,
\] 
and specifically the challenge of evaluating $\log h_\phi(\btheta)$ in the entropy term of the ELBO, the algorithm introduces an additional set of $J$ auxiliary samples $\{\bpsi_j\}_{j=1}^J$. For each $\bpsi_j$, a reparameterized sample $\btheta_j = f(\tilde{\epsilon}_j, \bpsi_j)$ is drawn with $\tilde{\epsilon}_j \sim p(\epsilon)$. The term $\log h_\phi(\btheta_j)$ is then approximated via Monte Carlo averaging over the $K$ previously drawn auxiliary samples and the current $\bpsi_j$, leading to the $(K+1)$-sample approximation:
\[
\log h_\phi(\btheta_j) \approx \log \left( \frac{1}{K+1} \Big[ \sum_{k=1}^K q(\btheta_j \mid \bpsi^{(k)}) + q(\btheta_j \mid \bpsi_j) \Big] \right).
\]
Including $\bpsi_j$ among the averaging terms tightens the bound and reduces the risk of degeneracy in the variational approximation.  

Finally, the surrogate lower bound $\underline{L}_K$ is updated using contributions from the log-likelihood and prior terms, and the variational parameter $\phi$ is updated via gradient ascent with step size $\eta_t$. This iterative procedure continues until convergence, yielding the optimized $\phi$ that defines the implicit variational distribution. Figure~\ref{Fig:SIVIalgorithm} provides an overview of the SIVI workflow of Algorithm~\ref*{alg:sivi_basis}.

We use the lower bound $\underline{\mathcal{L}}$ of the ELBO $\mathcal{L}$ to optimize our algorithm. Algorithm~\ref*{alg:sivifull} describes the semi-implicit variational inference (SIVI) procedure when the variational parameters $\xi$ are updated for the explicit variational distribution $q(\btheta \mid \bpsi)$. Algorithm~\ref*{alg:sivi} corresponds to the case where the variational parameters $\xi$ are not included. These algorithms can also be combined with subsampling techniques \citep{yin2018semi}. In our study, we employ Algorithm~\ref*{alg:sivi} and do not use subsampling.

  \begin{algorithm}
  \caption{Semi-Implicit Variational Inference (SIVI) when $\xi$ are fixed \\ ~\citep{yin2018semi}}
  \label{alg:sivi}
  \begin{algorithmic}[1]
  \State \textbf{Input:} Data $\{\bZ_i\}_{1:N}$, joint likelihood $p(\bZ,\btheta)$, explicit variational distribution $q(\btheta \mid \bpsi)$ with reparameterization $\btheta=f(\epsilon,\bpsi)$, $\epsilon\sim p(\epsilon)$, implicit layer neural network $T_\phi(\epsilon)$ and source of randomness $q(\epsilon)$
    \State \textbf{Output:} Implicit variational parameter $\phi$ for the mixing distribution $q_\phi(\bpsi)$
    \Statex
  
  \State Initialize $\phi$ randomly
  \While{not converged}
  \State Set $\underline{L}_{K_t}=0$ and $\eta_t$ as step sizes, and $K_t\ge 0$ as a non-decreasing integer;
  \State Sample $\bpsi^{(k)}=T_\phi(\epsilon^{(k)})$, $\epsilon^{(k)}\sim q(\epsilon)$ for $k=1,\ldots,K_t$; 
    \For{$j=1$ to $J$}
  \State Sample $\bpsi_j=T_\phi(\epsilon_j)$, $\epsilon_j\sim q(\epsilon)$
    \State Sample $\btheta_j=f(\tilde{\epsilon}_j,\bpsi_j)$, $\tilde{\epsilon}_j\sim p(\epsilon)$
\State $\underline{L}_{K_t} \gets \underline{L}_{K_t} 
   + \frac{1}{J} \left\{
   \begin{aligned}[t]
      & -\log\!\frac{1}{K_t+1}\Bigg[\sum_{k=1}^{K_t} q(\btheta_j \mid \bpsi^{(k)}) 
      + q(\btheta_j \mid \bpsi_j)\Bigg] \\
      &+ \log p(\bZ \mid \btheta_j) + \log p(\btheta_j)
   \end{aligned}
   \right\}$
      \EndFor
    \State $t \gets t+1$
      \State $\phi \gets \phi + \eta_t \nabla_\phi \underline{L}_{K_t}\!\left(\{\bpsi^{(k)}\}_{1,K_t}, \{\bpsi_j\}_{1,J}, \{\btheta_j\}_{1,J}\right)$
      \EndWhile
    \end{algorithmic}
    \end{algorithm}

\begin{algorithm}
\caption{Semi-Implicit Variational Inference (SIVI) when $\xi$ are updated ~\citep{yin2018semi}}
\label{alg:sivifull}
\begin{algorithmic}[1]
\State \textbf{Input:} Data $\{\bZ_i\}_{1:N}$, joint likelihood $p(\bZ,\btheta)$, explicit variational distribution $q_\xi(\btheta \mid \bpsi)$ with reparameterization $\btheta=f(\epsilon,\xi,\bpsi)$, $\epsilon\sim p(\epsilon)$, implicit layer neural network $T_\phi(\epsilon)$ and source of randomness $q(\epsilon)$
  \State \textbf{Output:} Explicit Variational parameter $\xi$ for the conditional distribution $q_\xi(\btheta \mid \bpsi)$,
\Statex \hspace{\algorithmicindent} implicit variational parameter $\phi$ for the mixing distribution $q_\phi(\bpsi)$
  \Statex

\State Initialize $\xi$ and $\phi$ randomly
\While{not converged}
\State Set $\underline{L}_{K_t}=0$, $\rho_t$ and $\eta_t$ as step sizes, and $K_t\ge 0$ as a non-decreasing integer;
\State Sample $\bpsi^{(k)}=T_\phi(\epsilon^{(k)})$, $\epsilon^{(k)}\sim q(\epsilon)$ for $k=1,\ldots,K_t$; 
  \For{$j=1$ to $J$}
\State Sample $\bpsi_j=T_\phi(\epsilon_j)$, $\epsilon_j\sim q(\epsilon)$
  \State Sample $\btheta_j=f(\tilde{\epsilon}_j,\xi,\bpsi_j)$, $\tilde{\epsilon}_j\sim p(\epsilon)$
\State $\underline{L}_{K_t} \gets \underline{L}_{K_t} 
   + \frac{1}{J} \left\{
   \begin{aligned}[t]
      & -\log\!\frac{1}{K_t+1}\Bigg[\sum_{k=1}^{K_t} q_\xi(\btheta_j \mid \bpsi^{(k)}) 
      + q_\xi(\btheta_j \mid \bpsi_j)\Bigg] \\
      &+ \log p(\bZ \mid \btheta_j) + \log p(\btheta_j)
   \end{aligned}
   \right\}$
    \EndFor
  \State $t \gets t+1$
    \State $\xi \gets \xi + \rho_t \nabla_\xi \underline{L}_{K_t}\!\left(\{\bpsi^{(k)}\}_{1,K_t}, \{\bpsi_j\}_{1,J}, \{\btheta_j\}_{1,J}\right)$
    \State $\phi \gets \phi + \eta_t \nabla_\phi \underline{L}_{K_t}\!\left(\{\bpsi^{(k)}\}_{1,K_t}, \{\bpsi_j\}_{1,J}, \{\btheta_j\}_{1,J}\right)$
    \EndWhile
  \end{algorithmic}
  \end{algorithm}

\clearpage
\section{Sensitivity Analysis for the stopping criteria in SIVI}
In this section, we examine the effect of different stopping criteria ($10^{-1}$, $10^{-3}$, and $10^{-4}$) on SIVI performance. For each setting, we report RMSPE, walltime (seconds), and speedup for MH, HMC, and SIVI across all data types considered -  negative binomial, gamma, Bernoulli, Poisson, and Gaussian.

\subsection{Tables by using stopping criterion (1e-1) for SIVI} \label{sec:VB2smallcriteria1e1} 

\begin{table}[ht]
\caption{Comparison of RMSPE, walltime (in seconds), and speedup for MH, HMC, and SIVI under different parameter settings for the negative binomial (NB) case, when we use the $10^{-1}$ stopping criterion for SIVI.}
\label{tab:NB_results_e1}
\centering
\resizebox{\textwidth}{!}{
\begin{tabular}{lcccccccc}
\toprule
 & \multicolumn{2}{c}{MH} & \multicolumn{2}{c}{HMC} & \multicolumn{2}{c}{SIVI} & \multicolumn{2}{c}{Speedup} \\
\cmidrule(lr){2-3}\cmidrule(lr){4-5}\cmidrule(lr){6-7}\cmidrule(lr){8-9}
 & RMSPE & Walltime & RMSPE & Walltime & RMSPE & Walltime & MH/SIVI & HMC/SIVI \\
\midrule
\rowcolor{yellow!30} \textbf{NB, $\nu=0.5$} & & & & & & & & \\
$\phi=0.1$ & 3.473 & 3194.339 & 3.473 & 138.831 & 3.475 & 38.054 & \textcolor{red}{83.943} & \textcolor{red}{3.648} \\
$\phi=0.3$ & 3.930 & 3308.680 & 3.930 & 148.911 & 3.938 & 34.994 & \textcolor{red}{94.550} & \textcolor{red}{4.255} \\
\midrule
\rowcolor{yellow!30} \textbf{NB, $\nu=1.5$} & & & & & & & & \\
$\phi=0.1$ & 3.971 & 3162.185 & 3.971 & 141.780 & 3.981 & 35.941 & \textcolor{red}{87.982} & \textcolor{red}{3.945} \\
$\phi=0.3$ & 3.887 & 3334.172 & 3.887 & 168.685 & 3.895 & 36.397 & \textcolor{red}{91.606} & \textcolor{red}{4.635} \\
\bottomrule
\end{tabular}
}
\end{table}

\begin{table}[ht]
\caption{Comparison of RMSPE, walltime (in seconds), and speedup for MH, HMC, and SIVI under different parameter settings for the gamma case, when we use the $10^{-1}$ stopping criterion for SIVI.}
\label{tab:Gamma_results_e1}
\centering
\resizebox{\textwidth}{!}{
\begin{tabular}{lcccccccc}
\toprule
 & \multicolumn{2}{c}{MH} & \multicolumn{2}{c}{HMC} & \multicolumn{2}{c}{SIVI} & \multicolumn{2}{c}{Speedup} \\
\cmidrule(lr){2-3}\cmidrule(lr){4-5}\cmidrule(lr){6-7}\cmidrule(lr){8-9}
 & RMSPE & Walltime & RMSPE & Walltime & RMSPE & Walltime & MH/SIVI & HMC/SIVI \\
\midrule
\rowcolor{yellow!30} \textbf{Gamma, $\nu=0.5$} & & & & & & & & \\
$\phi=0.1$ & 3.945 & 4291.942 & 3.945 & 122.830 & 3.961 & 12.702 & \textcolor{red}{337.907} & \textcolor{red}{9.670} \\
$\phi=0.3$ & 3.855 & 4071.220 & 3.855 & 129.538 & 3.869 & 11.801 & \textcolor{red}{345.002} & \textcolor{red}{10.977} \\
\midrule
\rowcolor{yellow!30} \textbf{Gamma, $\nu=1.5$} & & & & & & & & \\
$\phi=0.1$ & 4.614 & 4204.997 & 4.614 & 116.045 & 4.616 & 11.918 & \textcolor{red}{352.824} & \textcolor{red}{9.737} \\
$\phi=0.3$ & 3.559 & 4577.599 & 3.559 & 138.682 & 3.560 & 11.973 & \textcolor{red}{382.322} & \textcolor{red}{11.583} \\
\bottomrule
\end{tabular}
}
\end{table}

\begin{table}[ht]
\caption{Comparison of RMSPE, walltime (in seconds), and speedup for MH, HMC, and SIVI under different parameter settings for the Count case, when we use the $10^{-1}$ stopping criterion for SIVI.}
\label{tab:Count_results_e1}
\centering
\resizebox{\textwidth}{!}{
\begin{tabular}{lcccccccc}
\toprule
 & \multicolumn{2}{c}{MH} & \multicolumn{2}{c}{HMC} & \multicolumn{2}{c}{SIVI} & \multicolumn{2}{c}{Speedup} \\
\cmidrule(lr){2-3}\cmidrule(lr){4-5}\cmidrule(lr){6-7}\cmidrule(lr){8-9}
 & RMSPE & Walltime & RMSPE & Walltime & RMSPE & Walltime & MH/SIVI & HMC/SIVI \\
\midrule
\rowcolor{yellow!30} \textbf{Count, $\nu=0.5$} & & & & & & & & \\
$\phi=0.1$ & 1.374 & 1187.895 & 1.374 & 218.930 & 1.382 & 23.220 & \textcolor{red}{51.158} & \textcolor{red}{9.428} \\
$\phi=0.3$ & 1.489 & 1188.624 & 1.489 & 203.563 & 1.494 & 27.544 & \textcolor{red}{43.153} & \textcolor{red}{7.390} \\
\midrule
\rowcolor{yellow!30} \textbf{Count, $\nu=1.5$} & & & & & & & & \\
$\phi=0.1$ & 1.435 & 1304.939 & 1.435 & 201.499 & 1.502 & 25.905 & \textcolor{red}{50.375} & \textcolor{red}{7.778} \\
$\phi=0.3$ & 1.479 & 1341.094 & 1.479 & 205.189 & 1.502 & 28.665 & \textcolor{red}{46.785} & \textcolor{red}{7.158} \\
\bottomrule
\end{tabular}
}
\end{table}

\begin{table}[ht]
\caption{Comparison of AUC, walltime (in seconds), and speedup for MH, HMC, and SIVI under different parameter settings for the Binary case, when we use the $10^{-1}$ stopping criterion for SIVI.}
\label{tab:Binary_results_e1}
\centering
\resizebox{\textwidth}{!}{
\begin{tabular}{lcccccccc}
\toprule
 & \multicolumn{2}{c}{MH} & \multicolumn{2}{c}{HMC} & \multicolumn{2}{c}{SIVI} & \multicolumn{2}{c}{Speedup} \\
\cmidrule(lr){2-3}\cmidrule(lr){4-5}\cmidrule(lr){6-7}\cmidrule(lr){8-9}
 & AUC & Walltime & AUC & Walltime & AUC & Walltime & MH/SIVI & HMC/SIVI \\
\midrule
\rowcolor{yellow!30} \textbf{Binary, $\nu=0.5$} & & & & & & & & \\
$\phi=0.1$ & 0.756 & 1305.052 & 0.756 & 216.202 & 0.735 & 9.070 & \textcolor{red}{143.891} & \textcolor{red}{23.838} \\
$\phi=0.3$ & 0.750 & 1308.432 & 0.750 & 227.728 & 0.731 & 9.620 & \textcolor{red}{136.012} & \textcolor{red}{23.673} \\
\midrule
\rowcolor{yellow!30} \textbf{Binary, $\nu=1.5$} & & & & & & & & \\
$\phi=0.1$ & 0.768 & 1242.841 & 0.768 & 181.665 & 0.763 & 8.124 & \textcolor{red}{152.984} & \textcolor{red}{22.362} \\
$\phi=0.3$ & 0.751 & 1256.131 & 0.751 & 184.890 & 0.751 & 9.212 & \textcolor{red}{136.358} & \textcolor{red}{20.071} \\
\bottomrule
\end{tabular}
}
\end{table}

\begin{table}[ht]
\caption{Comparison of RMSPE, walltime (in seconds), and speedup for MH, HMC, and SIVI under different parameter settings for the Gaussian case, when we use the $10^{-1}$ stopping criterion for SIVI.}
\label{tab:Gaussian_results_e1}
\centering
\resizebox{\textwidth}{!}{
\begin{tabular}{lcccccccc}
\toprule
 & \multicolumn{2}{c}{MH} & \multicolumn{2}{c}{HMC} & \multicolumn{2}{c}{SIVI} & \multicolumn{2}{c}{Speedup} \\
\cmidrule(lr){2-3}\cmidrule(lr){4-5}\cmidrule(lr){6-7}\cmidrule(lr){8-9}
 & RMSPE & Walltime & RMSPE & Walltime & RMSPE & Walltime & MH/SIVI & HMC/SIVI \\
\midrule
\rowcolor{yellow!30} \textbf{Gaussian, $\nu=0.5$} & & & & & & & & \\
$\phi=0.1$ & 1.000 & 650.205 & 1.000 & 62.124 & 1.005 & 11.796 & \textcolor{red}{55.122} & \textcolor{red}{5.267} \\
$\phi=0.3$ & 1.001 & 650.551 & 1.001 & 66.697 & 1.002 & 13.270 & \textcolor{red}{49.026} & \textcolor{red}{5.026} \\
\midrule
\rowcolor{yellow!30} \textbf{Gaussian, $\nu=1.5$} & & & & & & & & \\
$\phi=0.1$ & 1.000 & 684.511 & 1.000 & 60.007 & 1.000 & 12.993 & \textcolor{red}{52.685} & \textcolor{red}{4.619} \\
$\phi=0.3$ & 1.001 & 660.625 & 1.001 & 76.154 & 1.015 & 11.907 & \textcolor{red}{55.481} & \textcolor{red}{6.396} \\
\bottomrule
\end{tabular}
}
\end{table}

\clearpage
\begin{table}[ht]
\caption{Walltime quantiles (25\%, 50\%, 75\%) for MH, HMC, and SIVI with SIVI stopping criterion $\epsilon=10^{-1}$, and corresponding speedups (MH/SIVI, HMC/SIVI). Red values highlight median speedups.}
\label{tab:quantile_speedups_eps1e-1}
\centering
\setlength{\tabcolsep}{5pt}    
\renewcommand{\arraystretch}{1.1}
\resizebox{\textwidth}{!}{
\begin{tabular}{l ccc ccc ccc ccc ccc}
\toprule
 & \multicolumn{3}{c}{\textbf{MH}} & \multicolumn{3}{c}{\textbf{HMC}} & \multicolumn{3}{c}{\textbf{SIVI}}
 & \multicolumn{3}{c}{\textbf{Speedup (MH/SIVI)}} & \multicolumn{3}{c}{\textbf{Speedup (HMC/SIVI)}}\\
\cmidrule(lr){2-4}\cmidrule(lr){5-7}\cmidrule(lr){8-10}\cmidrule(lr){11-13}\cmidrule(lr){14-16}
 & 25\% & 50\% & 75\% & 25\% & 50\% & 75\% & 25\% & 50\% & 75\% & 25\% & 50\% & 75\% & 25\% & 50\% & 75\% \\
\midrule

\rowcolor{yellow!30}\multicolumn{16}{l}{\textbf{NB}}\\
\rowcolor{yellow!15}\multicolumn{16}{l}{\textbf{$\nu=0.5$}}\\
$\phi=0.1$ &
3038.053 & 3205.406 & 3332.122 &
117.461  & 129.722  & 153.758  &
29.320   & 38.406   & 45.472   &
103.616  & \textcolor{red}{83.461} & 73.279 &
4.006    & \textcolor{red}{3.378}  & 3.381 \\
$\phi=0.3$ &
3205.791 & 3337.000 & 3431.428 &
128.285  & 138.807  & 167.047  &
26.425   & 33.540   & 44.680   &
121.317  & \textcolor{red}{99.495} & 76.800 &
4.855    & \textcolor{red}{4.139}  & 3.739 \\
\rowcolor{yellow!15}\multicolumn{16}{l}{\textbf{$\nu=1.5$}}\\
$\phi=0.1$ &
3041.410 & 3165.704 & 3265.998 &
119.776  & 138.726  & 159.894  &
27.977   & 34.950   & 42.384   &
108.712  & \textcolor{red}{90.579} & 77.057 &
4.281    & \textcolor{red}{3.969}  & 3.772 \\
$\phi=0.3$ &
3212.261 & 3385.727 & 3450.251 &
152.501  & 162.734  & 182.837  &
26.386   & 33.917   & 47.816   &
121.743  & \textcolor{red}{99.822} & 72.157 &
5.780    & \textcolor{red}{4.798}  & 3.824 \\
\midrule

\rowcolor{yellow!30}\multicolumn{16}{l}{\textbf{Gamma}}\\
\rowcolor{yellow!15}\multicolumn{16}{l}{\textbf{$\nu=0.5$}}\\
$\phi=0.1$ &
3435.886 & 3540.562 & 3684.057 &
105.957  & 122.399  & 140.450  &
8.334   & 12.532   & 16.276   &
412.281  & \textcolor{red}{282.514} & 226.355 &
12.714    & \textcolor{red}{9.767}  & 8.630 \\
$\phi=0.3$ &
3216.671 & 3337.640 & 3478.441 &
111.208  & 130.591  & 148.576  &
6.471   & 10.054   & 15.445   &
497.121  & \textcolor{red}{331.970} & 225.220 &
17.187    & \textcolor{red}{12.989}  & 9.620 \\
\rowcolor{yellow!15}\multicolumn{16}{l}{\textbf{$\nu=1.5$}}\\
$\phi=0.1$ &
3287.160 & 3479.287 & 3607.622 &
95.533  & 116.644  & 133.472  &
7.110   & 12.225   & 15.106   &
462.350  & \textcolor{red}{284.596} & 238.824 &
13.437    & \textcolor{red}{9.541}  & 8.836 \\
$\phi=0.3$ &
2975.072 & 3150.037 & 3357.022 &
121.353  & 139.674  & 153.938  &
6.781   & 10.831   & 17.114   &
438.749  & \textcolor{red}{290.844} & 196.159 &
17.897    & \textcolor{red}{12.896}  & 8.995 \\
\midrule

\rowcolor{yellow!30}\multicolumn{16}{l}{\textbf{Binary}}\\
\rowcolor{yellow!15}\multicolumn{16}{l}{\textbf{$\nu=0.5$}}\\
$\phi=0.1$ &
1102.228 & 1361.181 & 1534.596 &
207.740  & 213.407  & 226.810  &
3.516   & 7.821   & 13.080   &
313.509  & \textcolor{red}{174.041} & 117.324 &
59.088    & \textcolor{red}{27.286}  & 17.340 \\
$\phi=0.3$ &
1100.464 & 1371.999 & 1527.190 &
206.923  & 221.621  & 240.509  &
3.989   & 8.994   & 13.774   &
275.849  & \textcolor{red}{152.547} & 110.872 &
51.869    & \textcolor{red}{24.641}  & 17.461 \\
\rowcolor{yellow!15}\multicolumn{16}{l}{\textbf{$\nu=1.5$}}\\
$\phi=0.1$ &
1060.812 & 1158.577 & 1504.216 &
176.950  & 179.799  & 184.318  &
3.522   & 8.464   & 13.201   &
301.162  & \textcolor{red}{136.883} & 113.951 &
50.235    & \textcolor{red}{21.243}  & 13.963 \\
$\phi=0.3$ &
1048.858 & 1260.038 & 1503.935 &
179.101  & 182.640  & 189.276  &
4.359   & 9.485   & 15.032   &
240.619  & \textcolor{red}{132.852} & 100.052 &
41.088    & \textcolor{red}{19.257}  & 12.592 \\
\midrule

\rowcolor{yellow!30}\multicolumn{16}{l}{\textbf{Count}}\\
\rowcolor{yellow!15}\multicolumn{16}{l}{\textbf{$\nu=0.5$}}\\
$\phi=0.1$ &
1043.119 & 1199.590 & 1228.619 &
201.462  & 220.268  & 236.339  &
16.670   & 23.314   & 27.758   &
62.573  & \textcolor{red}{51.453} & 44.263 &
12.085    & \textcolor{red}{9.448}  & 8.514 \\
$\phi=0.3$ &
1045.639 & 1185.762 & 1239.167 &
186.200  & 207.146  & 221.766  &
18.964   & 24.873   & 34.772   &
55.139  & \textcolor{red}{47.673} & 35.637 &
9.819    & \textcolor{red}{8.328}  & 6.378 \\
\rowcolor{yellow!15}\multicolumn{16}{l}{\textbf{$\nu=1.5$}}\\
$\phi=0.1$ &
1089.074 & 1367.412 & 1515.595 &
175.381  & 196.032  & 220.526  &
16.896   & 27.213   & 34.846   &
64.457  & \textcolor{red}{50.249} & 43.494 &
10.380    & \textcolor{red}{7.204}  & 6.329 \\
$\phi=0.3$ &
1100.422 & 1461.657 & 1525.157 &
176.261  & 203.580  & 225.811  &
18.714   & 29.923   & 38.702   &
58.803  & \textcolor{red}{48.847} & 39.408 &
9.419    & \textcolor{red}{6.803}  & 5.835 \\
\midrule

\rowcolor{yellow!30}\multicolumn{16}{l}{\textbf{Gaussian}}\\
\rowcolor{yellow!15}\multicolumn{16}{l}{\textbf{$\nu=0.5$}}\\
$\phi=0.1$ &
560.900 & 678.775 & 692.222 &
51.765  & 56.302  & 72.026  &
8.763   & 12.609   & 15.088   &
64.010  & \textcolor{red}{53.833} & 45.880 &
5.907    & \textcolor{red}{4.465}  & 4.774 \\
$\phi=0.3$ &
559.822 & 679.687 & 692.743 &
56.092  & 62.514  & 76.800  &
10.482   & 13.363   & 16.101   &
53.407  & \textcolor{red}{50.863} & 43.024 &
5.351    & \textcolor{red}{4.678}  & 4.770 \\
\rowcolor{yellow!15}\multicolumn{16}{l}{\textbf{$\nu=1.5$}}\\
$\phi=0.1$ &
672.899 & 684.509 & 688.272 &
51.950  & 57.459  & 66.417  &
10.911   & 13.677   & 15.635   &
61.671  & \textcolor{red}{50.047} & 44.021 &
4.761    & \textcolor{red}{4.201}  & 4.248 \\
$\phi=0.3$ &
645.736 & 681.685 & 688.288 &
65.718  & 75.579  & 84.850  &
8.090   & 12.135   & 15.854   &
79.814  & \textcolor{red}{56.177} & 43.414 &
8.123    & \textcolor{red}{6.228}  & 5.352 \\
\bottomrule
\end{tabular}
}
\end{table}

\clearpage
\subsection{Tables by using stopping criterion (1e-3) for SIVI} \label{sec:VB2smallcriteria1e3} 

\begin{table}[ht]
\caption{Comparison of RMSPE, walltime (in seconds), and speedup for MH, HMC, and SIVI under different parameter settings for the negative binomial (NB) case, when we use the $10^{-3}$ stopping criterion for SIVI.}
\label{tab:NB_results_e3}
\centering
\resizebox{\textwidth}{!}{
\begin{tabular}{lcccccccc}
\toprule
 & \multicolumn{2}{c}{MH} & \multicolumn{2}{c}{HMC} & \multicolumn{2}{c}{SIVI} & \multicolumn{2}{c}{Speedup} \\
\cmidrule(lr){2-3}\cmidrule(lr){4-5}\cmidrule(lr){6-7}\cmidrule(lr){8-9}
 & RMSPE & Walltime & RMSPE & Walltime & RMSPE & Walltime & MH/SIVI & HMC/SIVI \\
\midrule
\rowcolor{yellow!30} \textbf{NB, $\nu=0.5$} & & & & & & & & \\
$\phi=0.1$ & 3.473 & 3194.339 & 3.473 & 142.706 & 3.473 & 126.521 & \textcolor{red}{25.247} & \textcolor{red}{1.128} \\
$\phi=0.3$ & 3.930 & 3308.680 & 3.930 & 152.007 & 3.930 & 146.967 & \textcolor{red}{22.513} & \textcolor{red}{1.034} \\
\midrule
\rowcolor{yellow!30} \textbf{NB, $\nu=1.5$} & & & & & & & & \\
$\phi=0.1$ & 3.971 & 3162.185 & 3.971 & 140.109 & 3.973 & 135.374 & \textcolor{red}{23.359} & \textcolor{red}{1.035} \\
$\phi=0.3$ & 3.887 & 3334.172 & 3.887 & 169.582 & 3.887 & 129.575 & \textcolor{red}{25.732} & \textcolor{red}{1.309} \\
\bottomrule
\end{tabular}
}
\end{table}

\vspace{3cm}

\begin{table}[ht]
\caption{Comparison of RMSPE, walltime (in seconds), and speedup for MH, HMC, and SIVI under different parameter settings for the gamma case, when we use the $10^{-3}$ stopping criterion for SIVI.}
\label{tab:Gamma_results_style_e3}
\centering
\resizebox{\textwidth}{!}{
\begin{tabular}{lcccccccc}
\toprule
 & \multicolumn{2}{c}{MH} & \multicolumn{2}{c}{HMC} & \multicolumn{2}{c}{SIVI} & \multicolumn{2}{c}{Speedup} \\
\cmidrule(lr){2-3}\cmidrule(lr){4-5}\cmidrule(lr){6-7}\cmidrule(lr){8-9}
 & RMSPE & Walltime & RMSPE & Walltime & RMSPE & Walltime & MH/SIVI & HMC/SIVI \\
\midrule
\rowcolor{yellow!30} \textbf{Gamma, $\nu=0.5$} & & & & & & & & \\
$\phi=0.1$ & 3.945 & 4291.942 & 3.945 & 127.911 & 3.944 & 60.680 & \textcolor{red}{70.731} & \textcolor{red}{2.108} \\
$\phi=0.3$ & 3.855 & 4071.220 & 3.855 & 137.453 & 3.855 & 62.252 & \textcolor{red}{65.399} & \textcolor{red}{2.208} \\
\midrule
\rowcolor{yellow!30} \textbf{Gamma, $\nu=1.5$} & & & & & & & & \\
$\phi=0.1$ & 4.614 & 4204.997 & 4.614 & 121.774 & 4.614 & 60.760 & \textcolor{red}{69.206} & \textcolor{red}{2.004} \\
$\phi=0.3$ & 3.559 & 4577.599 & 3.559 & 145.113 & 3.559 & 88.178 & \textcolor{red}{51.913} & \textcolor{red}{1.646} \\
\bottomrule
\end{tabular}
}
\end{table}

\begin{table}[ht]
\caption{Comparison of AUC, walltime (in seconds), and speedup for MH, HMC, and SIVI under different parameter settings for the Binary case, when we use the $10^{-3}$ stopping criterion for SIVI.}
\label{tab:Binary_results_style_e3}
\centering
\resizebox{\textwidth}{!}{
\begin{tabular}{lcccccccc}
\toprule
 & \multicolumn{2}{c}{MH} & \multicolumn{2}{c}{HMC} & \multicolumn{2}{c}{SIVI} & \multicolumn{2}{c}{Speedup} \\
\cmidrule(lr){2-3}\cmidrule(lr){4-5}\cmidrule(lr){6-7}\cmidrule(lr){8-9}
 & AUC & Walltime & AUC & Walltime & AUC & Walltime & MH/SIVI & HMC/SIVI \\
\midrule
\rowcolor{yellow!30} \textbf{Binary, $\nu=0.5$} & & & & & & & & \\
$\phi=0.1$ & 0.756 & 1305.052 & 0.756 & 239.870 & 0.756 & 72.414 & \textcolor{red}{18.022} & \textcolor{red}{3.312} \\
$\phi=0.3$ & 0.750 & 1308.432 & 0.750 & 226.196 & 0.749 & 63.186 & \textcolor{red}{20.708} & \textcolor{red}{3.580} \\
\midrule
\rowcolor{yellow!30} \textbf{Binary, $\nu=1.5$} & & & & & & & & \\
$\phi=0.1$ & 0.768 & 1242.841 & 0.768 & 235.282 & 0.766 & 69.785 & \textcolor{red}{17.810} & \textcolor{red}{3.372} \\
$\phi=0.3$ & 0.751 & 1256.131 & 0.751 & 228.975 & 0.751 & 76.491 & \textcolor{red}{16.422} & \textcolor{red}{2.993} \\
\bottomrule
\end{tabular}
}
\end{table}

\begin{table}[ht]
\caption{Comparison of RMSPE, walltime (in seconds), and speedup for MH, HMC, and SIVI under different parameter settings for the Count case, when we use the $10^{-3}$ stopping criterion for SIVI.}
\label{tab:Count_results_style_e3}
\centering
\resizebox{\textwidth}{!}{
\begin{tabular}{lcccccccc}
\toprule
 & \multicolumn{2}{c}{MH} & \multicolumn{2}{c}{HMC} & \multicolumn{2}{c}{SIVI} & \multicolumn{2}{c}{Speedup} \\
\cmidrule(lr){2-3}\cmidrule(lr){4-5}\cmidrule(lr){6-7}\cmidrule(lr){8-9}
 & RMSPE & Walltime & RMSPE & Walltime & RMSPE & Walltime & MH/SIVI & HMC/SIVI \\
\midrule
\rowcolor{yellow!30} \textbf{Count, $\nu=0.5$} & & & & & & & & \\
$\phi=0.1$ & 1.374 & 1187.895 & 1.374 & 203.727 & 1.374 & 79.767 & \textcolor{red}{14.892} & \textcolor{red}{2.554} \\
$\phi=0.3$ & 1.489 & 1188.624 & 1.489 & 199.552 & 1.490 & 81.929 & \textcolor{red}{14.508} & \textcolor{red}{2.436} \\
\midrule
\rowcolor{yellow!30} \textbf{Count, $\nu=1.5$} & & & & & & & & \\
$\phi=0.1$ & 1.435 & 1304.939 & 1.435 & 253.268 & 1.437 & 116.549 & \textcolor{red}{11.196} & \textcolor{red}{2.173} \\
$\phi=0.3$ & 1.479 & 1341.094 & 1.479 & 250.507 & 1.481 & 108.057 & \textcolor{red}{12.411} & \textcolor{red}{2.318} \\
\bottomrule
\end{tabular}
}
\end{table}

\begin{table}[ht]
\caption{Comparison of RMSPE, walltime (in seconds), and speedup for MH, HMC, and SIVI under different parameter settings for the Gaussian case, when we use the $10^{-3}$ stopping criterion for SIVI.}
\label{tab:Gaussian_results_eps1e-3}
\centering
\resizebox{\textwidth}{!}{
\begin{tabular}{lcccccccc}
\toprule
 & \multicolumn{2}{c}{MH} & \multicolumn{2}{c}{HMC} & \multicolumn{2}{c}{SIVI} & \multicolumn{2}{c}{Speedup} \\
\cmidrule(lr){2-3}\cmidrule(lr){4-5}\cmidrule(lr){6-7}\cmidrule(lr){8-9}
 & RMSPE & Walltime & RMSPE & Walltime & RMSPE & Walltime & MH/SIVI & HMC/SIVI \\
\midrule
\rowcolor{yellow!30} \textbf{Gaussian, $\nu=0.5$} & & & & & & & & \\
$\phi=0.1$ & 1.000 & 650.205 & 1.000 & 72.100 & 1.000 & 44.465 & \textcolor{red}{14.623} & \textcolor{red}{1.621} \\
$\phi=0.3$ & 1.001 & 650.551 & 1.001 & 78.032 & 1.001 & 59.823 & \textcolor{red}{10.875} & \textcolor{red}{1.304} \\
\midrule
\rowcolor{yellow!30} \textbf{Gaussian, $\nu=1.5$} & & & & & & & & \\
$\phi=0.1$ & 1.000 & 684.511 & 1.000 & 60.907 & 1.000 & 42.352 & \textcolor{red}{16.162} & \textcolor{red}{1.438} \\
$\phi=0.3$ & 1.001 & 660.625 & 1.001 & 78.869 & 1.001 & 68.319 & \textcolor{red}{9.670} & \textcolor{red}{1.154} \\
\bottomrule
\end{tabular}
}
\end{table}

\clearpage
\begin{table}[ht]
\caption{Walltime quantiles (25\%, 50\%, 75\%) for MH, HMC, and SIVI with SIVI stopping criterion $\epsilon=10^{-3}$, and corresponding speedups (MH/SIVI, HMC/SIVI). Red values highlight median speedups.
}
\label{tab:quantile_speedups_style_e3}
\centering
\setlength{\tabcolsep}{5pt}
\renewcommand{\arraystretch}{1.12}

\newcommand{\spdtri}[3]{#1 & \textcolor{red}{#2} & #3}

\resizebox{\textwidth}{!}{
\begin{tabular}{l ccc ccc ccc ccc ccc}
\toprule
& \multicolumn{3}{c}{\textbf{MH}} &
  \multicolumn{3}{c}{\textbf{HMC}} &
  \multicolumn{3}{c}{\textbf{SIVI}} &
  \multicolumn{3}{c}{\textbf{Speedup (MH/SIVI)}} &
  \multicolumn{3}{c}{\textbf{Speedup (HMC/SIVI)}} \\
\cmidrule(lr){2-4}\cmidrule(lr){5-7}\cmidrule(lr){8-10}\cmidrule(lr){11-13}\cmidrule(lr){14-16}
& 25\% & \textbf{50\%} & 75\% &
  25\% & \textbf{50\%} & 75\% &
  25\% & \textbf{50\%} & 75\% &
  25\% & \textbf{\textcolor{red}{50\%}} & 75\% &
  25\% & \textbf{\textcolor{red}{50\%}} & 75\% \\
\midrule

\rowcolor{yellow!30}\multicolumn{16}{l}{\textbf{NB}}\\
\rowcolor{yellow!15}\multicolumn{16}{l}{\textbf{$\nu=0.5$}}\\
\rowcolor{gray!10}
$\phi=0.1$ &
3038.053 & 3205.406 & 3332.122 &
120.350  & 135.022  & 159.146  &
86.440   & 109.643   & 149.906   &
\spdtri{35.146}{29.235}{22.228} &
\spdtri{1.392}{1.231}{1.062} \\
\rowcolor{gray!10}
$\phi=0.3$ &
3205.791 & 3337.000 & 3431.428 &
131.277  & 143.392  & 171.819  &
86.529   & 130.681   & 188.672   &
\spdtri{37.049}{25.535}{18.187} &
\spdtri{1.517}{1.097}{0.911} \\
\rowcolor{yellow!15}\multicolumn{16}{l}{\textbf{$\nu=1.5$}}\\
\rowcolor{gray!10}
$\phi=0.1$ &
3041.410 & 3165.704 & 3265.998 &
118.720  & 134.266  & 157.311  &
88.205   & 124.257   & 166.581   &
\spdtri{34.481}{25.477}{19.606} &
\spdtri{1.346}{1.081}{0.944} \\
\rowcolor{gray!10}
$\phi=0.3$ &
3212.261 & 3385.727 & 3450.251 &
154.199  & 161.309  & 182.403  &
79.395   & 118.681   & 166.529   &
\spdtri{40.459}{28.528}{20.719} &
\spdtri{1.942}{1.359}{1.095} \\
\midrule

\rowcolor{yellow!30}\multicolumn{16}{l}{\textbf{Gamma}}\\
\rowcolor{yellow!15}\multicolumn{16}{l}{\textbf{$\nu=0.5$}}\\
\rowcolor{gray!10}
$\phi=0.1$ &
3435.886 & 3540.562 & 3684.057 &
109.259  & 129.318  & 141.412  &
42.996   & 57.757   & 75.374   &
\spdtri{79.911}{61.301}{48.877} &
\spdtri{2.541}{2.239}{1.876} \\
\rowcolor{gray!10}
$\phi=0.3$ &
3216.671 & 3337.640 & 3478.441 &
115.914  & 138.752  & 156.596  &
45.259   & 56.480   & 74.066   &
\spdtri{71.073}{59.095}{46.964} &
\spdtri{2.561}{2.457}{2.114} \\
\rowcolor{yellow!15}\multicolumn{16}{l}{\textbf{$\nu=1.5$}}\\
\rowcolor{gray!10}
$\phi=0.1$ &
3287.160 & 3479.287 & 3607.622 &
97.908  & 122.440  & 141.675  &
43.365   & 60.556   & 76.850   &
\spdtri{75.803}{57.455}{46.944} &
\spdtri{2.258}{2.022}{1.844} \\
\rowcolor{gray!10}
$\phi=0.3$ &
2975.072 & 3150.037 & 3357.022 &
124.933  & 143.993  & 161.639  &
50.909   & 74.058   & 127.901   &
\spdtri{58.439}{42.535}{26.247} &
\spdtri{2.454}{1.944}{1.264} \\
\midrule

\rowcolor{yellow!30}\multicolumn{16}{l}{\textbf{Binary}}\\
\rowcolor{yellow!15}\multicolumn{16}{l}{\textbf{$\nu=0.5$}}\\
\rowcolor{gray!10}
$\phi=0.1$ &
1102.228 & 1361.181 & 1534.596 &
223.487  & 239.678  & 253.233  &
43.574   & 58.575   & 92.058   &
\spdtri{25.296}{23.238}{16.670} &
\spdtri{5.129}{4.092}{2.751} \\
\rowcolor{gray!10}
$\phi=0.3$ &
1100.464 & 1371.999 & 1527.190 &
207.539  & 221.961  & 238.598  &
44.133   & 55.674   & 74.460   &
\spdtri{24.935}{24.643}{20.510} &
\spdtri{4.703}{3.987}{3.204} \\
\rowcolor{yellow!15}\multicolumn{16}{l}{\textbf{$\nu=1.5$}}\\
\rowcolor{gray!10}
$\phi=0.1$ &
1060.812 & 1158.577 & 1504.216 &
220.718  & 231.518  & 249.354  &
46.946   & 64.335   & 86.280   &
\spdtri{22.596}{18.009}{17.434} &
\spdtri{4.702}{3.599}{2.890} \\
\rowcolor{gray!10}
$\phi=0.3$ &
1048.858 & 1260.038 & 1503.935 &
206.463  & 218.830  & 248.145  &
51.856   & 68.235   & 95.003   &
\spdtri{20.226}{18.466}{15.830} &
\spdtri{3.981}{3.207}{2.612} \\
\midrule

\rowcolor{yellow!30}\multicolumn{16}{l}{\textbf{Count}}\\
\rowcolor{yellow!15}\multicolumn{16}{l}{\textbf{$\nu=0.5$}}\\
\rowcolor{gray!10}
$\phi=0.1$ &
1043.119 & 1199.590 & 1228.619 &
183.375  & 203.553  & 221.501  &
53.751   & 90.772   & 106.684   &
\spdtri{19.407}{13.215}{11.516} &
\spdtri{3.412}{2.242}{2.076} \\
\rowcolor{gray!10}
$\phi=0.3$ &
1045.639 & 1185.762 & 1239.167 &
184.518  & 203.658  & 215.995  &
58.996   & 91.017   & 107.471   &
\spdtri{17.724}{13.028}{11.530} &
\spdtri{3.128}{2.238}{2.010} \\
\rowcolor{yellow!15}\multicolumn{16}{l}{\textbf{$\nu=1.5$}}\\
\rowcolor{gray!10}
$\phi=0.1$ &
1089.074 & 1367.412 & 1515.595 &
200.664  & 255.867  & 306.601  &
81.532   & 108.852   & 150.878   &
\spdtri{13.358}{12.562}{10.045} &
\spdtri{2.461}{2.351}{2.032} \\
\rowcolor{gray!10}
$\phi=0.3$ &
1100.422 & 1461.657 & 1525.157 &
215.875  & 254.626  & 278.025  &
78.812   & 107.563   & 145.245   &
\spdtri{13.963}{13.589}{10.501} &
\spdtri{2.739}{2.367}{1.914} \\
\midrule

\rowcolor{yellow!30}\multicolumn{16}{l}{\textbf{Gaussian}}\\
\rowcolor{yellow!15}\multicolumn{16}{l}{\textbf{$\nu=0.5$}}\\
\rowcolor{gray!10}
$\phi=0.1$ &
560.900 & 678.775 & 692.222 &
61.237  & 67.676  & 81.171  &
28.500   & 41.047   & 60.140   &
\spdtri{19.681}{16.536}{11.510} &
\spdtri{2.149}{1.649}{1.350} \\
\rowcolor{gray!10}
$\phi=0.3$ &
559.822 & 679.687 & 692.743 &
67.691  & 74.976  & 89.590  &
33.497   & 50.201   & 80.250   &
\spdtri{16.712}{13.539}{8.632} &
\spdtri{2.021}{1.493}{1.116} \\
\rowcolor{yellow!15}\multicolumn{16}{l}{\textbf{$\nu=1.5$}}\\
\rowcolor{gray!10}
$\phi=0.1$ &
672.899 & 684.509 & 688.272 &
50.823  & 56.334  & 70.857  &
28.423   & 37.229   & 49.793   &
\spdtri{23.675}{18.386}{13.823} &
\spdtri{1.788}{1.513}{1.423} \\
\rowcolor{gray!10}
$\phi=0.3$ &
645.736 & 681.685 & 688.288 &
68.573  & 75.666  & 90.374  &
40.147   & 63.436   & 93.278   &
\spdtri{16.084}{10.746}{7.379} &
\spdtri{1.708}{1.193}{0.969} \\
\bottomrule
\end{tabular}
}
\end{table}

\clearpage
\subsection{Tables by using stopping criterion (1e-4) for SIVI} \label{sec:VB2smallcriteria1e4} 

\begin{table}[ht]
\caption{Comparison of RMSPE, walltime (in seconds), and speedup for MH, HMC, and SIVI under different parameter settings for the negative binomial (NB) case, when we use the $10^{-4}$ stopping criterion for SIVI.}
\label{tab:NB_results_eps1e-4}
\centering
\resizebox{\textwidth}{!}{
\begin{tabular}{lcccccccc}
\toprule
 & \multicolumn{2}{c}{MH} & \multicolumn{2}{c}{HMC} & \multicolumn{2}{c}{SIVI} & \multicolumn{2}{c}{Speedup} \\
\cmidrule(lr){2-3}\cmidrule(lr){4-5}\cmidrule(lr){6-7}\cmidrule(lr){8-9}
 & RMSPE & Walltime & RMSPE & Walltime & RMSPE & Walltime & MH/SIVI & HMC/SIVI \\
\midrule
\rowcolor{yellow!30} \textbf{NB, $\nu=0.5$} & & & & & & & & \\
$\phi=0.1$ & 3.473 & 3194.339 & 3.473 & 127.039 & 3.473 & 124.081 & \textcolor{red}{25.744} & \textcolor{red}{1.024} \\
$\phi=0.3$ & 3.930 & 3308.680 & 3.930 & 136.123 & 3.930 & 144.391 & \textcolor{red}{22.915} & \textcolor{red}{0.943} \\
\midrule
\rowcolor{yellow!30} \textbf{NB, $\nu=1.5$} & & & & & & & & \\
$\phi=0.1$ & 3.971 & 3162.185 & 3.971 & 147.604 & 3.972 & 157.521 & \textcolor{red}{20.075} & \textcolor{red}{0.937} \\
$\phi=0.3$ & 3.887 & 3334.172 & 3.887 & 159.560 & 3.887 & 149.573 & \textcolor{red}{22.291} & \textcolor{red}{1.067} \\
\bottomrule
\end{tabular}
}
\end{table}

\vspace{3cm}

\begin{table}[ht]
\caption{Comparison of RMSPE, walltime (in seconds), and speedup for MH, HMC, and SIVI under different parameter settings for the gamma case, when we use the $10^{-4}$ stopping criterion for SIVI.}
\label{tab:Gamma_results_eps1e-4}
\centering
\resizebox{\textwidth}{!}{
\begin{tabular}{lcccccccc}
\toprule
 & \multicolumn{2}{c}{MH} & \multicolumn{2}{c}{HMC} & \multicolumn{2}{c}{SIVI} & \multicolumn{2}{c}{Speedup} \\
\cmidrule(lr){2-3}\cmidrule(lr){4-5}\cmidrule(lr){6-7}\cmidrule(lr){8-9}
 & RMSPE & Walltime & RMSPE & Walltime & RMSPE & Walltime & MH/SIVI & HMC/SIVI \\
\midrule
\rowcolor{yellow!30} \textbf{Gamma, $\nu=0.5$} & & & & & & & & \\
$\phi=0.1$ & 3.945 & 4291.942 & 3.945 & 112.035 & 3.945 & 48.140 & \textcolor{red}{89.156} & \textcolor{red}{2.327} \\
$\phi=0.3$ & 3.855 & 4071.220 & 3.855 & 118.661 & 3.855 & 63.923 & \textcolor{red}{63.689} & \textcolor{red}{1.856} \\
\midrule
\rowcolor{yellow!30} \textbf{Gamma, $\nu=1.5$} & & & & & & & & \\
$\phi=0.1$ & 4.614 & 4204.997 & 4.614 & 120.981 & 4.614 & 54.001 & \textcolor{red}{77.868} & \textcolor{red}{2.240} \\
$\phi=0.3$ & 3.559 & 4577.599 & 3.559 & 145.155 & 3.559 & 81.555 & \textcolor{red}{56.129} & \textcolor{red}{1.780} \\
\bottomrule
\end{tabular}
}
\end{table}

\begin{table}[ht]
\caption{Comparison of RMSPE, walltime (in seconds), and speedup for MH, HMC, and SIVI under different parameter settings for the Count case, when we use the $10^{-4}$ stopping criterion for SIVI.}
\label{tab:Count_results_eps1e-4}
\centering
\resizebox{\textwidth}{!}{
\begin{tabular}{lcccccccc}
\toprule
 & \multicolumn{2}{c}{MH} & \multicolumn{2}{c}{HMC} & \multicolumn{2}{c}{SIVI} & \multicolumn{2}{c}{Speedup} \\
\cmidrule(lr){2-3}\cmidrule(lr){4-5}\cmidrule(lr){6-7}\cmidrule(lr){8-9}
 & RMSPE & Walltime & RMSPE & Walltime & RMSPE & Walltime & MH/SIVI & HMC/SIVI \\
\midrule
\rowcolor{yellow!30} \textbf{Count, $\nu=0.5$} & & & & & & & & \\
$\phi=0.1$ & 1.374 & 1187.895 & 1.374 & 198.726 & 1.374 & 76.099 & \textcolor{red}{15.610} & \textcolor{red}{2.611} \\
$\phi=0.3$ & 1.489 & 1188.624 & 1.489 & 183.361 & 1.489 & 82.873 & \textcolor{red}{14.343} & \textcolor{red}{2.213} \\
\midrule
\rowcolor{yellow!30} \textbf{Count, $\nu=1.5$} & & & & & & & & \\
$\phi=0.1$ & 1.435 & 1304.939 & 1.435 & 185.134 & 1.450 & 75.955 & \textcolor{red}{17.181} & \textcolor{red}{2.437} \\
$\phi=0.3$ & 1.479 & 1341.094 & 1.479 & 185.299 & 1.480 & 84.551 & \textcolor{red}{15.861} & \textcolor{red}{2.192} \\
\bottomrule
\end{tabular}
}
\end{table}

\begin{table}[ht]
\caption{Comparison of AUC, walltime (in seconds), and speedup for MH, HMC, and SIVI under different parameter settings for the Binary case, when we use the $10^{-4}$ stopping criterion for SIVI.}
\label{tab:Binary_results_eps1e-4}
\centering
\resizebox{\textwidth}{!}{
\begin{tabular}{lcccccccc}
\toprule
 & \multicolumn{2}{c}{MH} & \multicolumn{2}{c}{HMC} & \multicolumn{2}{c}{SIVI} & \multicolumn{2}{c}{Speedup} \\
\cmidrule(lr){2-3}\cmidrule(lr){4-5}\cmidrule(lr){6-7}\cmidrule(lr){8-9}
 & AUC & Walltime & AUC & Walltime & AUC & Walltime & MH/SIVI & HMC/SIVI \\
\midrule
\rowcolor{yellow!30} \textbf{Binary, $\nu=0.5$} & & & & & & & & \\
$\phi=0.1$ & 0.756 & 1305.052 & 0.756 & 221.527 & 0.756 & 58.770 & \textcolor{red}{22.206} & \textcolor{red}{3.769} \\
$\phi=0.3$ & 0.750 & 1308.432 & 0.750 & 244.570 & 0.750 & 66.535 & \textcolor{red}{19.665} & \textcolor{red}{3.676} \\
\midrule
\rowcolor{yellow!30} \textbf{Binary, $\nu=1.5$} & & & & & & & & \\
$\phi=0.1$ & 0.768 & 1242.841 & 0.768 & 181.665 & 0.763 & 69.110 & \textcolor{red}{17.984} & \textcolor{red}{2.629} \\
$\phi=0.3$ & 0.751 & 1256.131 & 0.751 & 184.890 & 0.751 & 79.191 & \textcolor{red}{15.862} & \textcolor{red}{2.335} \\
\bottomrule
\end{tabular}
}
\end{table}

\begin{table}[ht]
\caption{Comparison of RMSPE, walltime (in seconds), and speedup for MH, HMC, and SIVI under different parameter settings for the Gaussian case, when we use the $10^{-4}$ stopping criterion for SIVI.}
\label{tab:Gaussian_results_eps1e-4}
\centering
\resizebox{\textwidth}{!}{
\begin{tabular}{lcccccccc}
\toprule
 & \multicolumn{2}{c}{MH} & \multicolumn{2}{c}{HMC} & \multicolumn{2}{c}{SIVI} & \multicolumn{2}{c}{Speedup} \\
\cmidrule(lr){2-3}\cmidrule(lr){4-5}\cmidrule(lr){6-7}\cmidrule(lr){8-9}
 & RMSPE & Walltime & RMSPE & Walltime & RMSPE & Walltime & MH/SIVI & HMC/SIVI \\
\midrule
\rowcolor{yellow!30} \textbf{Gaussian, $\nu=0.5$} & & & & & & & & \\
$\phi=0.1$ & 1.000 & 650.205 & 1.000 & 73.118 & 1.000 & 38.826 & \textcolor{red}{16.747} & \textcolor{red}{1.883} \\
$\phi=0.3$ & 1.001 & 650.551 & 1.001 & 75.269 & 1.001 & 49.616 & \textcolor{red}{13.112} & \textcolor{red}{1.517} \\
\midrule
\rowcolor{yellow!30} \textbf{Gaussian, $\nu=1.5$} & & & & & & & & \\
$\phi=0.1$ & 1.000 & 684.511 & 1.000 & 67.510 & 1.000 & 45.505 & \textcolor{red}{15.043} & \textcolor{red}{1.484} \\
$\phi=0.3$ & 1.001 & 660.625 & 1.001 & 76.521 & 1.015 & 55.625 & \textcolor{red}{11.876} & \textcolor{red}{1.376} \\
\bottomrule
\end{tabular}
}
\end{table}

\clearpage
\clearpage
\begin{table}[ht]
\caption{Walltime quantiles (25\%, 50\%, 75\%) for MH, HMC, and SIVI with SIVI stopping criterion $\epsilon=10^{-4}$, and corresponding speedups (MH/SIVI, HMC/SIVI). Red values highlight median speedups.}
\label{tab:quantile_speedups_eps_e4}
\centering
\setlength{\tabcolsep}{5pt}    
\renewcommand{\arraystretch}{1.1}
\resizebox{\textwidth}{!}{
\begin{tabular}{l ccc ccc ccc ccc ccc}
\toprule
 & \multicolumn{3}{c}{\textbf{MH}} & \multicolumn{3}{c}{\textbf{HMC}} & \multicolumn{3}{c}{\textbf{SIVI}}
 & \multicolumn{3}{c}{\textbf{Speedup (MH/SIVI)}} & \multicolumn{3}{c}{\textbf{Speedup (HMC/SIVI)}}\\
\cmidrule(lr){2-4}\cmidrule(lr){5-7}\cmidrule(lr){8-10}\cmidrule(lr){11-13}\cmidrule(lr){14-16}
 & 25\% & 50\% & 75\% & 25\% & 50\% & 75\% & 25\% & 50\% & 75\% & 25\% & 50\% & 75\% & 25\% & 50\% & 75\% \\
\midrule

\rowcolor{yellow!30}\multicolumn{16}{l}{\textbf{NB}}\\
\rowcolor{yellow!15}\multicolumn{16}{l}{\textbf{$\nu=0.5$}}\\
$\phi=0.1$ &
3038.053 & 3205.406 & 3332.122 &
107.641  & 120.830  & 140.437  &
83.232   & 106.617   & 148.620   &
36.501   & \textcolor{red}{30.065} & 22.420 &
1.293    & \textcolor{red}{1.133}  & 0.945 \\
$\phi=0.3$ &
3205.791 & 3337.000 & 3431.428 &
117.791  & 126.542  & 153.687  &
84.014   & 126.849   & 182.183   &
38.158   & \textcolor{red}{26.307} & 18.835 &
1.402    & \textcolor{red}{0.998}  & 0.844 \\
\rowcolor{yellow!15}\multicolumn{16}{l}{\textbf{$\nu=1.5$}}\\
$\phi=0.1$ &
3041.410 & 3165.704 & 3265.998 &
123.971  & 144.374  & 167.344  &
99.140   & 146.179   & 188.494   &
30.678   & \textcolor{red}{21.656} & 17.327 &
1.250    & \textcolor{red}{0.988}  & 0.888 \\
$\phi=0.3$ &
3212.261 & 3385.727 & 3450.251 &
139.664  & 151.710  & 174.620  &
81.366   & 125.747   & 180.512   &
39.479   & \textcolor{red}{26.925} & 19.114 &
1.716    & \textcolor{red}{1.206}  & 0.967 \\
\midrule

\rowcolor{yellow!30}\multicolumn{16}{l}{\textbf{Gamma}}\\
\rowcolor{yellow!15}\multicolumn{16}{l}{\textbf{$\nu=0.5$}}\\
$\phi=0.1$ &
3435.886 & 3540.562 & 3684.057 &
96.562  & 112.361  & 127.280  &
36.227   & 47.815   & 54.909   &
94.844   & \textcolor{red}{74.047} & 67.094 &
2.665    & \textcolor{red}{2.350}  & 2.318 \\
$\phi=0.3$ &
3216.671 & 3337.640 & 3478.441 &
101.161  & 121.122  & 135.219  &
35.844   & 64.731   & 80.045   &
89.741   & \textcolor{red}{51.562} & 43.456 &
2.822    & \textcolor{red}{1.871}  & 1.689 \\
\rowcolor{yellow!15}\multicolumn{16}{l}{\textbf{$\nu=1.5$}}\\
$\phi=0.1$ &
3287.160 & 3479.287 & 3607.622 &
100.009  & 123.216  & 138.951  &
33.891   & 53.893   & 68.950   &
96.991   & \textcolor{red}{64.559} & 52.322 &
2.951    & \textcolor{red}{2.286}  & 2.015 \\
$\phi=0.3$ &
2975.072 & 3150.037 & 3357.022 &
128.366  & 146.534  & 165.024  &
51.430   & 76.504   & 119.755   &
57.847   & \textcolor{red}{41.175} & 28.032 &
2.496    & \textcolor{red}{1.915}  & 1.378 \\
\midrule

\rowcolor{yellow!30}\multicolumn{16}{l}{\textbf{Binary}}\\
\rowcolor{yellow!15}\multicolumn{16}{l}{\textbf{$\nu=0.5$}}\\
$\phi=0.1$ &
1102.228 & 1361.181 & 1534.596 &
208.746  & 220.937  & 232.028  &
36.550   & 49.865   & 69.015   &
30.157   & \textcolor{red}{27.297} & 22.236 &
5.711    & \textcolor{red}{4.431}  & 3.362 \\
$\phi=0.3$ &
1100.464 & 1371.999 & 1527.190 &
230.291  & 243.403  & 256.529  &
45.188   & 58.132   & 81.778   &
24.353   & \textcolor{red}{23.601} & 18.675 &
5.096    & \textcolor{red}{4.187}  & 3.137 \\
\rowcolor{yellow!15}\multicolumn{16}{l}{\textbf{$\nu=1.5$}}\\
$\phi=0.1$ &
1060.812 & 1158.577 & 1504.216 &
176.950  & 179.799  & 184.318  &
35.759   & 57.203   & 79.527   &
29.666   & \textcolor{red}{20.254} & 18.915 &
4.948    & \textcolor{red}{3.143}  & 2.318 \\
$\phi=0.3$ &
1048.858 & 1260.038 & 1503.935 &
179.101  & 182.640  & 189.276  &
39.700   & 56.906   & 78.108   &
26.420   & \textcolor{red}{22.142} & 19.254 &
4.511    & \textcolor{red}{3.210}  & 2.423 \\
\midrule

\rowcolor{yellow!30}\multicolumn{16}{l}{\textbf{Count}}\\
\rowcolor{yellow!15}\multicolumn{16}{l}{\textbf{$\nu=0.5$}}\\
$\phi=0.1$ &
1043.119 & 1199.590 & 1228.619 &
183.542  & 199.553  & 215.915  &
58.354   & 81.956   & 95.518   &
17.876   & \textcolor{red}{14.637} & 12.863 &
3.145    & \textcolor{red}{2.435}  & 2.260 \\
$\phi=0.3$ &
1045.639 & 1185.762 & 1239.167 &
169.421  & 184.509  & 199.852  &
63.100   & 89.074   & 108.949   &
16.571   & \textcolor{red}{13.312} & 11.374 &
2.685    & \textcolor{red}{2.071}  & 1.834 \\
\rowcolor{yellow!15}\multicolumn{16}{l}{\textbf{$\nu=1.5$}}\\
$\phi=0.1$ &
1089.074 & 1367.412 & 1515.595 &
159.561  & 180.962  & 202.700  &
55.787   & 80.965   & 106.868   &
19.522   & \textcolor{red}{16.889} & 14.182 &
2.860    & \textcolor{red}{2.235}  & 1.897 \\
$\phi=0.3$ &
1100.422 & 1461.657 & 1525.157 &
159.815  & 181.356  & 205.011  &
59.847   & 87.139   & 108.912   &
18.387   & \textcolor{red}{16.774} & 14.004 &
2.670    & \textcolor{red}{2.081}  & 1.882 \\
\midrule

\rowcolor{yellow!30}\multicolumn{16}{l}{\textbf{Gaussian}}\\
\rowcolor{yellow!15}\multicolumn{16}{l}{\textbf{$\nu=0.5$}}\\
$\phi=0.1$ &
560.900 & 678.775 & 692.222 &
61.827  & 68.491  & 84.581  &
28.449   & 36.796   & 49.246   &
19.716   & \textcolor{red}{18.447} & 14.056 &
2.173    & \textcolor{red}{1.861}  & 1.718 \\
$\phi=0.3$ &
559.822 & 679.687 & 692.743 &
66.164  & 72.995  & 84.423  &
35.377   & 46.384   & 61.083   &
15.824   & \textcolor{red}{14.653} & 11.341 &
1.870    & \textcolor{red}{1.574}  & 1.382 \\
\rowcolor{yellow!15}\multicolumn{16}{l}{\textbf{$\nu=1.5$}}\\
$\phi=0.1$ &
672.899 & 684.509 & 688.272 &
51.203  & 61.916  & 77.210  &
34.893   & 42.841   & 54.852   &
19.285   & \textcolor{red}{15.978} & 12.548 &
1.467    & \textcolor{red}{1.445}  & 1.408 \\
$\phi=0.3$ &
645.736 & 681.685 & 688.288 &
65.757  & 70.681  & 87.197  &
34.525   & 51.577   & 70.540   &
18.704   & \textcolor{red}{13.217} & 9.757 &
1.905    & \textcolor{red}{1.370}  & 1.236 \\
\bottomrule
\end{tabular}
}
\end{table}

\clearpage

\clearpage

\clearpage
\section{Sensitivity Analysis for the key tuning parameters in SIVI}

We provide sensitivity analysis for key tuning parameters here. We show sensitivity to (1) the number of basis functions m, (2) auxiliary sample size K, (3) stopping threshold, (4) maximum number of iterations, (5) different activation functions for MLP, and (6) different number of hidden layers for MLP.

\noindent \textbf{Auxiliary sample size}: In Algorithm~1 of the main manuscript, $K$ denotes the number of auxiliary samples drawn from the mixing distribution to approximate the semi-implicit variational density. A larger $K$ yields a tighter surrogate ELBO bound, and we set $K = 1{,}000$ in our experiments to balance approximation accuracy with computational cost. We conduct a sensitivity analysis using $K = 500$, $1{,}000$, and $2{,}000$ for the negative binomial model under smoothness parameter $\nu = 0.5$ and spatial range parameter $\phi = 0.1$. As shown in Table~\ref{tab:NB_auxiliary_K}, with $K = 1{,}000$, SIVI requires 51.198 seconds and yields an RMSPE of 3.401, corresponding to a 63.0-fold speedup over MH. Increasing to $K = 2{,}000$ raises the computation time to 67.544 seconds with a nearly identical RMSPE of 3.400, reducing the speedup to 47.8-fold. As expected, larger values of $K$ increase computation time with negligible improvement in RMSPE, confirming that $K = 1{,}000$ provides a reasonable trade-off between approximation accuracy and computational efficiency.

\noindent \textbf{Number of basis functions}: In the main manuscript, we use $m = 50$ basis functions. To assess sensitivity to this choice, we conduct an additional analysis with $m = 20, 50, 100$ for the negative binomial case, comparing Metropolis--Hastings (MH), Hamiltonian Monte Carlo (HMC), and Semi-Implicit Variational Inference (SIVI) in Table~\ref{tab:NB_basis_functions} under smoothness parameter $\nu = 0.5$ and spatial range parameter $\phi = 0.1$. When $m = 20$ MH requires 2916.634 seconds and yields an RMSPE of 3.621, whereas SIVI achieves a slightly lower RMSPE of 3.619 in only 40.766 seconds, corresponding to a 71.5-fold speedup. With fewer basis functions, the speedup of SIVI relative to MH is more pronounced, while the speedup relative to HMC is comparatively modest. As the number of basis functions increases, RMSPE decreases across all methods at the cost of greater computation time.

\noindent \textbf{Stopping Criteria}: In the main manuscript, we use a stopping criterion of $10^{-2}$. To assess sensitivity to this choice, we previously conducted an additional analysis using stopping criteria of $10^{-1}$, $10^{-2}$, $10^{-3}$, and $10^{-4}$ across the negative binomial, gamma, Gaussian, count, and binary data models; full results are reported in Supplementary Material Section~\textbf{S.6}. In Table~\ref{tab:NB_Gamma_stopping_criteria}, we present results for the gamma and negative binomial distributions under smoothness parameter $\nu = 0.5$ and spatial range parameter $\phi = 0.1$. As expected, smaller stopping criteria lead to slightly improved predictive accuracy (lower RMSPE) but require additional computation time. For example, the gamma model with stopping criterion $10^{-1}$ yields an RMSPE of 3.961, which is slightly higher than the RMSPE of 3.945 obtained by MH and HMC, but requires only 12.702 seconds, corresponding to a 337.9-fold speedup over MH. With the stopping criterion of $10^{-2}$ used in the main manuscript, SIVI produces an RMSPE of 3.945, matching MH and HMC, while still achieving a 145.2-fold speedup. When smaller stopping criteria such as $10^{-3}$ or $10^{-4}$ are used, computation time increases with negligible improvement in RMSPE. These results suggest that the stopping criterion should be chosen to balance predictive accuracy and computational efficiency, and that $10^{-2}$ provides a reasonable default in practice.

\begin{table}[ht]
    \caption{Comparison of RMSPE, walltime (in seconds), and speedup for MH, HMC, and SIVI under different numbers of auxiliary samples $K$ for the negative binomial model when smoothness parameter $\nu = 0.5$ and spatial range parameter $\phi = 0.1$. MH and HMC do not depend on $K$.}
    \label{tab:NB_auxiliary_K}
    \centering
    \small
    \begin{tabular}{llcccc}
    \toprule
    \rowcolor{yellow!30} \multicolumn{6}{l}{\textbf{negative binomial}} \\[2pt]
    Method & $K$ & RMSPE & Walltime & \multicolumn{2}{c}{Speedup over SIVI} \\
    \midrule
    MH   & ---  & 3.401 & 3225.591 & \multicolumn{2}{c}{---} \\
    HMC  & ---  & 3.401 & 128.912  & \multicolumn{2}{c}{---} \\
    \midrule
     & & & & MH/SIVI & HMC/SIVI \\
    \cmidrule(lr){5-6}
    SIVI & 500  & \textcolor{red}{\textbf{3.400}} & 46.232  & \textcolor{red}{\textbf{69.770}} & \textcolor{red}{\textbf{2.788}} \\
    SIVI & 1000 & \textcolor{red}{\textbf{3.401}} & 51.198  & \textcolor{red}{\textbf{63.002}} & \textcolor{red}{\textbf{2.518}} \\
    SIVI & 2000 & \textcolor{red}{\textbf{3.400}} & 67.544  & \textcolor{red}{\textbf{47.755}} & \textcolor{red}{\textbf{1.909}} \\
    \bottomrule
    \end{tabular}
\end{table}

    \begin{table}[ht]
    \caption{Comparison of RMSPE, walltime (in seconds), and speedup for MH, HMC, and SIVI under different numbers of basis functions for the negative binomial model when smoothness parameter $\nu = 0.5$ and spatial range parameter $\phi = 0.1$.}
    \label{tab:NB_basis_functions}
    \centering
    \resizebox{\textwidth}{!}{
    \begin{tabular}{lcccccccc}
    \toprule
     & \multicolumn{2}{c}{MH} & \multicolumn{2}{c}{HMC} & \multicolumn{2}{c}{SIVI} & \multicolumn{2}{c}{Speedup} \\
    \cmidrule(lr){2-3}\cmidrule(lr){4-5}\cmidrule(lr){6-7}\cmidrule(lr){8-9}
    Number of Basis Functions & RMSPE & Walltime & RMSPE & Walltime & RMSPE & Walltime & MH/SIVI & HMC/SIVI \\
    \midrule
    \rowcolor{yellow!30} \textbf{negative binomial} & & & & & & & & \\
    20  & 3.621 & 2916.634 & 3.621 & 101.329 & \textcolor{red}{\textbf{3.619}} & 40.766 & \textcolor{red}{\textbf{71.546}} & \textcolor{red}{\textbf{2.486}} \\
    50  & 3.401 & 3225.591 & 3.401 & 128.912 & \textcolor{red}{\textbf{3.401}} & 51.198 & \textcolor{red}{\textbf{63.002}} & \textcolor{red}{\textbf{2.518}} \\
    100 & 3.273 & 3483.409 & 3.275 & 231.710 & \textcolor{red}{\textbf{3.274}} & 57.420 & \textcolor{red}{\textbf{60.665}} & \textcolor{red}{\textbf{4.035}} \\
    \bottomrule
    \end{tabular}
    }
    \end{table}

    \begin{table}[ht]
    \caption{Comparison of RMSPE, walltime (in seconds), and speedup for MH, HMC, and SIVI under different stopping criteria for the negative binomial and gamma models. MH and HMC do not depend on the stopping criteria.}\label{tab:NB_Gamma_stopping_criteria}
    \centering
    \small
    \begin{tabular}{llcccc}
    \toprule
    \rowcolor{yellow!30} \multicolumn{6}{l}{\textbf{negative binomial}} \\[2pt]
    Method & Stopping Criteria & RMSPE & Walltime & \multicolumn{2}{c}{Speedup over SIVI} \\
    \midrule
    MH   & ---  & 3.473 & 3194.339 & \multicolumn{2}{c}{---} \\
    HMC  & ---  & 3.473 & 141.085  & \multicolumn{2}{c}{---} \\
    \midrule
     & & & & MH/SIVI & HMC/SIVI \\
    \cmidrule(lr){5-6}
    SIVI & $10^{-1}$ & \textcolor{red}{\textbf{3.475}} & 38.054  & \textcolor{red}{\textbf{83.943}}  & \textcolor{red}{\textbf{3.707}} \\
    SIVI & $10^{-2}$ & \textcolor{red}{\textbf{3.473}} & 71.909  & \textcolor{red}{\textbf{44.422}}  & \textcolor{red}{\textbf{1.962}} \\
    SIVI & $10^{-3}$ & \textcolor{red}{\textbf{3.473}} & 126.521 & \textcolor{red}{\textbf{25.247}}  & \textcolor{red}{\textbf{1.115}} \\
    SIVI & $10^{-4}$ & \textcolor{red}{\textbf{3.473}} & 124.081 & \textcolor{red}{\textbf{25.744}}  & \textcolor{red}{\textbf{1.137}} \\
    \midrule
    \rowcolor{yellow!30} \multicolumn{6}{l}{\textbf{Gamma}} \\[2pt]
    Method & Stopping Criteria & RMSPE & Walltime & \multicolumn{2}{c}{Speedup over SIVI} \\
    \midrule
    MH   & ---  & 3.945 & 4291.942 & \multicolumn{2}{c}{---} \\
    HMC  & ---  & 3.945 & 121.865  & \multicolumn{2}{c}{---} \\
    \midrule
     & & & & MH/SIVI & HMC/SIVI \\
    \cmidrule(lr){5-6}
    SIVI & $10^{-1}$ & \textcolor{red}{\textbf{3.961}} & 12.702  & \textcolor{red}{\textbf{337.907}} & \textcolor{red}{\textbf{9.594}} \\
    SIVI & $10^{-2}$ & \textcolor{red}{\textbf{3.945}} & 29.560  & \textcolor{red}{\textbf{145.193}} & \textcolor{red}{\textbf{4.123}} \\
    SIVI & $10^{-3}$ & \textcolor{red}{\textbf{3.944}} & 60.680  & \textcolor{red}{\textbf{70.731}}  & \textcolor{red}{\textbf{2.008}} \\
    SIVI & $10^{-4}$ & \textcolor{red}{\textbf{3.945}} & 48.140  & \textcolor{red}{\textbf{89.156}}  & \textcolor{red}{\textbf{2.531}} \\
    \bottomrule
    \end{tabular}
\end{table}

\clearpage
\noindent \textbf{Choices of $q(\btheta \mid \bpsi)$ and $q_\phi(\bpsi)$}: In the proposed SIVI framework, the variational distribution is constructed as a semi-implicit mixture in which $q(\btheta \mid \bpsi)$ is chosen as a simple explicit distribution and $q_\phi(\bpsi)$ is generated implicitly through a neural network transformation. In our implementation, we adopt a Gaussian form for $q(\btheta \mid \bpsi)$ with diagonal covariance because it provides a stable and computationally efficient conditional distribution while allowing flexibility through the implicit mixing distribution $q_\phi(\bpsi)$. In addition, both the explicit distribution $q(\btheta \mid \bpsi)$ and the implicit mixing distribution $q_\phi(\bpsi)$ must be reparameterizable in order to compute low-variance stochastic gradients of the ELBO. This requirement restricts the choice of variational families and motivates the use of a Gaussian conditional distribution together with a neural network transformation of a noise variable. The role of $q_\phi(\bpsi)$ is to enrich the expressiveness of the variational family. By generating $\bpsi$ through a neural network transformation of a noise variable, the resulting marginal distribution $h_\phi(\btheta)$ can represent complex non-Gaussian shapes while maintaining tractable optimization of the ELBO. In practice, we found that this combination provides a good balance between flexibility and computational stability. Sensitivity analyses for the maximum number of iterations and the neural network architecture are presented below.

\noindent \textbf{Sensitivity analysis for the maximum number of iterations}: Here, we present the sensitivity analysis for the maximum number of iterations; Table~\ref{tab:NB_max_iterations} compares the RMSPE, walltime (in seconds), and speedup for MH, HMC, and SIVI under different maximum iteration limits for the negative binomial model with smoothness parameter $\nu = 0.5$ and spatial range parameter $\phi = 0.1$. We use $5{,}000$ as the maximum number of iterations in our main manuscript and the results show that the maximum number of iterations has negligible effect on both RMSPE and walltime. This is because SIVI converges before reaching the iteration cap, with the loss function (negative ELBO) typically stabilizing well before the maximum number of iterations is reached.

\noindent \textbf{Activation function and number of hidden layers}: We employ a multilayer perceptron (MLP) architecture consisting of three hidden layers of sizes 40, 60, and 40, respectively, with a linear activation function for the output layer. For the negative binomial, binary, count, and Gaussian data models, we use the ReLU activation function for the hidden layers, whereas for the gamma data model, we use the tanh activation function. As shown in Table~\ref{tab:NB_activation_function}, the ReLU activation function yields an RMSPE of 3.401 in 51.198 seconds, corresponding to a 63.0-fold speedup over MH, while tanh produces an RMSPE of 3.410 with a 240.6-fold speedup. Although the tanh activation function achieves a substantially greater speedup due to faster convergence, its RMSPE is slightly higher than that of MH and HMC, suggesting that the choice of activation function for the hidden layers can affect predictive accuracy. We also evaluate the effect of network depth by implementing architectures with 6 and 9 hidden layers, as reported in Table~\ref{tab:NB_hidden_layers}. Compared to the 3-hidden-layer architecture, the 9-hidden-layer network achieves the same RMSPE of 3.401 but requires only 41.290 seconds, compared to 51.198 seconds for 3 hidden layers. Although increasing the number of hidden layers increases the per-iteration computational cost, the deeper network has greater expressive capacity, which enables it to reach the stopping criterion ($\text{tol} = 10^{-2}$) in fewer iterations. As a result, the total walltime decreases despite the added network depth, since the reduction in the number of iterations more than compensates for the increased cost per iteration.

\begin{table}[ht]
    \caption{Comparison of RMSPE, walltime (in seconds), and speedup for MH, HMC, and SIVI under different maximum numbers of iterations for the negative binomial model when smoothness parameter $\nu = 0.5$ and spatial range parameter $\phi = 0.1$. MH and HMC do not depend on the maximum number of iterations.}
    \label{tab:NB_max_iterations}
    \centering
    \small
    \begin{tabular}{llcccc}
    \toprule
    \rowcolor{yellow!30} \multicolumn{6}{l}{\textbf{NB}} \\[2pt]
    Method & Max Iterations & RMSPE & Walltime & \multicolumn{2}{c}{Speedup over SIVI} \\
    \midrule
    MH   & ---  & 3.401 & 3225.591 & \multicolumn{2}{c}{---} \\
    HMC  & ---  & 3.401 & 128.912  & \multicolumn{2}{c}{---} \\
    \midrule
     & & & & MH/SIVI & HMC/SIVI \\
    \cmidrule(lr){5-6}
    SIVI & 2{,}500  & \textcolor{red}{\textbf{3.401}} & 50.126  & \textcolor{red}{\textbf{64.349}} & \textcolor{red}{\textbf{2.572}} \\
    SIVI & 5{,}000  & \textcolor{red}{\textbf{3.401}} & 51.198  & \textcolor{red}{\textbf{63.002}} & \textcolor{red}{\textbf{2.518}} \\
    SIVI & 10{,}000 & \textcolor{red}{\textbf{3.401}} & 51.818  & \textcolor{red}{\textbf{62.248}} & \textcolor{red}{\textbf{2.488}} \\
    \bottomrule
    \end{tabular}
\end{table}

\begin{table}[ht]
    \caption{Comparison of RMSPE, walltime (in seconds), and speedup for MH, HMC, and SIVI under different activation functions for the negative binomial model when smoothness parameter $\nu = 0.5$ and spatial range parameter $\phi = 0.1$. MH and HMC do not depend on the activation function.}
    \label{tab:NB_activation_function}
    \centering
    \small
    \begin{tabular}{llcccc}
    \toprule
    \rowcolor{yellow!30} \multicolumn{6}{l}{\textbf{NB}} \\[2pt]
    Method & Activation Function & RMSPE & Walltime & \multicolumn{2}{c}{Speedup over SIVI} \\
    \midrule
    MH   & ---  & 3.401 & 3225.591 & \multicolumn{2}{c}{---} \\
    HMC  & ---  & 3.401 & 128.912  & \multicolumn{2}{c}{---} \\
    \midrule
     & & & & MH/SIVI & HMC/SIVI \\
    \cmidrule(lr){5-6}
    SIVI & ReLU & \textcolor{red}{\textbf{3.401}} & 51.198  & \textcolor{red}{\textbf{63.002}}  & \textcolor{red}{\textbf{2.518}} \\
    SIVI & Tanh & \textcolor{red}{\textbf{3.410}} & 13.408  & \textcolor{red}{\textbf{240.578}} & \textcolor{red}{\textbf{9.615}} \\
    \bottomrule
    \end{tabular}
\end{table}

\begin{table}[ht]
    \caption{Comparison of RMSPE, walltime (in seconds), and speedup for MH, HMC, and SIVI under different numbers of hidden layers for the negative binomial model when smoothness parameter $\nu = 0.5$ and spatial range parameter $\phi = 0.1$. MH and HMC do not depend on the number of hidden layers.}
    \label{tab:NB_hidden_layers}
    \centering
    \small
    \begin{tabular}{llcccc}
    \toprule
    \rowcolor{yellow!30} \multicolumn{6}{l}{\textbf{NB}} \\[2pt]
    Method & Hidden Layers & RMSPE & Walltime & \multicolumn{2}{c}{Speedup over SIVI} \\
    \midrule
    MH   & ---  & 3.401 & 3225.591 & \multicolumn{2}{c}{---} \\
    HMC  & ---  & 3.401 & 128.912  & \multicolumn{2}{c}{---} \\
    \midrule
     & & & & MH/SIVI & HMC/SIVI \\
    \cmidrule(lr){5-6}
    SIVI & 3 & \textcolor{red}{\textbf{3.401}} & 51.198  & \textcolor{red}{\textbf{63.002}} & \textcolor{red}{\textbf{2.518}} \\
    SIVI & 6 & \textcolor{red}{\textbf{3.402}} & 43.153  & \textcolor{red}{\textbf{74.749}} & \textcolor{red}{\textbf{2.987}} \\
    SIVI & 9 & \textcolor{red}{\textbf{3.401}} & 41.290  & \textcolor{red}{\textbf{78.120}} & \textcolor{red}{\textbf{3.122}} \\
    \bottomrule
    \end{tabular}
\end{table}

\clearpage
\section{Comparison of Variational Bayes work in spatial statistics}

    \begin{table}[!ht]
    \centering
    \caption{Comparison of VB spatial work}
    \label{tab:comparisonSIVI}
    \resizebox{\textwidth}{!}{%
    \begin{tabular}{l p{4.2cm} p{4.2cm} p{4.2cm}}
    \hline
    \textbf{Feature} & \textbf{\cite{lee2025scalable}} & \textbf{\cite{garneau2025semi}} & \tblue{\textbf{Proposed Method}} \\
    \hline
    
    VI Method &
    Mean-Field Variational Bayes (MFVB), Integrated non-factorized variational Bayes (INFVB) &
    Semi-Implicit Variational Inference (SIVI) &
    \tblue{Semi-Implicit Variational Inference (SIVI)} \\
    \hline
    
    Response Types &
    Gaussian, Bernoulli, Poisson &
    Gaussian, Poisson (small-scale) &
    \tblue{Gaussian, Bernoulli, Poisson, gamma, negative binomial} \\
    \hline
    
    Dispersion Parameter &
    Not supported &
    Not supported &
    \tblue{Supported (gamma, NB)} \\
    \hline
    
    Conjugacy Dependence &
    Relies on conjugate / near-conjugate structures (Laplace, Jaakkola approximations) &
    Not required (SIVI) &
    \tblue{Not required (SIVI)} \\
    \hline
    
    Modeling and Scalability&
    Full SGLMMs and Basis expansion (semi-parametric) &
    GP / NNGP prior &
    \tblue{Basis expansion (semi-parametric)} \\
    \hline
    
    
    Max.\ Simulation Size &
    $N=25$K &
    Poisson: $N{=}500$; Gaussian: $N{=}150$K &
    \tblue{$N{=}50$K for all response types}  \\
    \hline
    
    Real Data Application &
    Binary, Count data &
    Gaussian data; latent spatial effects integrated out, substantially reducing the number of estimable parameters &
    \tblue{gamma and negative binomial data with full latent effect estimation} \\
    \hline
    
    Key Limitation &
    Restricted to conjugate / near-conjugate likelihoods; gamma and NB not supported &
    Restricted to Gaussian and small-scale Poisson; NB, gamma, Bernoulli not supported &
    --- \\
    \hline
    
    \end{tabular}%
    }
    \end{table}

      In Table~\ref{tab:comparisonSIVI}, the method of \citet{lee2025scalable} is developed for spatial data arising from Gaussian, Bernoulli, and Poisson distributions and relies on conjugate or near-conjugate structures, using approximations such as the Laplace and Jaakkola methods. In contrast, our approach extends variational inference to non-conjugate spatial models where such approximations are not directly applicable, particularly for distributions such as the gamma and negative binomial with dispersion parameters. \citet{garneau2025semi} incorporates a Nearest Neighbor Gaussian Process (NNGP) with SIVI; however, their framework is restricted to Gaussian responses and small-scale Poisson simulations ($N = 500$). It does not consider other response types such as negative binomial models with dispersion parameters, gamma responses, or Bernoulli outcomes. In addition, their real-data application focuses on Gaussian spatial data, where the latent spatial random effects can be integrated out analytically, substantially reducing the number of parameters that must be estimated. In contrast, our approach accommodates gamma and negative binomial data while explicitly estimating the full latent spatial effects. Overall, the proposed framework broadens the applicability of SIVI to spatial generalized linear mixed models by supporting non-conjugate likelihoods, including gamma and negative binomial distributions with dispersion parameters, while maintaining computational scalability for large spatial datasets.

\subsection{Compare SIVI with hybrid MFVB}

\begin{figure}[H]
     \begin{center}
    \includegraphics[width=1.0\linewidth]{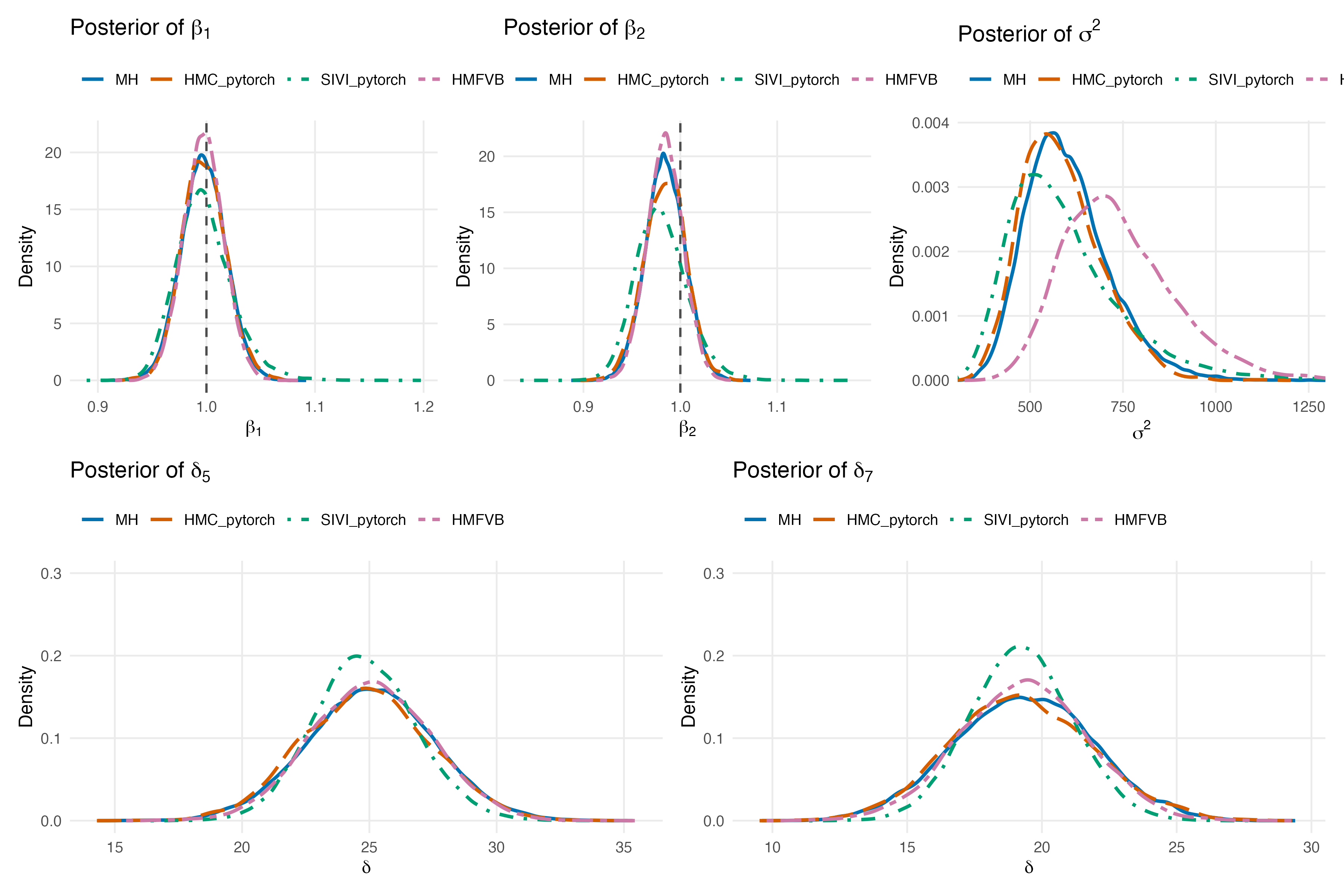}
    \caption{Posterior density estimates of selected model parameters under four inference methods 
        for the binary spatial model with $\nu = 1.5$ and $\phi = 0.1$: 
        Metropolis--Hastings (MH), Hamiltonian Monte Carlo (HMC), Semi-Implicit Variational 
        Inference (SIVI), and Hybrid Mean-Field Variational Bayes (HMFVB). 
        Panels display the regression coefficients $\beta_{1}$ and $\beta_{2}$, 
        the variance component $\sigma^{2}$, and two representative spatial basis coefficients 
        $\delta_{5}$ and $\delta_{7}$. Dashed vertical lines indicate true parameter values. 
        MH, HMC, and SIVI produce closely aligned posteriors for all parameters, 
        while HMFVB exhibits a noticeable discrepancy in $\sigma^{2}$.}\label{Fig:Binary_HMFVBcomparison}
    \end{center}
\end{figure}

\begin{figure}[H]
     \begin{center}
    \includegraphics[width=1.0\linewidth]{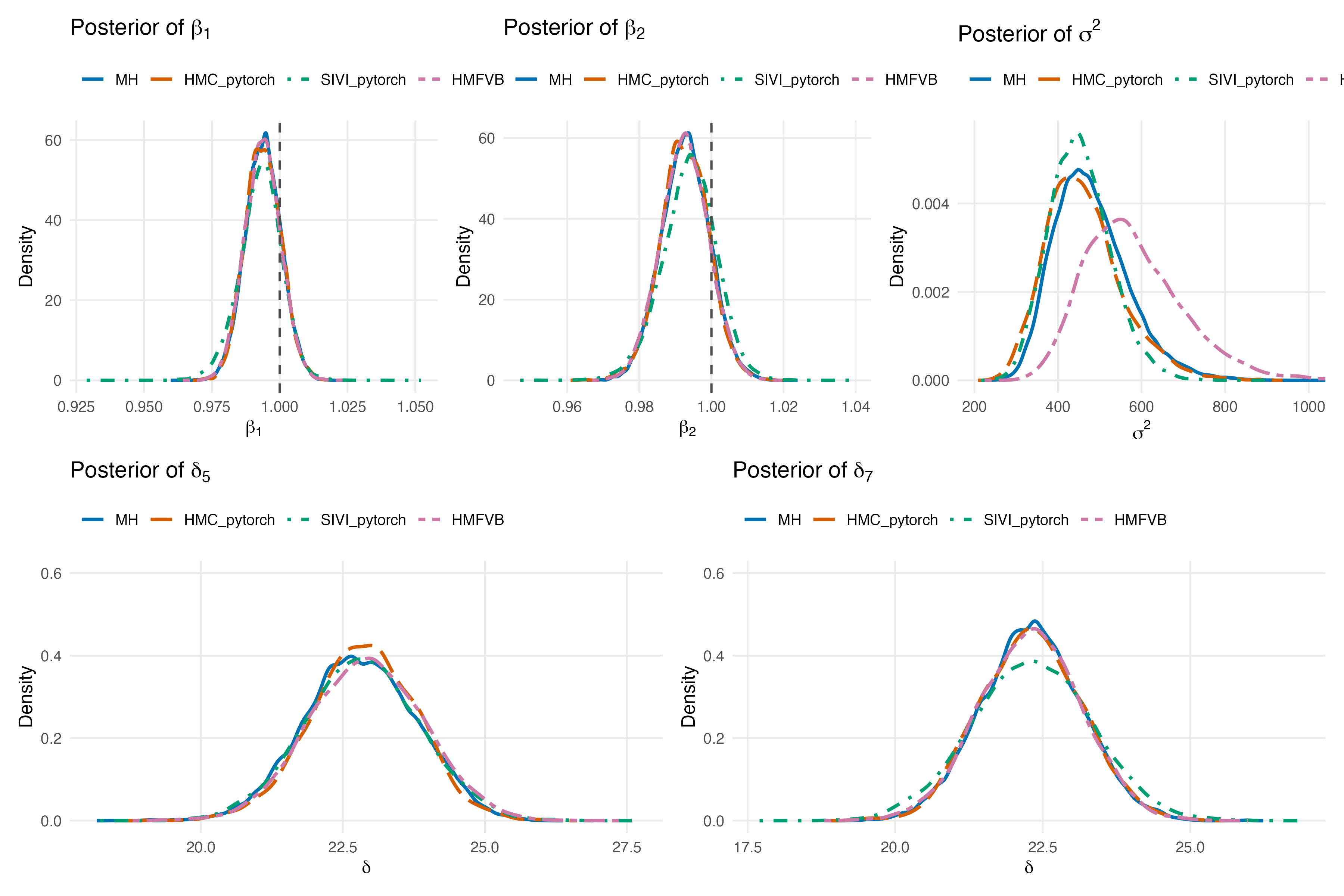}
    \caption{Posterior density estimates of selected model parameters under four inference methods 
        for the count spatial model with $\nu = 0.5$ and $\phi = 0.1$: 
        Metropolis--Hastings (MH), Hamiltonian Monte Carlo (HMC), Semi-Implicit Variational 
        Inference (SIVI), and Hybrid Mean-Field Variational Bayes (HMFVB). 
        Panels display the regression coefficients $\beta_{1}$ and $\beta_{2}$, 
        the variance component $\sigma^{2}$, and two representative spatial basis coefficients 
        $\delta_{5}$ and $\delta_{7}$. Dashed vertical lines indicate true parameter values. 
        MH, HMC, and SIVI produce closely aligned posteriors for all parameters, 
        while HMFVB exhibits a noticeable discrepancy in $\sigma^{2}$.}\label{Fig:Count_HMFVBcomparison}
    \end{center}
\end{figure}

We have conducted additional simulation studies comparing our proposed SIVI-based approach with a Hybrid Mean-Field Variational Bayes (HMFVB) method for both binary and count data settings. We refer to this approach as ``hybrid'' MFVB because standard mean-field variational Bayes requires conjugacy, which does not hold for the Bernoulli or Poisson likelihoods considered here~\citep{wu2018fast,lee2025scalable}. Specifically, the HMFVB implementation employs the Jaakkola--Jordan bound for binary data and a Laplace approximation for count data to achieve tractable closed-form updates~\citep{jaakkola1997variational, tierney1986accurate}. Tables~\ref{tab:binary_comparison} and~\ref{tab:count_comparison} summarize the predictive accuracy and computational cost across four inference methods: Metropolis--Hastings (MH), Hamiltonian Monte Carlo (HMC), SIVI, and HMFVB. For the binary setting (Table~\ref{tab:binary_comparison}), all four methods achieve an identical AUC of 0.77, yet HMFVB completes inference in only 0.37 seconds---a speedup of approximately 2{,}384$\times$ over MH and 501$\times$ over HMC. By comparison, SIVI achieves a speedup of roughly 32$\times$ over MH. Similarly, for the count setting (Table~\ref{tab:count_comparison}), all methods produce comparable RMSPE values of 1.43, while HMFVB is approximately 165$\times$ faster than MH, compared to SIVI's 15$\times$ speedup. However, despite the substantial computational advantage of HMFVB, Figures~\ref{Fig:Binary_HMFVBcomparison} and~\ref{Fig:Count_HMFVBcomparison} reveal an important limitation in posterior inference quality. While the posterior distributions of regression coefficients ($\beta_1$, $\beta_2$) and spatial basis coefficients ($\delta_5$, $\delta_7$) are closely aligned across all four methods, HMFVB exhibits noticeable discrepancies in the posterior of $\sigma^2$ for both binary and count data. This can be attributed to the mean-field independence assumption inherent in HMFVB, which fails to adequately capture posterior dependencies among model parameters in the Bayesian hierarchical framework~\citep{blei2006variational, blei2017variational}. In contrast, SIVI leverages a mixing distribution that can represent posterior dependence structure, yielding posterior estimates for $\sigma^2$ that are substantially closer to those obtained by MH and HMC. Furthermore, the HMFVB approach relies on model-specific analytic approximations---the Jaakkola--Jordan bound for binary data and the Laplace approximation for count data---that are not readily available for all likelihoods. In particular, for gamma and negative binomial response distributions, no such standard conjugate or analytic bounds exist, rendering the HMFVB approach inapplicable. Our proposed SIVI method, by contrast, does not depend on conjugacy~\citep{yin2018semi}. This generality enables SIVI to be applied to a broader class of spatial generalized linear mixed models, as demonstrated in our gamma and negative binomial simulation studies. In summary, while HMFVB offers superior computational speed, our proposed SIVI approach provides two key advantages: (1) more accurate posterior inference, particularly for variance components where posterior dependencies are important, and (2) broader applicability to non-conjugate likelihoods such as the gamma and negative binomial distributions, where HMFVB is not feasible.

\begin{table}[ht]
\centering
\caption{Comparison of AUC, walltime (in seconds), and speedup for MH, HMC, SIVI, and HMFVB for the Binary model when smoothness parameter $\nu = 1.5$ and spatial range parameter $\phi = 0.1$.}
\label{tab:binary_comparison}
\begin{tabular}{cc cc cc cc}
\hline
\multicolumn{2}{c}{MH} & \multicolumn{2}{c}{HMC} & \multicolumn{2}{c}{SIVI} & \multicolumn{2}{c}{HMFVB} \\
\cmidrule(lr){1-2} \cmidrule(lr){3-4} \cmidrule(lr){5-6} \cmidrule(lr){7-8}
AUC & Walltime & AUC & Walltime & AUC & Walltime & AUC & Walltime \\
\hline
0.77 & 891.72 & 0.77 & 187.38 & \textcolor{red}{0.77} & 28.19 & \textcolor{red}{0.77} & 0.37 \\
\hline
\multicolumn{8}{c}{\tblue{Speedup}} \\
\cmidrule(lr){1-2} \cmidrule(lr){3-4} \cmidrule(lr){5-6} \cmidrule(lr){7-8}
\multicolumn{2}{c}{MH/SIVI} & \multicolumn{2}{c}{HMC/SIVI} & \multicolumn{2}{c}{MH/HMFVB} & \multicolumn{2}{c}{HMC/HMFVB} \\
\hline
\multicolumn{2}{c}{\tblue{31.63}} & \multicolumn{2}{c}{\tblue{6.65}} & \multicolumn{2}{c}{\tblue{2384.28}} & \multicolumn{2}{c}{\tblue{501.02}} \\
\hline
\end{tabular}
\end{table}

\begin{table}[ht]
\centering
\caption{Comparison of RMSPE, walltime (in seconds), and speedup for MH, HMC, SIVI, and HMFVB for the Count model when smoothness parameter $\nu = 0.5$ and spatial range parameter $\phi = 0.1$.}
\label{tab:count_comparison}
\begin{tabular}{cc cc cc cc}
\hline
\multicolumn{2}{c}{MH} & \multicolumn{2}{c}{HMC} & \multicolumn{2}{c}{SIVI} & \multicolumn{2}{c}{HMFVB} \\
\cmidrule(lr){1-2} \cmidrule(lr){3-4} \cmidrule(lr){5-6} \cmidrule(lr){7-8}
RMSPE & Walltime & RMSPE & Walltime & RMSPE & Walltime & RMSPE & Walltime \\
\hline
1.43 & 1199.14 & 1.43 & 148.81 & \textcolor{red}{1.43} & 79.77 & \textcolor{red}{1.43} & 7.26 \\
\hline
\multicolumn{8}{c}{\tblue{Speedup}} \\
\cmidrule(lr){1-2} \cmidrule(lr){3-4} \cmidrule(lr){5-6} \cmidrule(lr){7-8}
\multicolumn{2}{c}{MH/SIVI} & \multicolumn{2}{c}{HMC/SIVI} & \multicolumn{2}{c}{MH/HMFVB} & \multicolumn{2}{c}{HMC/HMFVB} \\
\hline
\multicolumn{2}{c}{\tblue{15.03}} & \multicolumn{2}{c}{\tblue{1.87}} & \multicolumn{2}{c}{\tblue{165.26}} & \multicolumn{2}{c}{\tblue{20.51}} \\
\hline
\end{tabular}
\end{table}

\clearpage
\section{Subsampling method with SIVI}
        SIVI can be further extended to large-scale datasets using subsampling or mini-batching. Stochastic variational inference (SVI) allows one or more observations to be randomly subsampled from the dataset, and the evidence lower bound (ELBO) is then estimated iteratively using mini-batches until convergence \citep{hoffman2013stochastic, blei2017variational, zhang2018advances, bleistein1975asymptotic}. Algorithm~\ref{alg:siviwithsubsampling} outlines how subsampling can be incorporated into the SIVI framework, following the approach suggested by \citet{yin2018semi}. In this setting, the log-likelihood term in the ELBO is evaluated using a mini-batch of observations rather than the full dataset. For independent observations $\{\bZ_i\}_{i=1}^N$, the full log-likelihood is expressed as
    \[
    \log p(\bZ \mid \btheta) = \sum_{i=1}^N \log p(\bZ_i \mid \btheta).
    \]
    However, computing this summation at every iteration is computationally prohibitive when $N$ is large. To reduce the cost, we approximate the log-likelihood using a random mini-batch of size $\tblue{M} \ll N$:
    \[
    \widehat{\log p(\bZ \mid \btheta)} 
    = \tblue{\frac{N}{M}} \sum_{i \in \text{batch}} \log p(\bZ_i \mid \btheta).
    \]
    Here, the factor $\tblue{\frac{N}{M}}$ serves as an importance weight that rescales the contribution of the mini-batch to approximate the full data log-likelihood. This correction ensures that the resulting estimator is unbiased, i.e.,
    \[
    \mathbb{E}_{\text{batch}} \left[ \tblue{\frac{N}{M}} \sum_{i \in \text{batch}} \log p(\bZ_i \mid \btheta) \right] 
    = \sum_{i=1}^N \log p(\bZ_i \mid \btheta).
    \]
    If we were to use $\tfrac{M}{N}$ instead, the mini-batch contribution would systematically underestimate the total log-likelihood, thereby biasing the ELBO and leading to incorrect variational updates. Thus, the scaling factor $\tblue{\tfrac{N}{M}}$, highlighted in Algorithm~\ref{alg:siviwithsubsampling} in blue, is essential to ensure an unbiased estimate of the full log-likelihood and thereby maintain theoretical correctness in stochastic variational inference. This subsampling method could substantially broaden the scope of variational inference for large-scale spatial statistics.

  \begin{algorithm}
  \caption{Semi-Implicit Variational Inference (SIVI) when $\xi$ are fixed with \tblue{subsampling} \\ ~\citep{yin2018semi}}
  \label{alg:siviwithsubsampling}
  \begin{algorithmic}[1]
  \State \textbf{Input:} Data $\{\bZ_i\}_{1:N}$, joint likelihood $p(\bZ,\btheta)$, explicit variational distribution $q(\btheta \mid \bpsi)$ with reparameterization $\btheta=f(\epsilon,\bpsi)$, $\epsilon\sim p(\epsilon)$, implicit layer neural network $T_\phi(\epsilon)$ and source of randomness $q(\epsilon)$
    \State \textbf{Output:} Implicit variational parameter $\phi$ for the mixing distribution $q_\phi(\bpsi)$
    \Statex
  
  \State Initialize $\phi$ randomly
  \While{not converged}
  \State Set $\underline{L}_{K_t}=0$ and $\eta_t$ as step sizes, and $K_t\ge 0$ as a non-decreasing integer;
  \State Sample $\bpsi^{(k)}=T_\phi(\epsilon^{(k)})$, $\epsilon^{(k)}\sim q(\epsilon)$ for $k=1,\ldots,K_t$; 
  \tblue{take subsample $\mathbf{x}=\{\bZ_i\}_{i_{1}:i_{M}}$}
    \For{$j=1$ to $J$}
  \State Sample $\bpsi_j=T_\phi(\epsilon_j)$, $\epsilon_j\sim q(\epsilon)$
    \State Sample $\btheta_j=f(\tilde{\epsilon}_j,\bpsi_j)$, $\tilde{\epsilon}_j\sim p(\epsilon)$
\State $\underline{L}_{K_t} \gets \underline{L}_{K_t} 
   + \frac{1}{J} \left\{
   \begin{aligned}[t]
      & -\log\!\frac{1}{K_t+1}\Bigg[\sum_{k=1}^{K_t} q(\btheta_j \mid \bpsi^{(k)}) 
      + q(\btheta_j \mid \bpsi_j)\Bigg] \\
      &+ \tblue{\frac{N}{M}\log p(\bZ \mid \btheta_j)} + \log p(\btheta_j)
   \end{aligned}
   \right\}$      
      \EndFor
    \State $t \gets t+1$
      \State $\phi \gets \phi + \eta_t \nabla_\phi \underline{L}_{K_t}\!\left(\{\bpsi^{(k)}\}_{1,K_t}, \{\bpsi_j\}_{1,J}, \{\btheta_j\}_{1,J}\right)$
      \EndWhile
    \end{algorithmic}
    \end{algorithm}

\section{Larger scale analysis for MODIS data}\label{sec:largeMODIS}

\begin{figure}[H]
 \begin{center}
\includegraphics[width=0.8\linewidth]{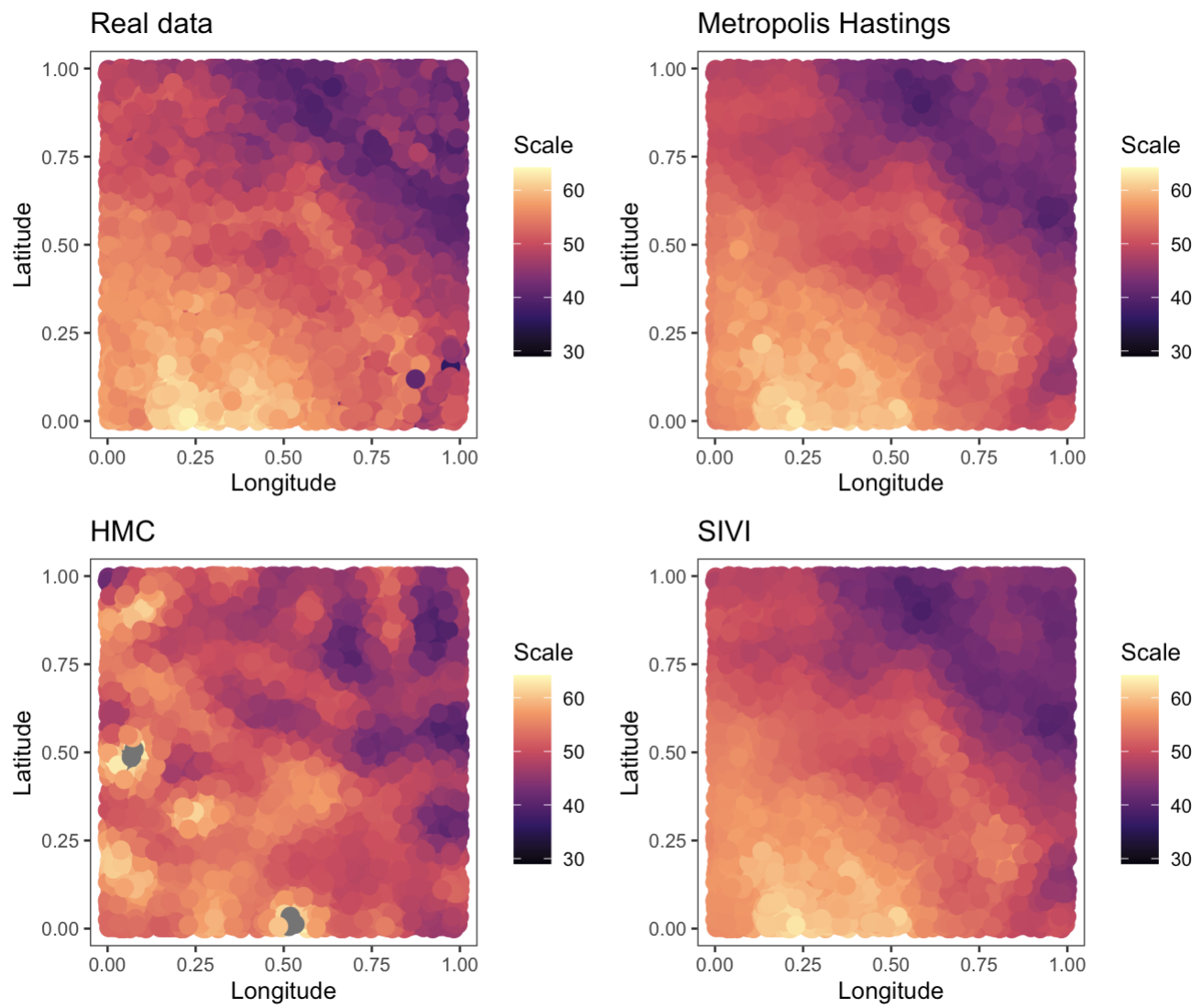}
\caption{Interpolation results from modeling the MODIS land surface temperature data. Implementations of Metropolis--Hastings (MH) MCMC, Hamiltonian Monte Carlo (HMC), and Semi-Implicit Variational Inference (SIVI) with basis function representations are shown for 100,000 spatial locations}\label{Fig:SIVILST100k}
\end{center}
\end{figure}

Our proposed method SIVI is not subject to scalability limitations. To demonstrate this, we apply SIVI to 100,000 locations. Figure~\ref{Fig:SIVILST100k} presents the interpolation results from modeling MODIS land surface temperature data at 100,000 locations, showing that SIVI produces results comparable to those of the Metropolis--Hastings method. These findings are consistent with those in Figure~4 in the main manuscript, which reports analogous results for 50,000 locations. In contrast, Hamiltonian Monte Carlo (HMC) exhibits inferior performance compared to both MH and our SIVI method in Figure~\ref{Fig:SIVILST100k}. Furthermore, Table~\ref{tab:SIVI_Difflocations} highlights that SIVI achieves greater computational gains as the data scale increases. Specifically, at 100,000 locations, SIVI attains an RMSPE of 1.36, outperforming MH (1.38) while achieving a $15.49\times$ speedup---a substantial improvement over the $10.46\times$ speedup observed at 50,000 locations. In addition, HMC requires more than 7 hours of computation and yields a higher RMSPE, indicating inferior performance relative to the other two methods.

\begin{table}[ht]
    \caption{Comparison of RMSPE, walltime (in seconds), and speedup for MH, HMC, and SIVI for the MODIS LST under different numbers of locations.}
    \label{tab:SIVI_Difflocations}
    \centering
    \resizebox{\textwidth}{!}{
    \begin{tabular}{lcccccccc}
    \toprule
     & \multicolumn{2}{c}{MH} & \multicolumn{2}{c}{HMC} & \multicolumn{2}{c}{SIVI} & \multicolumn{2}{c}{Speedup} \\
    \cmidrule(lr){2-3}\cmidrule(lr){4-5}\cmidrule(lr){6-7}\cmidrule(lr){8-9}
    Number of Locations & RMSPE & Walltime & RMSPE & Walltime & RMSPE & Walltime & MH/SIVI & HMC/SIVI \\
    \midrule
    50k & 1.42 & 2741.00 & 1.85 & 12826.73 & 1.43 & 262.05 & \textcolor{red}{\textbf{10.46}} & \textcolor{red}{\textbf{48.95}} \\
    100k & 1.38 & 5172.07 & 4.17 & 25507.07 & 1.36 & 333.95 & \textcolor{red}{\textbf{15.49}} & \textcolor{red}{\textbf{76.38}} \\
    \bottomrule
    \end{tabular}
    }
\end{table}

\clearpage
\section{Detailed discussion of surrogate ELBO}

Below we provide a detailed discussion of (i) the relationship between the surrogate lower bound and the true ELBO, (ii) the regularized surrogate that our algorithm actually optimizes, (iii) practical implications for finite $K$.

\textcolor{black}{\noindent \textbf{(i) Lower bound of the ELBO and the gap $\mathcal{L} - \underline{\mathcal{L}}$.}
  Recall that the semi-implicit variational density is defined hierarchically as
  $\btheta \sim q(\btheta \mid \bpsi)$, $\bpsi \sim q_\phi(\bpsi)$,
  with the marginal variational density
  \begin{equation}\label{eq:h_phi}
  h_\phi(\btheta)
  = \mathbb{E}_{\bpsi \sim q_\phi(\bpsi)}\bigl[q(\btheta \mid \bpsi)\bigr].
  \end{equation}
  The true ELBO under this marginal is
  $\mathcal{L} = \mathbb{E}_{\btheta \sim h_\phi(\btheta)}\bigl[\log \frac{p(\mathbf{Z},\btheta)}{h_\phi(\btheta)}\bigr]$.
  Because $q_\phi(\bpsi)$ is implicit, $h_\phi(\btheta)$ is generally intractable \citep{huszar2017variational, mohamed2016learning}, making $\mathcal{L}$ impossible to evaluate directly. The surrogate lower bound $\underline{\mathcal{L}}$ circumvents this by relying only on the explicit conditional $q(\btheta \mid \bpsi)$~\citep{yin2018semi}:
    \begin{equation}\label{eq:surrogate_lb}
  \underline{\mathcal{L}}
  \;=\; \mathbb{E}_{\bpsi \sim q_\phi(\bpsi)}\,
  \mathbb{E}_{\btheta \sim q(\btheta \mid \bpsi)}
  \Bigl[\log \tfrac{p(\mathbf{Z},\btheta)}{q(\btheta \mid \bpsi)}\Bigr]
  \;\leq\; \mathcal{L}.
  \end{equation}
  This inequality follows from Jensen's inequality and the convexity of the KL divergence \citep{cover1999elements}:
\begin{equation}
    \mathbb{E}_{\bpsi \sim q_\phi(\bpsi)}\,\mathrm{KL}\!\bigl(q(\btheta \mid \bpsi)\,\|\,p(\btheta \mid \mathbf{Z})\bigr)
    \;\geq\; \mathrm{KL}\!\bigl(h_\phi(\btheta)\,\|\,p(\btheta \mid \mathbf{Z})\bigr),
\end{equation}
which, after rearranging, yields $\underline{\mathcal{L}} \leq \mathcal{L}$.}
 
\textcolor{black}{The gap between the true ELBO and the surrogate can be derived explicitly. Expanding the definitions of $\mathcal{L}$ and $\underline{\mathcal{L}}$:
\begin{align}
    \mathcal{L} - \underline{\mathcal{L}}
    &= \mathbb{E}_{\btheta \sim h_\phi}\!\Big[\log \frac{p(\mathbf{Z},\btheta)}{h_\phi(\btheta)}\Big]
       - \mathbb{E}_{\bpsi}\,\mathbb{E}_{\btheta \sim q(\btheta|\bpsi)}\!\Big[\log \frac{p(\mathbf{Z},\btheta)}{q(\btheta|\bpsi)}\Big]. \label{eq:gap_step1}
\end{align}
Since sampling $\btheta \sim h_\phi(\btheta)$ is equivalent to first drawing $\bpsi \sim q_\phi(\bpsi)$ and then $\btheta \sim q(\btheta \mid \bpsi)$, by the law of iterated expectations we have
\begin{equation}\label{eq:iterated}
    \mathbb{E}_{\btheta \sim h_\phi}\!\big[\log p(\mathbf{Z},\btheta)\big]
    = \mathbb{E}_{\bpsi \sim q_\phi(\bpsi)}\,\mathbb{E}_{\btheta \sim q(\btheta|\bpsi)}\!\big[\log p(\mathbf{Z},\btheta)\big].
\end{equation}
Therefore the $\log p(\mathbf{Z},\btheta)$ terms cancel in \eqref{eq:gap_step1}, leaving only the entropy-related terms:
\begin{align}
    \mathcal{L} - \underline{\mathcal{L}}
    &= -\,\mathbb{E}_{\btheta \sim h_\phi}\!\big[\log h_\phi(\btheta)\big]
       + \mathbb{E}_{\bpsi}\,\mathbb{E}_{\btheta \sim q(\btheta|\bpsi)}\!\big[\log q(\btheta|\bpsi)\big]. \label{eq:gap_step2}
\end{align}
Applying the law of iterated expectations in reverse to the first term,
$\mathbb{E}_{\btheta \sim h_\phi}[\log h_\phi(\btheta)]
= \mathbb{E}_{\bpsi}\,\mathbb{E}_{\btheta \sim q(\btheta|\bpsi)}[\log h_\phi(\btheta)]$,
we can combine both terms under a single expectation:
\begin{align}
    \mathcal{L} - \underline{\mathcal{L}}
    &= \mathbb{E}_{\bpsi \sim q_\phi(\bpsi)}\,
       \mathbb{E}_{\btheta \sim q(\btheta|\bpsi)}\!\Big[\log \frac{q(\btheta|\bpsi)}{h_\phi(\btheta)}\Big] \nonumber \\[4pt]
    &= \mathbb{E}_{\bpsi \sim q_\phi(\bpsi)}\bigl[
        \mathrm{KL}\!\bigl(q(\btheta \mid \bpsi)\,\|\,h_\phi(\btheta)\bigr)
    \bigr]. \label{eq:gap}
\end{align}
This expression has a clear interpretation: the gap measures the \emph{average KL divergence} between each conditional component $q(\btheta \mid \bpsi)$ and the overall mixture $h_\phi(\btheta)$. }
 
\medskip
\textcolor{black}{\noindent \textbf{(ii) Regularized surrogate ELBO $\underline{\mathcal{L}}_K$ and finite-$K$ behavior.}
Directly maximizing $\underline{\mathcal{L}}$ can drive $q_\phi(\bpsi)$ toward a point mass, collapsing SIVI to standard VI \citep{yin2018semi}. To prevent this degeneracy, our algorithm (Algorithm~1) optimizes the \emph{regularized} surrogate ELBO:
\begin{equation}\label{eq:L_K}
    \underline{\mathcal{L}}_K
    \;=\; \underline{\mathcal{L}} + B_K,
\end{equation}
where the regularization term $B_K$ is defined as
\begin{equation}\label{eq:B_K}
    B_K
    \;=\; \mathbb{E}_{\bpsi, \bpsi^{(1)},\ldots,\bpsi^{(K)} \sim q_\phi(\bpsi)}\,
          \mathrm{KL}\!\bigl(q(\btheta \mid \bpsi)\,\|\,\tilde{h}_K(\btheta)\bigr),
\end{equation}
with the finite-sample mixture approximation
\begin{equation}\label{eq:h_tilde_K}
    \tilde{h}_K(\btheta)
    \;=\; \frac{q(\btheta \mid \bpsi) + \sum_{k=1}^{K} q(\btheta \mid \bpsi^{(k)})}{K+1}.
\end{equation}
Note that $B_K \geq 0$ with equality if and only if $K = 0$ or $q_\phi(\bpsi)$ is a point mass. For $K \geq 1$, maximizing $\underline{\mathcal{L}}_K$ encourages $B_K > 0$, thereby actively preventing degeneracy of the mixing distribution.}
 
\textcolor{black}{In practice, the surrogate ELBO computed in Step~9 of Algorithm~1 in the main manuscript takes the form:
\begin{equation}
    \underline{\mathcal{L}}_{K_t}
    \;=\; \frac{1}{J}\sum_{j=1}^{J}\Bigl\{
        -\log\frac{1}{K_t+1}\Bigl[\sum_{k=1}^{K_t} q(\btheta_j \mid \bpsi^{(k)}) + q(\btheta_j \mid \bpsi_j)\Bigr]
        + \log p(\mathbf{Z} \mid \btheta_j) + \log p(\btheta_j)
    \Bigr\},
\end{equation}
which is precisely $\underline{\mathcal{L}}_{K_t}$ evaluated via Monte Carlo with $J$ samples.}
 
\textcolor{black}{As $K \to \infty$, $\tilde{h}_K(\btheta) \to h_\phi(\btheta)$ by the law of large numbers, and thus
\begin{equation}
    \lim_{K \to \infty} B_K
    \;=\; \mathbb{E}_{\bpsi \sim q_\phi(\bpsi)}\,\mathrm{KL}\!\bigl(q(\btheta \mid \bpsi)\,\|\,h_\phi(\btheta)\bigr)
    \;=\; \mathcal{L} - \underline{\mathcal{L}},
\end{equation}
which, combined with \eqref{eq:L_K}, gives $\lim_{K\to\infty}\underline{\mathcal{L}}_K = \underline{\mathcal{L}} + B_K = \underline{\mathcal{L}} + (\mathcal{L} - \underline{\mathcal{L}}) =\mathcal{L}$ (Proposition~2 in \citet{yin2018semi}).}
 
\medskip
\textcolor{black}{\noindent \textbf{(iii) Practical implications for our framework.}
In our implementation (Section~4.1) in the main manuscript, we set $K = 1{,}000$ auxiliary samples with $J = 20$ Monte Carlo batches. Several considerations support this choice:}
 
\textcolor{black}{\begin{itemize}
    \item \textbf{Computational cost:} Each auxiliary sample $\bpsi^{(k)} = T_\phi(\epsilon^{(k)})$ requires only a forward pass through the MLP, which is computationally inexpensive relative to the likelihood evaluation over $N$ observations. Thus, increasing $K$ adds minimal overhead compared to the overall per-iteration cost.
    \item \textbf{Empirical evidence:} As shown in our simulation study (Section~5), SIVI with $K = 1{,}000$ achieves posterior distributions and predictive accuracy (RMSPE, AUC) comparable to MCMC methods (MH and HMC) across all 20 simulation scenarios, suggesting that the finite-$K$ gap does not materially affect inferential quality in our setting. In addition, Table~\ref{tab:NB_auxiliary_K} compares the predictive accuracy (RMSPE) and walltime across $K \in \{500, 1000, 2000\}$ for the negative binomial model with smoothness $\nu = 0.5$ and range $\phi = 0.1$. The RMSPE remains virtually unchanged across all three values of $K$ (3.400--3.401), closely matching MH (3.401) and HMC (3.401). By contrast, walltime increases from 46.2 seconds at $K = 500$ to 67.5 seconds at $K = 2{,}000$, reflecting the additional cost of evaluating more mixture components in $\tilde{h}_K$. Even at $K = 500$, the SIVI estimate is already indistinguishable from MCMC in terms of predictive performance, while achieving a speedup of roughly $70\times$ over MH and $2.8\times$ over HMC. These results confirm that, in our setting, the finite-$K$ discrepancy between $\underline{\mathcal{L}}_K$ and $\mathcal{L}$ has negligible impact on inferential quality, and $K = 1{,}000$ strikes a practical balance between approximation tightness and computational cost.    
  \end{itemize}}

\textcolor{black}{\noindent In summary, the discrepancy between the surrogate $\underline{\mathcal{L}}_K$ and the true ELBO $\mathcal{L}$ arises from approximating the intractable marginal $h_\phi(\btheta)$ with a finite mixture $\tilde{h}_K(\btheta)$ of $(K+1)$ components. Theoretically, this gap vanishes as $K \to \infty$ (Proposition~2 in \citet{yin2018semi}), and empirically, our results demonstrate that $K = 1{,}000$ is sufficient to achieve MCMC-comparable inference across all settings considered in this study, with predictive accuracy insensitive to the choice of $K$ (Table~\ref{tab:NB_auxiliary_K}).}

\clearpage
\section{Stability to Initialization and Convergence to Local Optima}

The optimization procedure may be sensitive to its starting values and could converge to a local optimum. To examine this, we conducted a sensitivity analysis in which the algorithm was initialized at different random starting values, and we compared the resulting posterior inference and predictive performance across these initializations.

\textcolor{black}{ As a representative case, we refit SIVI to a dataset with negative binomial responses generated from a latent Gaussian process model with smoothness $\nu = 0.5$ and spatial range parameter $\phi = 0.1$. We executed $R = 20$ independent runs that differ only in their random initialization. We hold the basis representation settings and tuning parameters fixed across runs. For each run, we record both the predictive accuracy (RMSPE) and the converged ELBO.}

\begin{figure}[H]
     \begin{center}
    \includegraphics[width=0.6\linewidth]{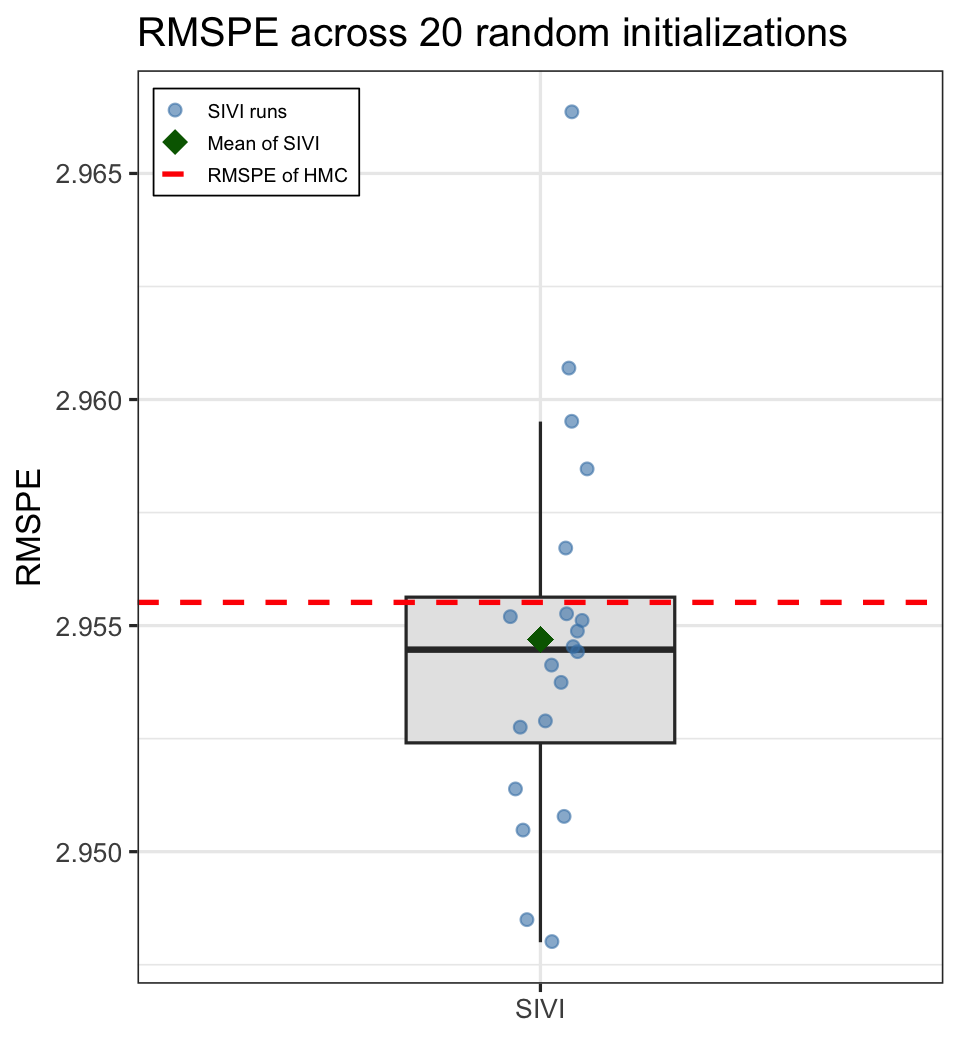}
    \caption{RMSPE across $R = 20$ SIVI runs that differ only in their random
initialization, on a negative binomial dataset generated from a latent Gaussian process model with smoothness $\nu = 0.5$ and spatial range parameter $\phi = 0.1$.
Blue points are individual runs, the green diamond is their mean, and the red
dashed line is the HMC benchmark ($\text{RMSPE} = 2.9555$). The runs cluster
tightly around the HMC value, indicating that predictive accuracy is insensitive
to initialization.}
\label{Fig:RMSPE_boxplot_ALL}
    \end{center}
    \end{figure}

\begin{figure}[H]
     \begin{center}
    \includegraphics[width=1.0\linewidth]{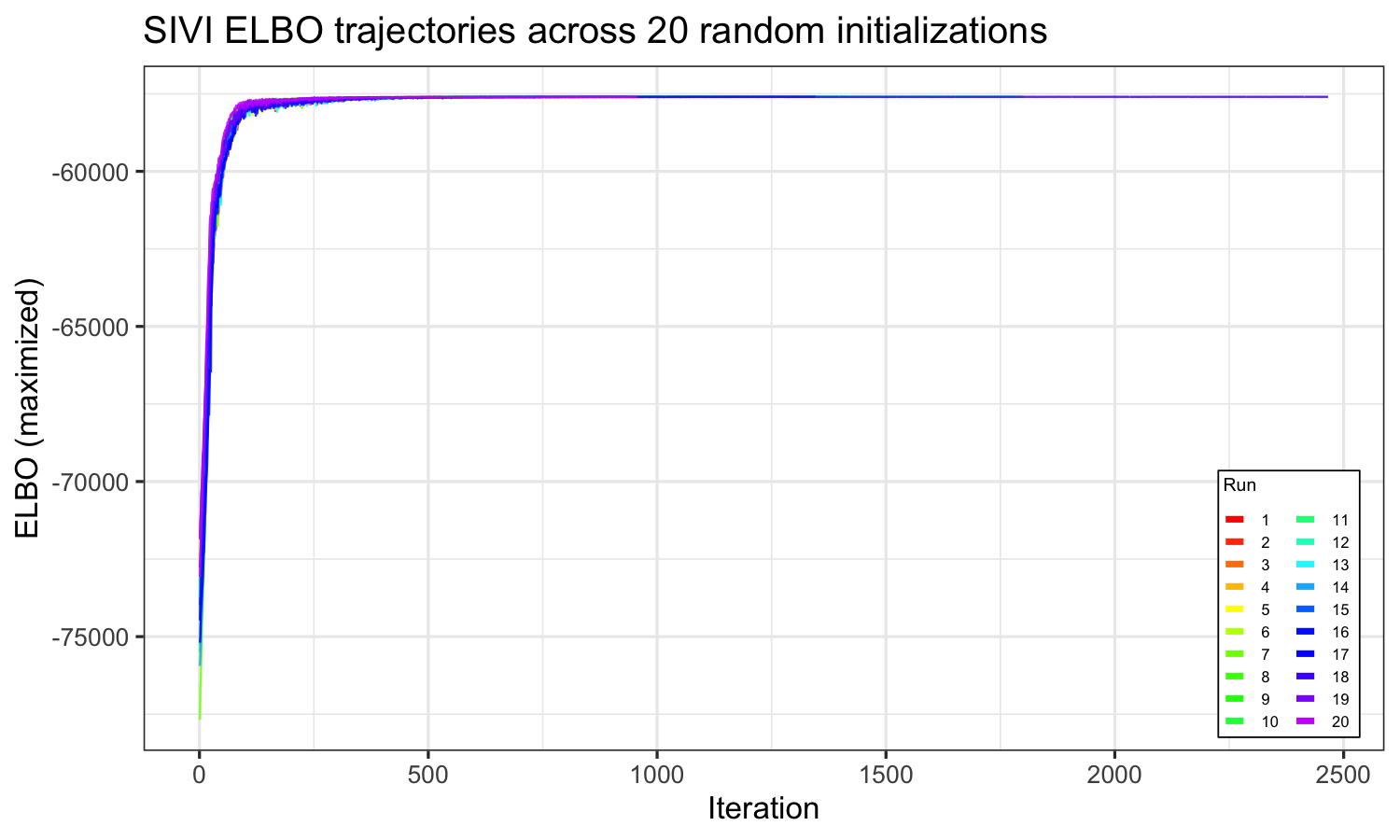}
    \caption{Optimization trajectories of the (maximized) ELBO for the $R = 20$ SIVI
runs in Figure~\ref{Fig:RMSPE_boxplot_ALL}. All runs level off at a common ELBO value.}\label{Fig:ELBO_trajectories_max}
    \end{center}
    \end{figure}

\textcolor{black}{%
The results are summarized in Table~\ref{tab:diffseed_stability} and
Figures~\ref{Fig:RMSPE_boxplot_ALL}--\ref{Fig:ELBO_trajectories_max}. Across all
$20$ initializations, the inference and predictive performance are nearly
identical. The RMSPE has a coefficient of variation of only $0.15\%$ (mean
$2.9547$, SD $0.0043$), and every run falls within $0.4\%$ of the HMC benchmark
($\text{RMSPE} = 2.9555$), shown as the red dashed line in
Figure~\ref{Fig:RMSPE_boxplot_ALL}. The converged ELBO is equally stable, varying
by only $0.02\%$ across runs (Table~\ref{tab:diffseed_stability}).
Figure~\ref{Fig:ELBO_trajectories_max} further shows that all $20$ runs flatten out at a similar ELBO level rather than at distinct local optima.
}

\textcolor{black}{%
These results indicate that the proposed SIVI approach shows no strong sensitivity to convergence to local optima, at least in this representative setting. Independent random initializations yield comparable RMSPE and ELBO values across runs, with predictive accuracy comparable to HMC.}

\begin{table}[ht]
\centering
\caption{Stability of SIVI to random initialization. The model was fit to a
negative binomial dataset generated from a latent Gaussian process model with smoothness $\nu = 0.5$ and spatial range parameter $\phi = 0.1$. We executed $R = 20$
independent runs that differ only in their random initialization, with the basis representation and all tuning parameters held fixed. For each run, we record the predictive accuracy (RMSPE) and the
converged ELBO (reported on the maximized scale, i.e., $-$loss). The table gives
the across-run mean, standard deviation (SD), and coefficient of variation
(CV~$=100\times\text{SD}/|\text{mean}|$). For reference, the HMC benchmark on the
same dataset gives $\text{RMSPE} = 2.9555$. The very small CVs (RMSPE
$0.15\%$, ELBO $0.02\%$) indicate that the optimization is empirically stable:
independent initializations converge to effectively the same solution, with
predictive accuracy within $0.4\%$ of HMC.}
\label{tab:diffseed_stability}
\begin{tabular}{lccc}
\toprule
Metric & Mean & SD & CV (\%) \\
\midrule
RMSPE                    & $2.9547$    & $0.0043$ & $0.1452$ \\
Final ELBO (maximized)   & $-57606.32$ & $12.18$  & $0.0211$ \\
\bottomrule
\end{tabular}
\end{table}

\end{document}